\documentclass[twocolumn]{aastex631}

\usepackage{xspace}

\newcommand{\thisgrba}{GRB~201015A\xspace}
\newcommand{\thisgrbb}{GRB~201216C\xspace}
\newcommand{\fermi}{{\em Fermi}\xspace}
\newcommand{\fermiT}{{T$_{0}$}\xspace}
\newcommand{\AstroSat}{{\em AstroSat}\xspace}
\newcommand{\kw}{{\em Konus}-Wind\xspace}
\newcommand{\keV}{{\rm ~keV}\xspace}

\newcommand{\s}{{\rm ~s}\xspace}
\newcommand{\swift}{{\em Swift}\xspace}
\newcommand{\swiftT}{{T$_{0}$}\xspace}
\newcommand{\tninty}{{T$_{90}$}\xspace}
\newcommand{\Ep}{$E_{\rm p}$\xspace}              
\newcommand{\sw}[1]{\texttt{#1}}
\graphicspath{{./}{figures/}}
\usepackage{color}
\usepackage{amsmath}
\usepackage[figuresright]{rotating}
\usepackage{threeparttable}
\usepackage{tablefootnote}
\usepackage{multirow,longtable,supertabular}
\usepackage{xcolor}

\usepackage{placeins}
\usepackage{chngcntr}

\shorttitle{VHE detected bursts: \thisgrba and \thisgrbb}
\shortauthors{A. K. Ror et al.}

\usepackage{multirow,longtable,supertabular}
\usepackage{float}
\graphicspath{{./}{figures/}}

\begin{document}

\title{Prompt emission and early optical afterglow of VHE detected \thisgrba and \thisgrbb: onset of the external forward shock}

\author[0000-0003-3164-8056]{Amit Kumar Ror$^\dagger$}
\affiliation{Aryabhatta Research Institute of Observational Sciences (ARIES), Manora Peak, Nainital-263002, India}
\author[0000-0003-4905-7801]{Rahul Gupta$^{*}$}
\affiliation{Aryabhatta Research Institute of Observational Sciences (ARIES), Manora Peak, Nainital-263002, India}
\affiliation{Department of Physics, Deen Dayal Upadhyaya Gorakhpur University, Gorakhpur-273009, India}
\author[0000-0003-3922-7416]{Martin Jel\'{\i}nek}
\affiliation{Astronomical Institute of the Czech Academy of Sciences (ASU-CAS), Fri\v{c}ova 298, 251 65 Ond\v{r}ejov, CZ}
\author[0000-0003-4905-7801]{Shashi Bhushan Pandey$^{**}$}
\affiliation{Aryabhatta Research Institute of Observational Sciences (ARIES), Manora Peak, Nainital-263002, India}
\author[0000-0003-4905-7801]{A. J. Castro-Tirado}
\affiliation{Instituto de Astrof\'isica de Andaluc\'ia (IAA-CSIC), Glorieta de la Astronom\'ia s/n, E-18008, Granada, Spain}
\affiliation{Unidad Asociada al CSIC Departamento de Ingenier\'ia de Sistemas y Autom\'atica, Escuela de Ingenier\'ias, Universidad de M\'alaga, C\/. Dr. Ortiz Ramos s\/n, E-29071, M\'alaga, Spain}
\author[0000-0002-7400-4608]{Y.-D. Hu}
\affiliation{Instituto de Astrof\'isica de Andaluc\'ia (IAA-CSIC), Glorieta de la Astronom\'ia s/n, E-18008, Granada, Spain}
\author[0000-0003-4905-7801]{Alžběta Maleňáková}
\affiliation{Astronomical Institute of the Czech Academy of Sciences (ASU-CAS), Fri\v{c}ova 298, 251 65 Ond\v{r}ejov, CZ}
\author[0000-0002-4147-2878]{Jan \v{S}trobl}
\affiliation{Astronomical Institute of the Czech Academy of Sciences (ASU-CAS), Fri\v{c}ova 298, 251 65 Ond\v{r}ejov, CZ}
\author[0000-0002-7978-7648]{Christina C. Thöne}
\affiliation{Astronomical Institute of the Czech Academy of Sciences (ASU-CAS), Fri\v{c}ova 298, 251 65 Ond\v{r}ejov, CZ}
\author[0000-0003-4905-7801]{René Hudec}
\affiliation{Astronomical Institute of the Czech Academy of Sciences (ASU-CAS), Fri\v{c}ova 298, 251 65 Ond\v{r}ejov, CZ}
\author[0000-0003-0035-651X]{Sergey Karpov}
\affiliation{Institute of Physics of the Czech Academy of Sciences (FZU AV ČR), Na Slovance 2, Praha 8, CZ}
\author[0000-0003-4905-7801]{Amit Kumar}
\affiliation{Aryabhatta Research Institute of Observational Sciences (ARIES), Manora Peak, Nainital-263002, India}
\affiliation{Department of Physics, University of Warwick, Gibbet Hill Road, Coventry CV4 7AL, UK}
\author[0000-0003-4905-7801]{A. Aryan}
\affiliation{Aryabhatta Research Institute of Observational Sciences (ARIES), Manora Peak, Nainital-263002, India}
\affiliation{Department of Physics, Deen Dayal Upadhyaya Gorakhpur University, Gorakhpur-273009, India}
\author[0000-0003-4905-7801]{S. R. Oates}
\affiliation{School of Physics and Astronomy \& Institute for Gravitational Wave Astronomy, University of Birmingham, B15 2TT, UK}
\author[0000-0003-4905-7801]{E. Fern\'andez-Garc\'ia}
\affiliation{Instituto de Astrof\'isica de Andaluc\'ia (IAA-CSIC), Glorieta de la Astronom\'ia s/n, E-18008, Granada, Spain}
\author[0000-0003-4905-7801]{C. P\'erez del Pulgar}
\affiliation{Unidad Asociada al CSIC Departamento de Ingenier\'ia de Sistemas y Autom\'atica, Escuela de Ingenier\'ias, Universidad de M\'alaga, C\/. Dr. Ortiz Ramos s\/n, E-29071, M\'alaga, Spain}
\author[0000-0003-4905-7801]{M.~D. Caballero-Garc\'{i}a}
\affiliation{Instituto de Astrof\'isica de Andaluc\'ia (IAA-CSIC), Glorieta de la Astronom\'ia s/n, E-18008, Granada, Spain}
\author[0000-0003-4905-7801]{A. Castell\'on}
\affiliation{Facultad de Ciencias, Universidad de M\'alaga, 29010 M\'alaga, Spain}
\author[0000-0003-4905-7801]{I. M. Carrasco-Garc\'{i}a}
\affiliation{Facultad de Ciencias, Universidad de M\'alaga, 29010 M\'alaga, Spain}
\author[0000-0003-4905-7801]{I. P\'erez-Garc\'{i}a}
\affiliation{Instituto de Astrof\'isica de Andaluc\'ia (IAA-CSIC), Glorieta de la Astronom\'ia s/n, E-18008, Granada, Spain}
\author[0000-0003-4905-7801]{A. J. Reina Terol}
\affiliation{Unidad Asociada al CSIC Departamento de Ingenier\'ia de Sistemas y Autom\'atica, Escuela de Ingenier\'ias, Universidad de M\'alaga, C\/. Dr. Ortiz Ramos s\/n, E-29071, M\'alaga, Spain}
\author[0000-0003-4905-7801]{F. Rendon}
\affiliation{Instituto de Astrof\'isica de Andaluc\'ia (IAA-CSIC), Glorieta de la Astronom\'ia s/n, E-18008, Granada, Spain}
\email{$^\dagger$amitror@aries.res.in}
\email{$^{*}$rahulbhu.c157@gmail.com}
\email{$^{**}$shashi@aries.res.in}

\begin{abstract}
We present a detailed prompt emission and early optical afterglow analysis of the two very high energy (VHE) detected bursts \thisgrba and \thisgrbb, and their comparison with a subset of similar bursts. Time-resolved spectral analysis of multi-structured \thisgrbb using the Bayesian binning algorithm revealed that during the entire duration of the burst, the low energy spectral index ($\alpha_{\rm pt}$) remained below the limit of the synchrotron line of death. However, statistically some of the bins supported the additional thermal component. Additionally, the evolution of spectral parameters showed that both peak energy (\Ep) and $\alpha_{\rm pt}$ tracked the flux. These results were further strengthened using the values of the physical parameters obtained by synchrotron modeling of the data. Our earliest optical observations of both bursts using FRAM-ORM and BOOTES robotic telescopes displayed a smooth bump in their early optical light curves, consistent with the onset of the afterglow due to synchrotron emission from an external forward shock. Using the observed optical peak, we constrained the initial bulk Lorentz factors of \thisgrba and \thisgrbb to $\Gamma_0$ = 204 and $\Gamma_0$ = 310, respectively. The present early optical observations are the earliest known observations constraining outflow parameters and our analysis indicate that VHE-detected bursts could have a diverse range of observed luminosity within the detectable redshift range of present VHE facilities. 
\end{abstract}

\section{Introduction} 
\label{sec:intro}
Gamma-ray bursts (GRBs) are sudden intense explosions of electromagnetic radiation in the $\keV$–MeV energy range, releasing energy ($E_{\rm \gamma, iso}$) in the range of $10^{51-54}$ erg. GRBs emit radiation across the electromagnetic spectrum broadly into two successive phases, i.e., prompt emission (generally in gamma rays or hard X-ray band) and afterglow emission (from radio to gamma rays), respectively \citep{zhang}. These cosmic stellar explosions are traditionally classified into long (LGRBs) and short (SGRBs) depending on their observed prompt emission duration \citep{1993ApJ...413L.101K}, which can be traced back to different progenitors. A massive star collapse under certain physical conditions, a ``collapsar'', is expected to be the progenitor of LGRBs \citep{1993ApJ...405..273W, 2003Natur.423..847H}. On the other hand, SGRBs are believed to originate from the merging of compact binaries like two neutron stars or a neutron star and a black hole \citep{Perna_2002, Abbott_2017}. However, recent discoveries of a few GRBs exhibiting hybrid properties from the collapse of massive stars \citep{ahumada2021discovery} as well as from binary mergers \citep{2022arXiv220410864R, 2022arXiv220903363} challenge current understanding and provide valuable clues about the physical nature of progenitors of GRBs.

There are many open questions related to the physics behind the prompt emission of GRBs, such as their jet compositions, emission mechanisms, and emission radii (see \citealt{zhang, 2015AdAst2015E..22P} for a review). To understand the jet compositions, there are two widely accepted scenarios: (1) baryonic-dominated hot fireball \citep{1990ApJ...365L..55S}, and (2) Poynting flux-dominated outflow \citep{2018NatAs...2...69Z}. In addition, there is also the possibility of a hybrid model that includes both components, Poynting flux outflow moving along with a hot fireball \citep{2015AdAst2015E..22P}. For the emission mechanisms, there are two widely accepted scenarios: (1) synchrotron emission from a cooling population of particles \citep{2020NatAs...4..174B}, and (2) thermal photospheric emission \citep{2015AdAst2015E..22P}. Since the beginning of the GRB spectroscopy, prompt spectral analysis of a larger sample of BATSE GRBs suggests non-thermal dominance, and spectra are described by a smoothly connected power-law empirical function, known as the \sw{Band} function \citep{1993ApJ...413..281B}. The low energy spectral index of the \sw{Band} function is widely used to understand the possible radiation process. However, some authors used the physical synchrotron modeling and suggested that it is a more accurate method to constrain the radiation physics rather than empirical fitting \citep{2020NatAs...4..174B, oganesyan2019}. 

Contrary to those predicted within the framework of the external forward and reverse shock models \citep{Sari_1998, 1999ApJ...520..641S}, the early time broadband afterglow emission of some GRBs exhibit deviations from power-law behavior such as flares, bumps, and plateaus largely attributed to effects from the unknown central engine. The early optical afterglow light curve initially rises until the blast wave reaches the self-similar phase, and the bulk Lorentz factor remains almost constant to its initial value. When the light curve is at its peak, the blast wave carries enough matter for the bulk Lorentz factor to begin progressively decreasing following the self-similar solution \citep{1976PhFl...19.1130B}, which makes the light curve decay, a process known as the onset of afterglow \citep{1999ApJ...520..641S}. \cite{1999ApJ...520..641S} explored the early optical afterglow emission and noted that the detection of the onset of afterglow can be utilized to calculate the initial bulk Lorentz factor of the relativistic outflow. Fast slewing (within a few minutes) of optical space (\swift Ultra-Violet and Optical Telescope) and ground-based telescopes (robotic telescopes such as MASTER, BOOTES, FRAM, etc.) are required to discover the onset of optical afterglow. \cite{2010ApJ...725.2209L} carried out an extensive search for the onset signatures in the early afterglow and identified twenty GRBs (through the literature search up to 2009) with an initial bump in their optical light curves. Additionally, they studied correlations among the characteristic parameters of the optical bump, like peak time, FWHM, isotropic energy, etc., and noted that most of the parameters have strong correlations with each other.

In recent years, detections of very high energy gamma-ray emissions during the afterglow phase by the imaging atmospheric Cherenkov telescopes such as Major Atmospheric Gamma Imaging Cherenkov (MAGIC, \citealt{2019Natur.575..455M}), High Energy Stereoscopic System (H.E.S.S., \citealt{2021Sci...372.1081H}), and  Large High Altitude Air Shower Observatory (LHAASO, \citealt{LHAASO_2022}) including the recently discovered GRB 221009A have challenged our understanding of afterglows and has opened a new window to explore this phase in more detail. A few general characteristics of VHE detected GRBs are tabulated in Table \ref{tab:gcn}. Generally, the traditional synchrotron emission can not explain the spectral energy distributions (SEDs) of VHE detected bursts. The double bump features observed in the broadband SEDs of GRB 180720B and GRB 190114C demand a synchrotron emission mechanism for the first bump and synchrotron-self Compton (SSC) to account for the second bump \citep{2019Natur.575..464A, 2019Natur.575..455M}. However, in the case of the nearby VHE detected GRB 190829A, the spectral index calculated using H.E.S.S. data was similar to the one of the synchrotron emission observed in the X-ray band, indicating that a single synchrotron component is sufficient to model the observed broadband spectrum from radio to VHE energies \citep{2021Sci...372.1081H}. In addition, in the case of GRB 190829A and GRB 190114C, dusty environments (large values of optical extinction in the host galaxies) have been observed, indicating a possible relationship between the occurrence of VHE emission and dusty environments \citep{2020A&A...633A..68D, 2021ApJ...917...95Z, 2022arXiv220513940G}. 

In this paper, we present a detailed prompt emission, and early optical afterglow analysis of two of the VHE detected bursts, \thisgrba \citep{2022icrc.confE.797S} and \thisgrbb \citep{2022icrc.confE.788F}. The article has been organized in the following sections: \S~\ref{multiwavelenghtobservations} presents multi-wavelength observations of \thisgrba and \thisgrbb, followed by the prompt and afterglow data analysis. The main results are given in \S~\ref{sec:result}, and followed by the discussion in \S~\ref{discussion}. Finally, the summary and conclusions are given in \S~\ref{summaryandconclusion}. Unless otherwise stated, all the uncertainties are expressed in $1\sigma$ throughout this article. We consider the Hubble parameter $H_0=71 $ $\rm km$ $\rm s^{-1} Mpc^{-1}$, density parameters $\Omega_{\Lambda}=0.73,$ and $\Omega_m=0.27$.

\begin{table*}
\centering
\scriptsize
\caption{Characteristic properties of the VHE detected GRBs (GRB 160821B has no firm detection, there is evidence of a signal of VHE emission at the level of 3$\sigma$) obtained from our analysis along with those published in several papers given below: (1) \cite{2019ApJ...883...48L}, (2) \cite{2019MNRAS.489.2104T}, (3) \cite{2019ApJ...885...29F}, (4) \cite{2020ApJ...903L..26H}, (5) \cite{2019Natur.575..455M, 2019Natur.575..459M}, (6) \cite{2019ApJ...879L..26F}, (7) \cite{2021ApJ...918...12F}, (8) \cite{2021A&A...646A..50H}, (9) \cite{Dichiara_2022}, (10) \cite{LHAASO_2022}, (11) \cite{2022arXiv221010673R}, (12) \cite{Ugarte-Postigo_2022}, (13) \cite{2022ATel15712....1V}.}
\label{tab:gcn}
\begin{center}
 \begin{tabular}{|p{2.45cm}|p{3cm}|c|c|c|c|p{1.6cm}|c|p{1.4cm}|}\hline 
 VHE detected GRBs 	&Light curve morphology&$z$	&\Ep (\keV)	&$E_{\rm \gamma, iso}$ (erg)	&$L_{\rm \gamma, iso} $ (erg/s)	&Ambient-medium	&X-ray flare&Supernova connection\\ [2ex]\hline
 160821B$^{(1,2)}$ 	&Short and bright pulse	&0.162	&84$\pm$19	    &2.10$\times$10$^{50}$	&2.00$\times$10$^{50}$	    &ISM   & No &kilonova\\[2ex] \hline
 180720B$^{(3,4)}$ 	&Single broad multi-peak light curve	    &0.654	&451$\pm$49	&6.00$\times$10$^{53}$	&1.80$\times$10$^{53}$	&ISM	&Yes&No\\\hline
 190114C$^{(5,6)}$ 	&Bright multi-peak pulse followed by soft tail emission &0.424	&926$\pm$17	&2.50$\times$10$^{53}$	&1.67$\times$10$^{53}$	&wind/ISM	& No&Yes\\\hline
 190829A$^{(7,8)}$ 	&Two-episodes with 40 s quiescent gap&0.0785	&11.5$\pm$0.4	&3.00$\times$10$^{50}$	&3.00$\times$10$^{49}$	&ISM    &Yes&Yes\\\hline
 201015A$^{\rm Present ~work}$	& Short overlapping pulses followed by soft and weak tail&0.426	&41$\pm$14	    &3.86$\times$10$^{51}$	&3.86$\times$10$^{50}$	&ISM	& No&Yes\\ \hline
 201216C$^{\rm Present ~work}$	& Complex multi-pulsed structured light curve &1.1	&352$\pm$12	&6.32$\times$10$^{53}$	&8.78$\times$10$^{52}$	&wind	& No&No\\ \hline
221009A$^{(9,10,11,12,13)}$& Two emission episodes followed by a long tail, extraordinarily brightness  & 0.151	&1060 $\pm$ 30	&$>$ 3 $\times$10$^{54}$	& $>$ 1 $\times$10$^{52}$	&wind	&No& Yes\\ \hline
 \end{tabular}
\end{center}
\end{table*}

\section{Multi-wavelength Observations and data reduction}
\label{multiwavelenghtobservations}

In the present section, we report the prompt emission, afterglow observations and data reduction of \thisgrba and \thisgrbb taken from space and ground-based facilities and are part of the present analysis.

\subsection{Prompt gamma-ray observations}

The Burst Alert Telescope (BAT, \citealt{2005SSRv..120..143B}) onboard the {\em Neil Gehrels Swift observatory} (henceforth \swift) triggered \thisgrba at 22:50:13.00 UT on 2020 October 15 (\swiftT) with a total duration of $\sim$ 10 s in the 15-350 \keV energy range. The burst was localized to RA, Dec = 354.310, +53.446 degrees (J2000) with a BAT uncertainty circle of 2.9$'$ \citep{2020GCN.28632....1D}. At the time of \swift detection, the Gamma-ray Burst Monitor (GBM, \citealt{Meegan}) onboard \fermi was observing the field of view of the GRB, but was unable to trigger on the burst. However, the burst was identified in GBM data through a targeted search from $\pm$ 30 s around the \swift-BAT trigger time.

The \fermi-GBM triggered \thisgrbb at 23:07:25.75 UT on 2020 December 16 (\fermiT, \citealt{2020GCN.29073....1M}). At \fermiT, the burst location was outside the field of view (FoV) of the Large Area Telescope (LAT) onboard \fermi (boresight angle is 93.0 degrees). It came into the FoV of LAT at $\sim$ \fermiT+3500 s and remained visible until $\sim$ \fermiT+5500 s. However, no significant GeV emission associated with \thisgrbb was observed during this time window \citep{2020GCN.29076....1B}. In addition to \fermi, the \swift-BAT triggered \thisgrbb at 23:07:31.00 UT on 2020 December 16 with a \tninty duration of 48.0 $\pm$ 16.0 s in the 15-350 \keV range \citep{2020GCN.29061....1B, 2020GCN.29080....1U}. The burst was localized at RA, Dec = 16.358, +16.537 degrees (J2000) with a BAT uncertainty circle of radius 3 arcmin \citep{2020GCN.29061....1B}. The prompt emission of \thisgrbb was also detected by \AstroSat CZT-Imager \citep{2020GCN.29074....1N} and \kw \citep{2020GCN.29084....1F}.

{\bf\swift-BAT data analysis:} We retrieved the \swift-BAT observation data for both the bursts from their \swift archive pages (\thisgrba, obsID: 01000452000 and \thisgrbb, obsID: 01013243000\footnote{\url{https://www.swift.ac.uk/archive/selectseq.php?tid=01013243&source=obs}}), respectively. We performed general processing of the BAT data given in the \swift-BAT Software Guide\footnote{\url{https://swift.gsfc.nasa.gov/analysis/bat_swguide_v6_3.pdf}}. Further, we analyzed the standard temporal and spectral BAT data products following the BAT data analysis methods given in \cite{2021MNRAS.505.4086G, 2022arXiv220507790C}. In this work, we have utilized the Multi-Mission Maximum Likelihood framework \citep[\sw{3ML}\footnote{\url{https://threeml.readthedocs.io/en/latest/}}]{2015arXiv150708343V} software for the time-averaged spectral analysis. We considered the BAT spectrum over the 15-2022MNRAS.511.1694G150 \keV energy range for the spectral analysis of the BAT data. 

{\bf \fermi-GBM data analysis:} For the \fermi-GBM data analysis of \thisgrbb, we downloaded the GBM time-tagged events (TTE) mode data from the \fermi-GBM Burst Catalog\footnote{\url{https://heasarc.gsfc.nasa.gov/W3Browse/fermi/fermigbrst.html}}. We selected the three brightest sodium iodide (NaI) and brightest bismuth germanate (BGO) detectors with minimum observing angles for temporal and spectral analysis of GBM data. We followed the methodology given in \cite{2022arXiv220507790C, 2022MNRAS.511.1694G} for the spectral and temporal analysis of \fermi-GBM data. \\

In this work, we have utilized \sw{3ML} \citep{2015arXiv150708343V} software for the time-averaged and time-resolved\footnote{For \thisgrba, we could not perform the time-resolved spectral analysis as \fermi (having broad spectral coverage) was unable to trigger on the burst.} spectral analysis. We considered the GBM spectrum over 8-900 \keV (NaI detectors) and 200-40000 \keV (BGO detector) energy ranges. We initially used two empirical models, \sw{Band} and \sw{Cutoff power-law} (CPL), to fit the time-averaged spectrum. To find the best fit model among the empirical models, we compared the deviance information criterion (DIC) values and chose the best model with the least DIC value ($\Delta$ DIC $<$ -10). Furthermore, we checked whether the addition of a thermal component (\sw{Blackbody}) with \sw{Band}, or \sw{Cutoff power-law} in the spectrum improves the fitting or not. We have applied the following criterion to determine if the spectrum has a thermal component:\\ \\
$\Delta$DIC$_{\rm Band/CPL} = $DIC$_{\rm Band+BB/CPL+ BB} - $DIC$_{\rm Band/CPL}$\\ \\
The negative value of $\Delta$DIC suggests an improvement in the spectral fit. If the difference in DIC is less than -10, it shows the existence of a significant amount of thermal component in the spectrum. 

\cite{2020NatAs...4..174B} suggested that the empirical models may not be able to reveal the emission process of GRBs. They found that GRB spectra can be well modeled with a physical synchrotron model even if the low-energy spectral index of the same spectra exceeds the synchrotron line-of-death if modeled using a \sw{Band} model. Therefore, in addition to empirical models, we have also used the physical synchrotron to model the emission mechanism of \thisgrbb. In this article, we have used the same physical synchrotron model \sw{pynchrotron}\footnote{\url{https://github.com/grburgess/pynchrotron}} (synchrotron emission from cooling electrons) for the spectral modeling of \thisgrbb used in \cite{2020NatAs...4..174B}. The synchrotron model is based on a comprehensive electron acceleration mechanism assumption. According to that, electrons are continuously injected into a power-law spectrum: N($\gamma$) $\propto$ $\gamma^{-p}$ with $\gamma_{inj} \leq$ $\gamma \leq$ $\gamma_{max}$, where $p$ is the spectral index of the injected electrons, $\gamma_{inj}$ and $\gamma_{max}$ are the lower and upper boundary of the injected electron spectrum, respectively. The cool synchrotron physical model details have been explained in \cite{2020NatAs...4..174B}. This model is characterized by the six physical parameters: 1. Magnetic field strength (B), 2. Spectral index of electrons ($p$), 3. $\gamma_{inj}$, 4. $\gamma_{max}$, 5. Characteristic Lorentz factor corresponds to the electrons' cooling time ($\gamma_{cool}$), and 6. Bulk Lorentz factor ($\gamma_{bulk}$). 

During the spectral modeling, we fixed some of the physical parameters. We have fixed $\gamma_{inj}$ = 10$^{5}$ due to a strong degeneracy between magnetic field strength and $\gamma_{inj}$. The bulk Lorentz factor ($\gamma_{bulk}$) is fixed at {513}, obtained using the onset of optical afterglow (see \S~\ref{onset_afterglow}). Furthermore, we have fixed $\gamma_{max}$ = 10$^{8}$ since the fast cooling synchrotron physical model does not fit the prompt spectrum of GRBs well \citep{2020NatAs...4..174B}. 

{\bf Time-resolved spectral analysis of \thisgrbb:}
The time-resolved spectral analysis of the prompt emission using broad spectral coverage GRB detectors such as \fermi-GBM has been used as a promising tool to investigate the emission mechanism and to study the correlations between the spectral parameters of GRBs.
To constrain the radiation process and spectral evolution of \thisgrbb, we performed the time-resolved spectral analysis using \sw{3ML} software with the same number of detectors used for the time-averaged spectral analysis. \sw{3ML} provides four possible methods to bin the light curves of GRBs. 1. Constant cadence (Cc) binning. All bins are equally spaced with the initially chosen time-width $\Delta$T. One disadvantage is that the spectral shape may vary slower or faster than the specified cadence. 2. Signal-to-noise (S/N) binning: here, we predefined the S/N for each bin, which ensures enough photons in each time bin, but it may fail to recover the intrinsic spectral evolution behavior of GRBs, 3. Knuth binning, and 4. Bayesian Blocks binning, in this case, the time bins have unequal widths and a variable signal-to-noise ratio. \cite{2014MNRAS.445.2589B} studied all these methods and suggested that the Bayesian Block binning method is the best time-slicing method to correctly obtain the intrinsic spectral evolution of GRBs. However, this method has one limitation: some bins might not have enough photons needed for correct spectral modeling. In the case of \thisgrbb, we initially performed the Bayesian Block binning on the brightest GBM detector (energy range 8–900 \keV) considering the false alarm probability P = 0.01 \citep{Scargle_2013}, and the other GBM detectors used the same temporal binning information. This results in 37 time-sliced spectra for time-resolved analysis of \thisgrbb. Further, we measured each spectrum's statistical significance (S, \citealt{2018ApJS..236...17V}) to ensure enough photon counts for spectral analysis and considered temporal bins with statistical significance greater than 25. This results in 27 time-sliced (Bayesian Block) spectra with S $>$ 25. 

For the time-resolved spectral analysis of \thisgrbb, we initially used the empirical \sw{Band} and \sw{Cutoff power-law} models and then refitted each spectrum after adding a thermal component to the empirical models. Furthermore, we fitted each spectrum using the physical slow cooling \sw{Synchrotron} model. We have used the Bayesian fitting method for the spectral fitting, and the sampler is set to the \sw{multi-nest} with 10,000 iterations. The spectral parameters, along with the associated errors, obtained using the time-resolved spectral analysis of \thisgrbb using the empirical and physical models are given in Table \ref{TRS:cpl_band}, \ref{TRS:Band} and \ref{TRS:Synchrotron} of the appendix.

\subsection{Afterglow observations}

The afterglows of \thisgrba\footnote{\url{https://gcn.gsfc.nasa.gov/other/201015A.gcn3}} and \thisgrbb\footnote{\url{https://gcn.gsfc.nasa.gov/other/201216C.gcn3}} were discovered from VHE to radio wavelengths by various observational facilities over the globe, including our earliest optical afterglow observations using the robotic telescopes, FRAM-ORM \citep{2020GCN.28664....1J, 2020GCN.29070....1J} and BOOTES \citep{2020GCN.28645....1H}. The redshifts of \thisgrba ($z$ = 0.426) and \thisgrbb ($z$ = 1.1) were measured using spectroscopic observations (emission features) of 10.4m GTC \citep{2020GCN.28649....1D} and the Very Large Telescope \citep{2020GCN.29077....1V}, respectively. The redshift measurement of \thisgrbb places the burst as the most distant known source associated with VHE emission (see Table \ref{tab:gcn}).

\subsubsection{X-ray afterglows}

For \thisgrba, \swift could not slew until \swiftT+51.6 minutes due to an observing constraint \citep{2020GCN.28632....1D}. The \swift X-ray telescope (XRT, \citealt{2005SSRv..120..165B}) began observations of \thisgrba at 23:43:47.2 UT on 2020 October 15, $\sim$ 3214.1 s post burst. \swift-XRT detected an uncatalogued fading X-ray source at the following location: RA, Dec = 354.32067, +53.41460 degrees (J2000) with an uncertainty radius of 3.8$^{"}$ \citep{2020GCN.28635....1K}. \swift-XRT observed this source up to $\sim{1.8}$ $\times ~10^{3}$ ks after the BAT detection. All observations were obtained in Photon Counting (PC) mode.

For \thisgrbb, \swift-XRT began observations at 23:56:58.5 UT on 2020 Dec 16, $\sim$ 2966.8 s post burst. \swift-XRT detected a new fading X-ray source at the following location: RA, Dec = 16.37114, +16.51659 degrees (J2000) with an uncertainty circle of 3.5$^{"}$ \citep{2020GCN.29064....1C}. \swift-XRT observed this source up to 2.2 $\times ~10^{3}$ ks after the initial detection. Window timing (WT) mode was used for the first $\sim$ 25 s of observations, and the remaining observations were obtained in PC mode.

{\bf X-ray afterglow data analysis:}  
In this work, we obtained the \swift-XRT data products from the XRT repository provided by the University of Leicester \citep{eva07, eva09}. We modeled the X-ray afterglow light curves of both the bursts using power-law and broken power-law empirical models to constrain their decay rates. On the other hand, to constrain the spectral indices, we modeled the XRT spectra of both the bursts in the energy range of 0.3-10 \keV using the X-ray Spectral Fitting Package \citep[\sw{XSPEC};][]{1996ASPC..101...17A}. We fixed the Galactic hydrogen column density to be $\rm NH_{\rm Gal}$= $3.60 \times 10^{21}$, and $5.04 \times 10^{20}$ cm$^{-2}$ for \thisgrba and \thisgrbb, respectively \citep{2013MNRAS.431..394W}. A more detailed method for the standard temporal and spectral XRT analysis is given in \cite{2021MNRAS.505.4086G}.

\subsubsection{Optical afterglows of \thisgrba and \thisgrbb}
{\bf \thisgrba:} The optical afterglow of \thisgrba was first reported by the MASTER robotic telescope at the position RA= 23:37:16.42, DEC= +53:24:55.8 \citep{2020GCN.28633....1L}.
In this paper, we present our optical observations from the 25cm FRAM-ORM \citep{2020GCN.28664....1J}, Burst Observer and Optical Transient Exploring System (BOOTES) \citep{2020GCN.28645....1H}, and 3.6m Devasthal Optical Telescope (DOT), along with additional optical data from the GCN circulars. Our photometric observations are listed in Table \ref{table:15A_optical} of the appendix.

{\bf BOOTES-network:} The BOOTES-network followed \thisgrba with three robotic telescopes at BOOTES-1 (INTA-CEDEA) station in Mazagon (Huelva, Spain) and BOOTES-2/TELMA station in La Mayora (Malaga, Spain). The BOOTES-1B performed one epoch of early observations at 22:50:48 on 2020 October 15, and the afterglow is clearly seen in the images. The BOOTES-1A performed two epoch observations: the first was the quick follow-up to the trigger and the second was the late follow-up on the next day. The BOOTES-2/TELMA also followed this event on 2020-10-16T22:15:13. The afterglow was not visible in both BOOTES-1A's and BOOTES-2's images.\\ 
\begin{figure*}
\centering
\includegraphics[scale=0.31]{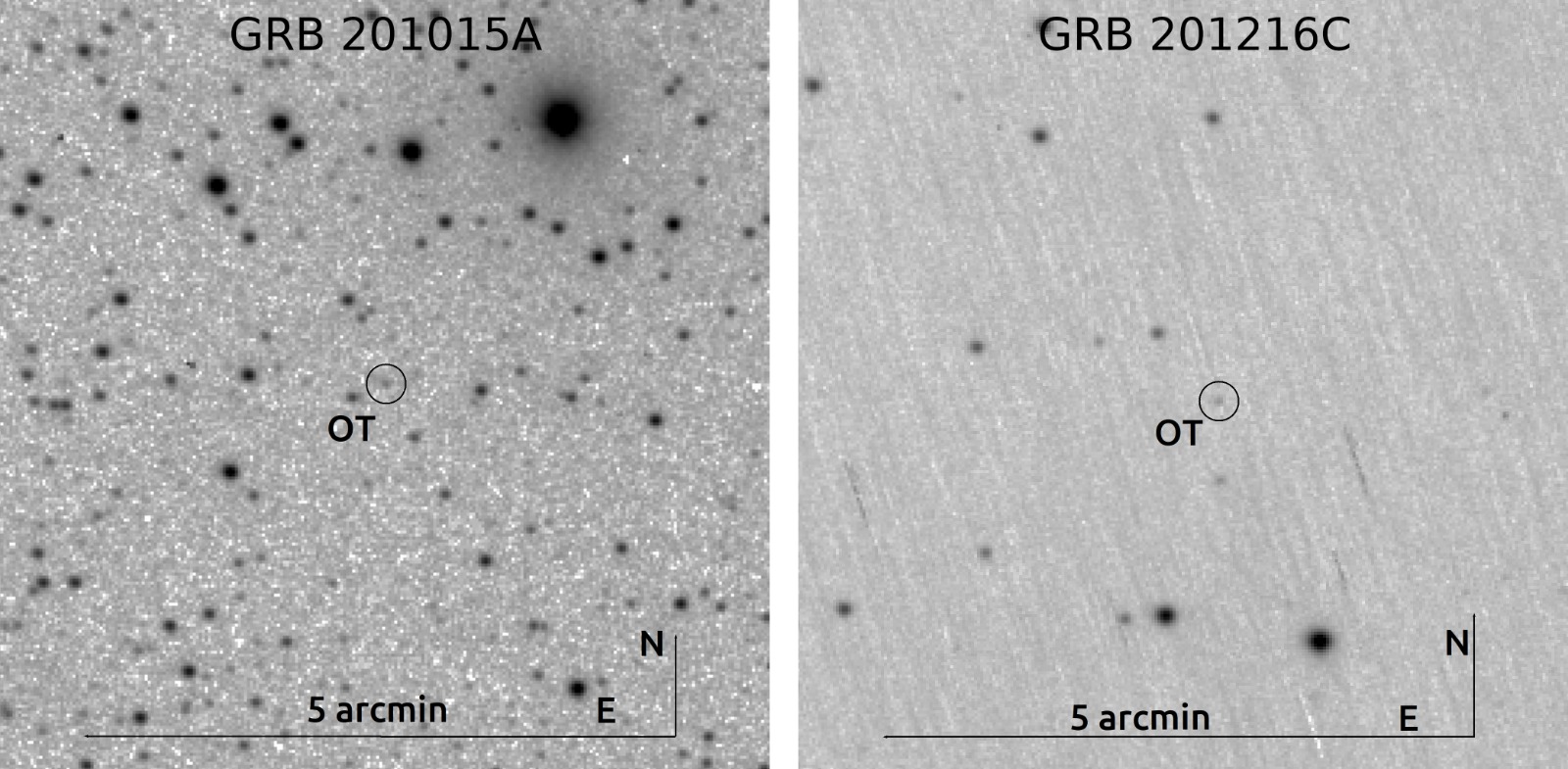}
\caption{The finding charts of \thisgrba (left), and \thisgrbb (right) observed utilizing the FRAM-ORM robotic telescope. Circles denote the positions of bursts. The horizontal and vertical lines indicate the directions (north and east). The images have a FoV of 5$'$ $\times$ 5$'$.} 
\label{fig:FC_FRAM}
\end{figure*}

A series of images were obtained with BOOTES-1B robotic telescope in the clear filter with exposures of 1 s, 10 s, and 60 s for \thisgrba. The image preprocessing (bias and dark-subtracted, flat-fielded, and cosmic-ray removal) was done using custom IRAF routines. The photometry was carried out using the standard IRAF package. The images were calibrated with nearby comparison stars from the Pan-STARRS catalog, which were imputed to R-band through the transformation equation from Lupton (2005)\footnote{\url{http://www.sdss.org/dr12/algoritms/sdssUBVRITransform/}} for the data in the clear filter. The obtained magnitudes are listed in Table \ref{table:15A_optical} of the appendix.\\

{\bf FRAM-ORM:} The FRAM-ORM is a 25cm f/6.3 robotic telescope facilitated with B, V, R, and z filters and a custom Moravian Instruments G2-1000B1 camera based on a back-illuminated CCD47-10 chip. We carried out the earliest optical observations of \thisgrba using the 25cm FRAM-ORM robotic telescope located at La Palma, Spain. We obtained a series of frames with exposure times of 20 s each in the clear filter, beginning on 2020 October 15 at 22:50:50.8 UT (37.6 s after the BAT trigger). We have clearly identified the source mentioned by \cite{2020GCN.28633....1L}. A finding chart of FRAM-ORM observation of \thisgrba is given in Figure \ref{fig:FC_FRAM}.\\
{\bf 3.6m DOT:} We performed the optical observations of the afterglow of \thisgrba starting on 2020 October 16 at 13:09:07.2 UT ($\sim$ 0.6 days post burst) using 3.6m DOT located at the Devasthal observatory, which is part of the Aryabhatta Research Institute of Observational Sciences (ARIES), India. We acquired multiple frames in B, V, R, and I filters (with an exposure time of 10 minutes in each) using the 4K $\times$ 4K CCD IMAGER \citep{2022JApA...43...27K} mounted on the main port of 3.6m DOT. In stacked DOT images, we clearly detect the optical afterglow of \thisgrba consistent with the error region of NOT observations \cite{2020GCN.28637....1M}. A finding chart of 3.6m DOT observation of \thisgrba is given in Figure \ref{fig:FC_DOT} of the appendix. We performed the DOT image reduction using the IRAF package. We first applied zero correction and flat fielding to the raw images taken from the telescope. After the removal of cosmic-ray hits, we stack the images to create a single image. We use the IRAF package to perform the aperture photometry. For the photometric calibration of \thisgrba , standard stars in the Landolt standard field PG 0231 were observed along with the GRB field in UBVRI bands. The R-band finding chart of \thisgrba  and the secondary stars marked with S1-S14 are shown in the Figure \ref{fig:FC_DOT}. The calibrated magnitudes of secondary stars are listed in Table \ref{tab:secondary_stars_aa} of the appendix, and the calibrated magnitudes of \thisgrba are listed in Table \ref{table:15A_optical} of the appendix.

{\bf \thisgrbb:} In the case of \thisgrbb, \cite{2020GCN.29066....1I} carried out the optical follow-up observations in the Sloan g$^{'}$, r$^{'}$, z$^{'}$ filters using VLT telescope, starting at 01:18:47 UT on 2020 December 17. They first reported the detection of an uncatalogued optical source at location RA: 01:05:28.980, Dec: +16:31:00.0 (J2000.0), consistent with the \swift-XRT enhanced location \citep{2020GCN.29064....1C}.

{\bf FRAM-ORM:} We performed the earliest optical follow-up observations to the alert of \thisgrbb using the FRAM-ORM telescope. We obtained a series of frames with an exposure times of 20 s each in the clear filter, starting on 2020 December 16 at 23:08:04.3 UT (31.6 s after the BAT trigger)  \citep{2020GCN.29070....1J}. We immediately detected an optical transient consistent with the location reported by \cite{2020GCN.29066....1I}. A finding chart of FRAM-ORM observation of \thisgrbb is given in Figure \ref{fig:FC_FRAM}. A log of photometric observations of the afterglow of \thisgrbb is presented in Table \ref{table:16C_optical} of the appendix. 

The observed frames for both the bursts have been processed through a difference imaging pipeline based on the HOTPANTS image subtraction code to remove the influence of nearby stars on the photometric measurements and then corrected for bias, flats, and cosmic-rays. The photometry has been carried out using the DAOPHOT package. The measured magnitudes have been calibrated using field stars in the PanSTARS DR1 catalog. Our optical observations reveal an early rise in the light curves of both bursts, reaching their maximum and followed by normal decay until the end of our observations. A log of our photometric observations is given in Table \ref{table:15A_optical} of the appendix.

\section{Results} 
\label{sec:result}
\subsection{Prompt emission}
In this section, we present the results of comprehensive analysis of prompt emission of \thisgrba and \thisgrbb. We have summarized our results in Table \ref{tab:combine_table}.

\subsubsection{Light curve and time-averaged spectra}
\label{lc_spectrum}
\begin{figure}
\centering
\includegraphics[scale=0.35]{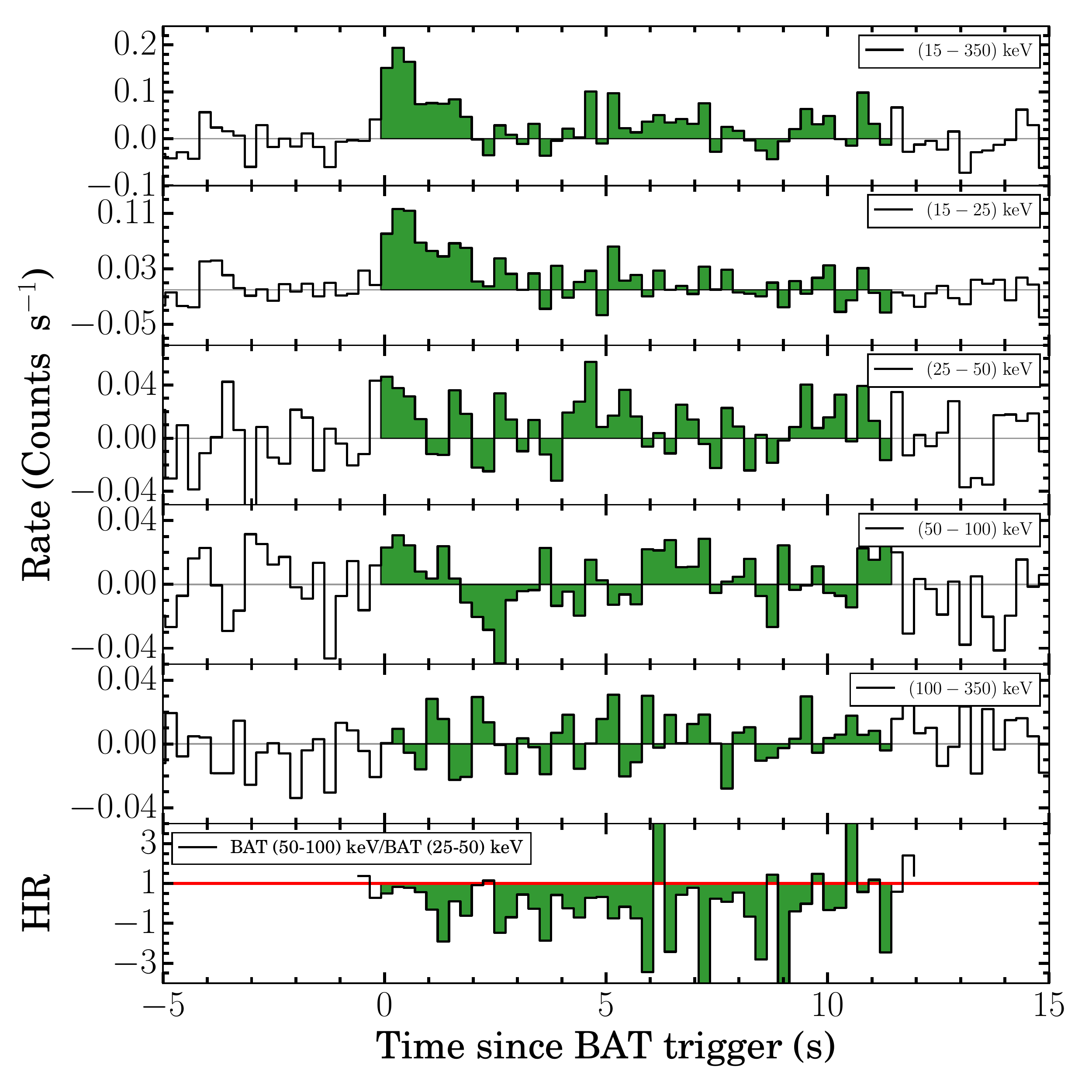}
\includegraphics[scale=0.35]{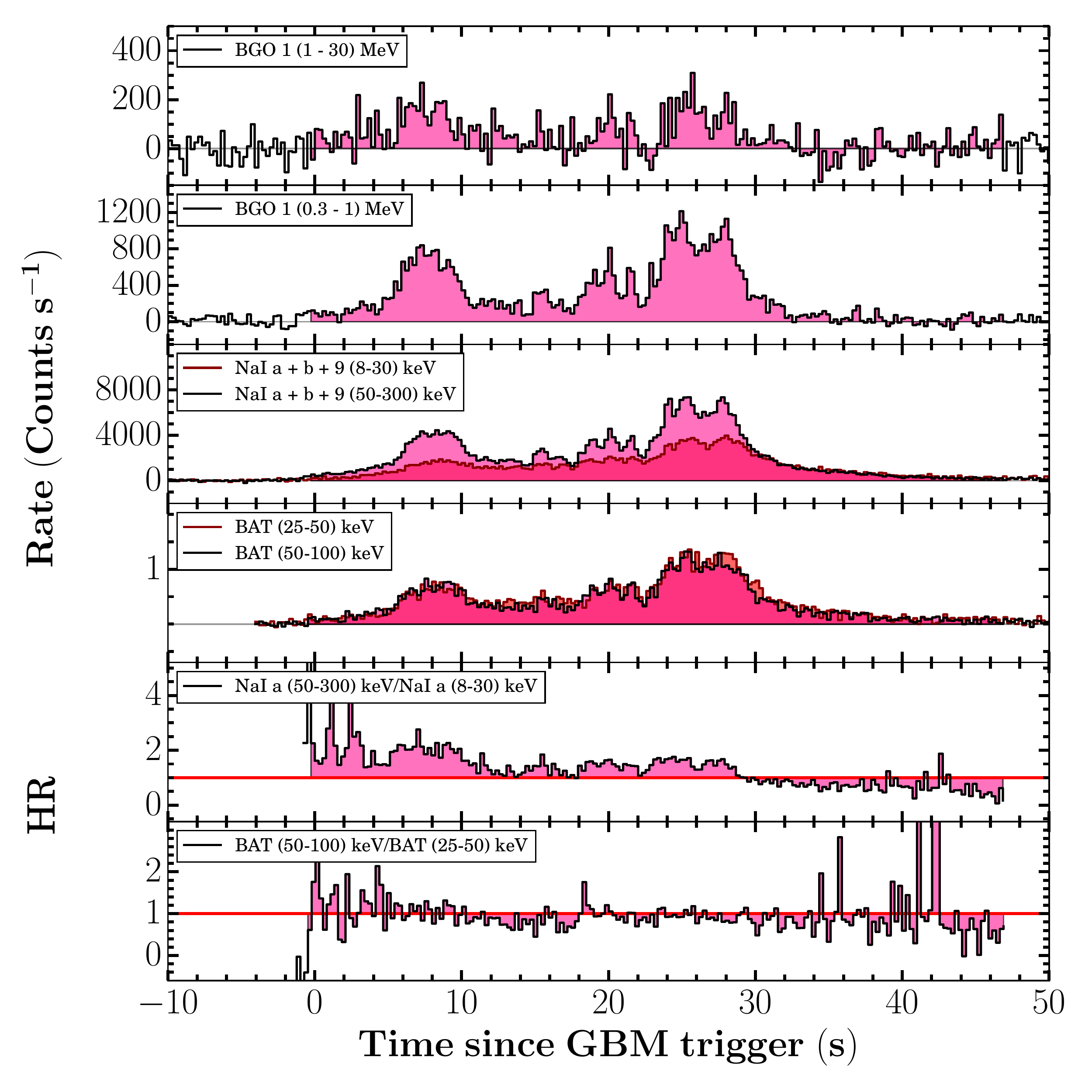}
\caption{Multi-channel prompt $\gamma$-ray/ hard X-ray light curves of \thisgrba (top) and \thisgrbb (bottom) along with hardness ratio evolution using \fermi-GBM and \swift-BAT observations. The shaded green and pink regions show the total time interval used for the time-averaged spectral analysis of \thisgrba and \thisgrbb, respectively.}
\label{fig:promptlc}
\end{figure}

The \swift-BAT energy-resolved mask-weighted light curve of \thisgrba along with HR evolution in 25-50 \keV and 50-100 \keV energy range is shown in the top panel of Figure \ref{fig:promptlc}. The \swift-BAT prompt emission light curve of \thisgrba has a short-soft emission starting from \swiftT and ending at \swiftT+1 s, followed by a weak-soft tail emission that lasts till $\sim$ \swiftT+10 s \citep{2020GCN.28658....1M}. The time-integrated spectrum (\swiftT-0.21 to \swiftT+11.57 s) of \thisgrba is explained using the simple power-law model with power index = -2.43 $\pm$ 0.25 (see Figure \ref{fig:TAS_band_bb} of the appendix). 

\begin{figure}
\centering
\includegraphics[scale=0.43]{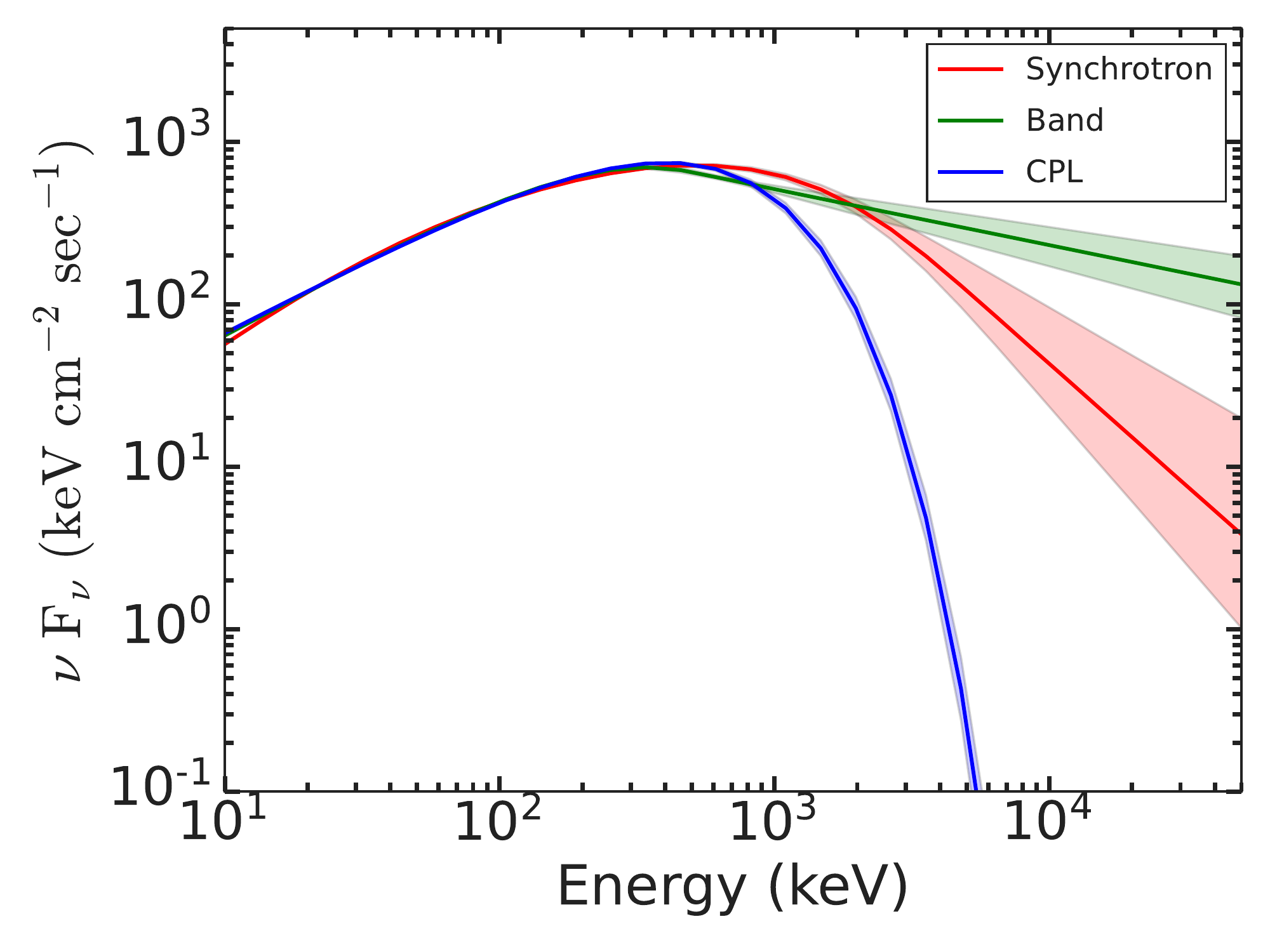}
\caption{Comparison between the time-averaged energy spectrum of empirical (\sw{Band}, \sw{CPL}) and physical (\sw{Synchrotron}) models for \thisgrbb. The red, green, and blue solid lines show the energy spectrum of the \sw{Synchrotron}, \sw{Band}, and \sw{Cutoff power-law} models in the model space, respectively. The corresponding shaded regions show the 95\% confidence interval.}
\label{fig:TAS}
\end{figure}

\begin{table}
\begin{center}
\caption{Observed properties of \thisgrba and \thisgrbb obtained from our comprehensive analysis of prompt emission, afterglow, and those reported on GCN circulars \citep{2020GCN.28658....1M, 2020GCN.29073....1M}. Parameters listed in the first column have their standard meaning.}
\label{tab:combine_table}
\begin{tabular}{|c| c c|}\hline 
{\bf Properties} &\bf \thisgrba &\bf \thisgrbb \\\hline
\tninty (s)  &     9.78 $\pm$ 3.47  & 29.95 $\pm$ 0.57\\
HR &  0.72    &  1.11 \\
$\alpha_{\rm pt}$ & & $-1.06 \pm 0.13$\\
$\beta_{\rm pt}$ & $-2.43 \pm 0.25$ & $-2.75 \pm 0.33$ \\
\Ep (\keV)&   41.37$^{+15.74}_{-11.41}$  & 352.31$^{+12.77}_{-12.74}$     \\
Fluence (erg $\rm cm^{-2}$) &        $2.25\pm0.38\times10^{-7}$&$2.00\pm0.10\times10^{-4}$\\
 $E_{\rm \gamma, iso}$ (erg) &3.86$\times10^{51}$ &$6.32\times10^{53}$\\
$L_{\rm \gamma, iso}$ (erg/s)&3.86$\times10^{50}$ & 8.78$\times10^{52}$\\
$\Gamma_0$  &$\sim{204}$ &$\sim{310}$ \\ 
$z$  & 0.426   & 1.1 \\ \hline
SN association &  Yes  & No \\ \hline
\end{tabular}
\end{center}
\end{table}

The background-subtracted energy-resolved \fermi-GBM light curve of \thisgrbb along with the hardness-ratio (HR) evolution is shown in the bottom panel of Figure \ref{fig:promptlc}. The \fermi-GBM prompt emission light curve consists of a broad, structured peak with a \tninty duration of $\sim$ 29.9 s (in 50-300 \keV). Similar to the \fermi-GBM, \swift-BAT observations also reveal multiple peaks in the mask-weighted prompt light curve. According to BAT observations, the energy fluence of the burst is 4.5 $\pm 0.1 \times 10^{-5}$ erg $\rm cm^{-2}$ in the 15-150 \keV energy range \citep{2020GCN.29080....1U}. Figure \ref{fig:promptlc} also shows the time interval used for the time-averaged spectral analysis. The time-averaged \fermi-GBM spectrum (from \fermiT-0.503 s to \fermiT+47.09 s) is best modeled using a \sw{Band+BB} function (see Figure \ref{fig:TAS_band_bb} of the appendix) and the best fit spectral parameters are presented in Table \ref{tab:TAS} of the appendix. The burst was significantly bright: the observed fluence is (2.00 $\pm$ 0.10) $\times 10^{-4}$ erg $\rm cm^{-2}$ in the 10-10$^4$ \keV energy band over the time-averaged interval. This fluence value is among the top 2 percent of the brightest GRBs observed by \fermi-GBM. A comparison between the time-averaged energy spectrum of the empirical (\sw{Band}, \sw{CPL}) and physical (\sw{Synchrotron}) models are shown in Figure \ref{fig:TAS}.

\subsubsection{Time-integrated \tninty-Spectral hardness distribution}
GRBs are classified into two families based on \tninty duration, however, there is one more fundamental difference between both the classes, i.e., spectral hardness. Long GRBs are expected to be softer (lower \Ep value), on the other hand, short bursts are usually harder (higher \Ep value). To know the true class of \thisgrba and \thisgrbb, we placed both bursts in the time-integrated \tninty-spectral hardness plane along with the other long and short GRBs obtained from the \fermi-GBM and \swift-BAT catalogs. 

We calculated the hardness ratio (HR) for \thisgrba using the ratio of fluence in hard (50-100 \keV) and soft (25-50 \keV) energy ranges and found its value equal to 0.72. Comparing the HR of \thisgrba with the \swift-BAT catalog, we note that \thisgrba is one of the softest bursts ever observed by the \swift mission (see Figure \ref{fig:prompt_properties} of the appendix). In the case of \thisgrbb, we calculated the HR using the ratio of counts in hard (50-300 \keV) and soft (8-30 \keV) energy ranges to be 1.11, this feature is similar to long/soft GRBs. The distribution of HR as a function of \tninty for both the bursts are given in Figure \ref{fig:prompt_properties} of the appendix.

Furthermore, we also placed both the bursts in the time-integrated \Ep-\tninty distribution of long and short GRBs. As for \thisgrba there was no onboard observation by \fermi and no public \fermi data is available for ground targeted search, we have used \swift-BAT observations of \thisgrba to constrain the time-integrated peak energy. However, the BAT time-integrated spectra of GRBs are usually modeled by simple power-law functions due to the instrument limited spectral coverage (15-150 \keV). In the case of \thisgrba, the time-integrated spectrum is also fitted using a power-law function (see \S~\ref{lc_spectrum}). We estimated the \Ep value using the known correlation between the \swift-BAT fluence and time-integrated \Ep \citep{2020ApJ...902...40Z}, i.e., \Ep = [fluence/($\rm 10^{-5} erg~cm^{-2}$)]$^{0.28}$ $\rm \times 117.5^{+44.7}_{-32.4}$ \keV $\approx$ 41.37$^{+15.74}_{-11.41}$ \keV. The estimated time-integrated \Ep value of \thisgrba is consistent with those observed for long GRBs. 
In the case of \thisgrbb, we fitted the time-integrated spectrum and calculated the peak energy (\Ep = 352.31$^{+12.77}_{-12.74}$ \keV) using the best fit model. The \tninty-\Ep distribution for both the GRBs is shown in Figure \ref{fig:prompt_properties} of the appendix along with other data points obtained from \fermi-GBM catalog \citep{Goldstein_2017}. We fitted the complete distribution obtained from \fermi-GBM catalog using the Bayesian Gaussian mixture model (BGMM) algorithm and calculated the probability of \thisgrbb being a long burst as 98.8 \%.

\subsubsection{Prompt correlations: Amati and Yonetoku} 
\label{subsec:Amati and Yonetoku correlations}

The cosmological corrected time-integrated peak energy $E_{p,z}$ = (1+z)\Ep of the prompt emission is correlated with the isotropic equivalent energy $E_{\rm \gamma, iso}$, and isotropic peak luminosity $L_{\rm \gamma, iso}$. The former is known as Amati correlation \citep{Amati} and the later is known as Yonetoku correlation \citep{Yonetoku_2004}. In the case of \thisgrba, we have used equation 6 of \cite{2015ApJ...815..102F} to calculate $E_{\rm \gamma, iso}$ due to the limitation of \swift-BAT spectral coverage. On the other hand, we have used the best fit \fermi-GBM time-integrated spectral model to calculate the $E_{\rm \gamma, iso}$ for \thisgrbb. The Amati correlation for both the GRBs along with a sample of long and short GRBs taken from \cite{2020MNRAS.492.1919M} is presented in Figure \ref{fig:prompt_properties} of the appendix. Similarly, we calculated the $L_{\rm \gamma, iso}$ values and placed \thisgrba and \thisgrbb in the $E_{p,z}$-$L_{\rm \gamma, iso}$ plane, as shown in Figure \ref{fig:prompt_properties} of the appendix. We noticed that both the bursts satisfied the Amati and Yonetoku correlations. The calculated isotropic equivalent luminosity values for both the bursts suggest that \thisgrba is an intermediate luminous GRB; on the other hand, \thisgrbb is a luminous GRB. 

\subsubsection{Time-resolved spectroscopy of \thisgrbb}
{\bf Distribution of spectral parameters:} 
The mean values and standard deviation for each spectral parameter obtained using \sw{Cutoff power-law}, \sw{Band}, \sw{Cutoff power-law + BB}, \sw{Band+ BB} and \sw{Synchrotron} models are listed in Table \ref{tab:mean_std} of the appendix.
The mean values of the low energy spectral indices $\alpha_{\rm CPL}$ obtained using \sw{Cutoff power-law} and $\alpha_{\rm pt}$ obtained using \sw{Band} spectral modeling of \thisgrbb are $-1.10 \pm 0.14$ and $-1.06 \pm 0.13$, respectively. These values are consistent with the typical average value of the low energy spectral index ($\alpha_{\rm pt}$ $\sim ~ -1$) of GRBs \citep{2000ApJS..126...19P}. Similarly, the averaged value of spectral peak energy \Ep and high energy photon index $\beta_{\rm pt}$ obtained using \sw{Band} are $339.43 \pm 119.39$ keV, and $-2.75 \pm 0.33$, respectively. These values are also consistent with the typical average value of \Ep and $\beta_{\rm pt}$ of GRBs. The averaged values of physical synchrotron spectral parameters for \thisgrbb are following: magnetic field (B) = $96.00 \pm 45.82$ G, index (p) = $4.18 \pm 0.54$, and Lorentz factor corresponds to the electron's cooling time $\gamma_{cool}$ = $(6.08 \pm 4.81) \times 10^{5}$.\\

{\bf Evidence for thermal component:} In our time-resolved spectral analysis, we first fitted each binned spectra using empirical \sw{Band} and \sw{Cutoff power-law} models individually. To search for the presence of an additional thermal component in the spectrum, we added the \sw{Blackbody} model along with \sw{Band} and \sw{Cutoff power-law} models, and calculated the difference of the DIC values. Negative values of $\rm \Delta DIC_{\rm Band}$ and $\rm \Delta DIC_{\rm CPL}$, implies that the additional \sw{Blackbody} component improves the fit statistic. Furthermore, in the case of any particular bin, if $\rm \Delta DIC < - 10$, it suggests a significant amount of thermal component present in that particular bin. Figure \ref{fig:DIC} of the appendix shows the evolution of $\rm \Delta DIC$ within the burst interval for \sw{Band + Blackbody} model. For all the bins (except the last bin), $\rm \Delta DIC$ values are negative, indicating that the addition of thermal components improves the spectral fitting. There are 10 bins ($\sim$ 37 \%) for which $\rm \Delta DIC < - 10$, suggesting for presence of thermal component in the spectrum. In light of this, we suggest that \thisgrbb is a hybrid (non-thermal+ quasi-thermal) burst. The thermal components are dominating during the initial and bright phases of the burst.

\begin{figure}
\centering
\includegraphics[scale=0.32]{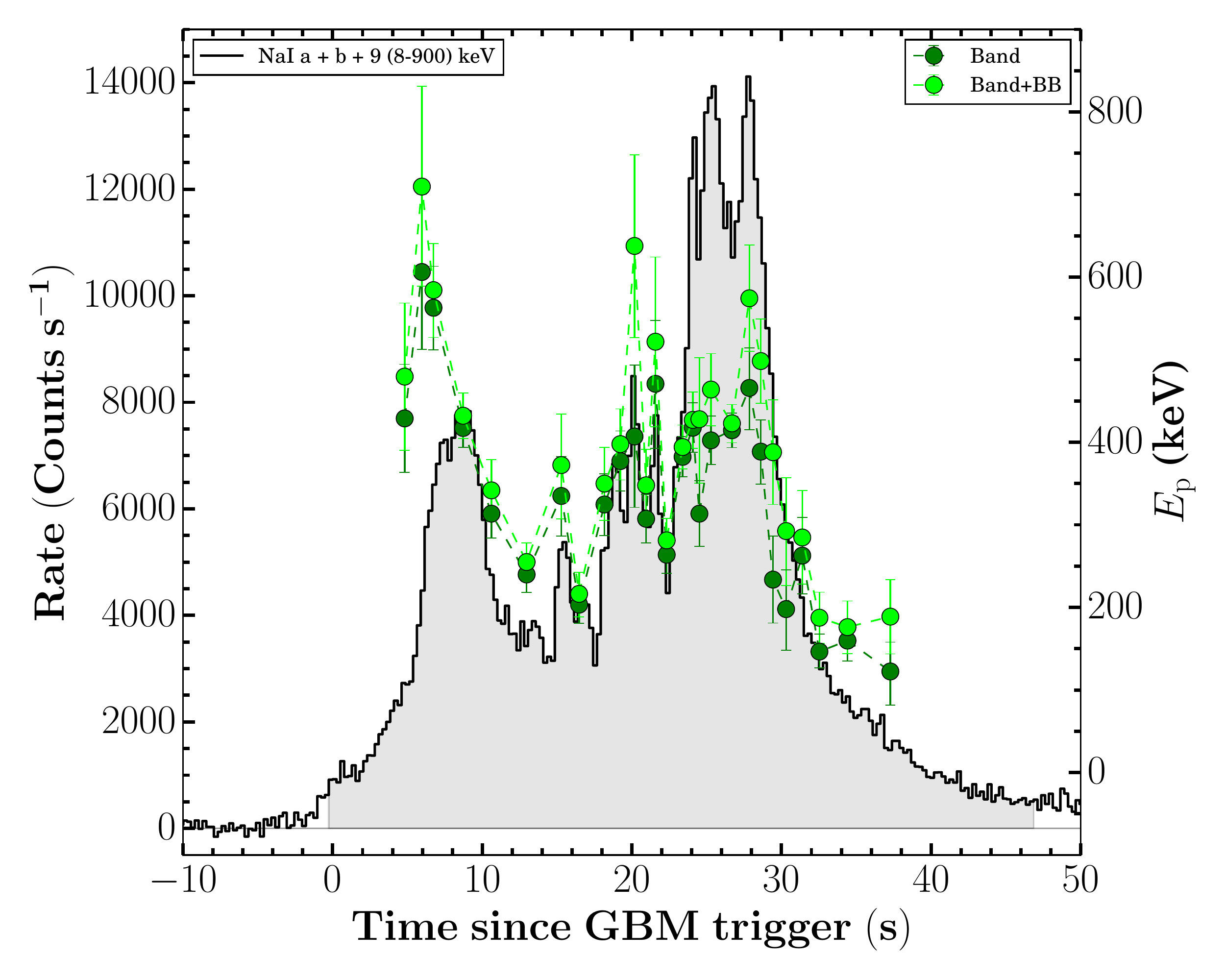}
\includegraphics[scale=0.32]{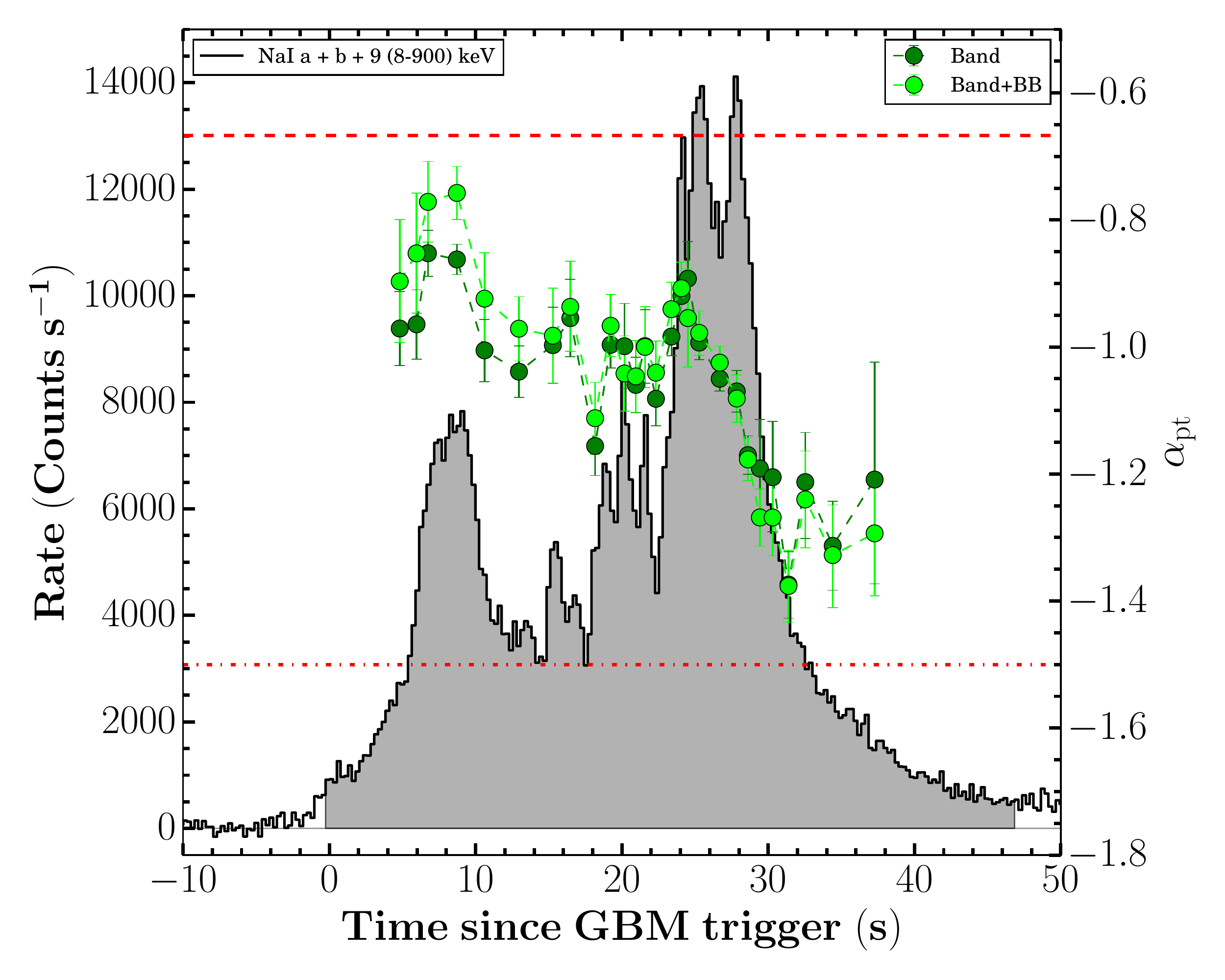}
\includegraphics[scale=0.32]{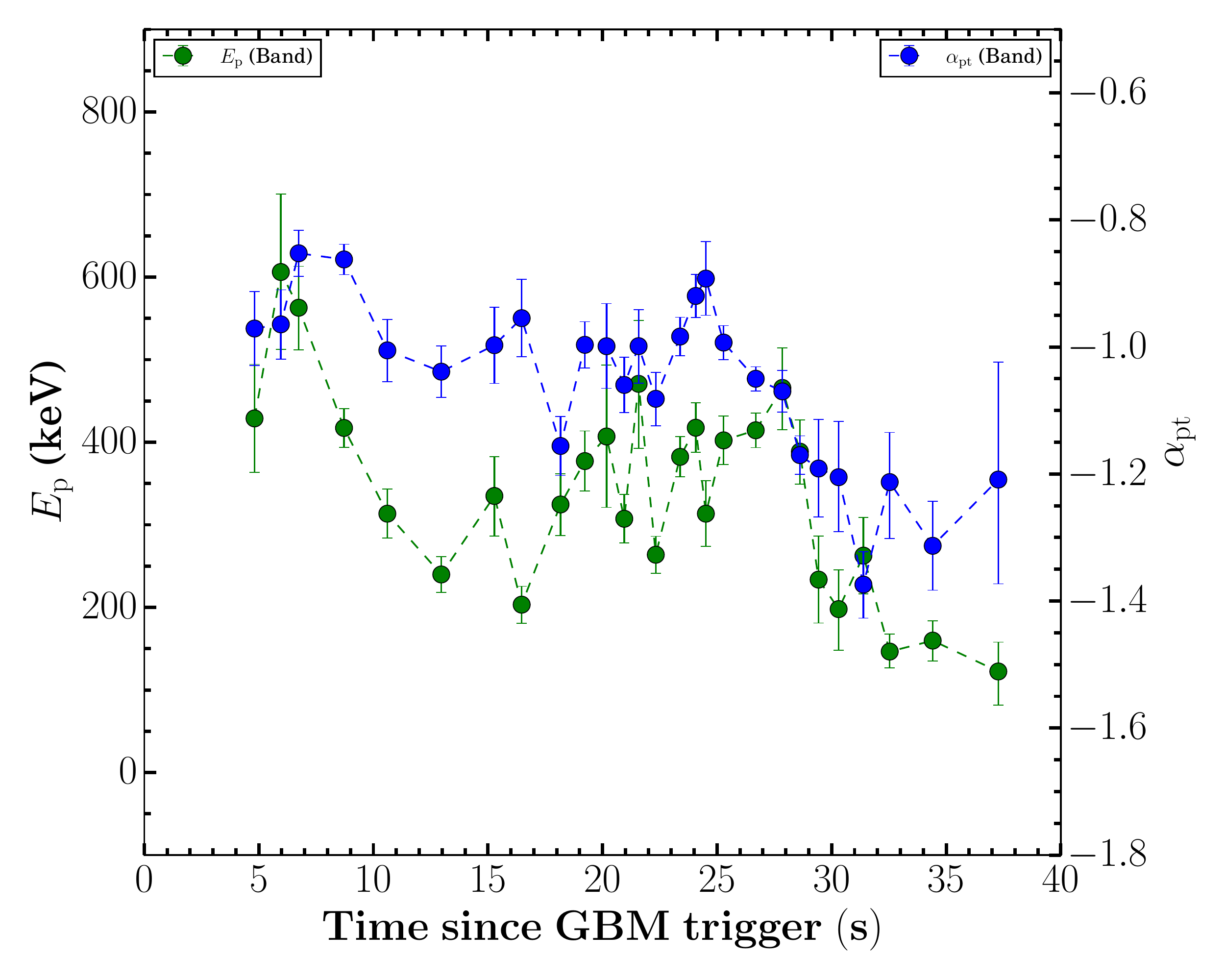}
\caption{Represents the evolution of spectral parameters, obtained from fitting of \sw{Band} function. Top panel: evolution of \Ep obtained from \sw{Band} (dark-green) and \sw{Band + Blackbody}(green) functions along with the count rate light curve. Middle panel: represents the same for $\alpha_{\rm pt}$, two red dotted lines represent the synchrotron death line at -2/3 and -3/2. Bottom panel: represents the simultaneous plot to compare the evolution of \Ep (green) and the $\alpha_{\rm pt}$ (blue).}
\label{fig:Band_para_evolution}
\end{figure}

\begin{figure}
\centering
\includegraphics[scale=0.32]{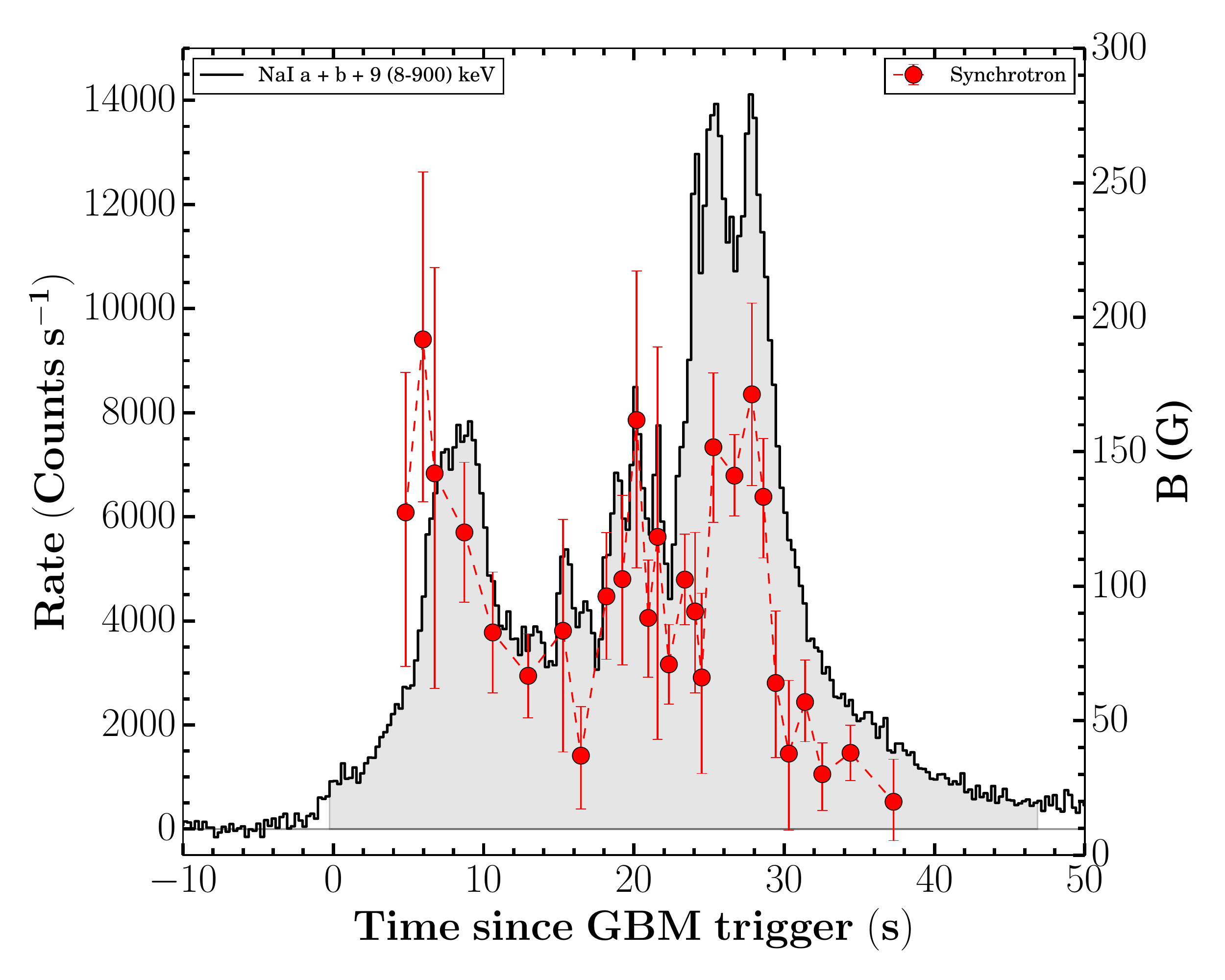}
\includegraphics[scale=0.32]{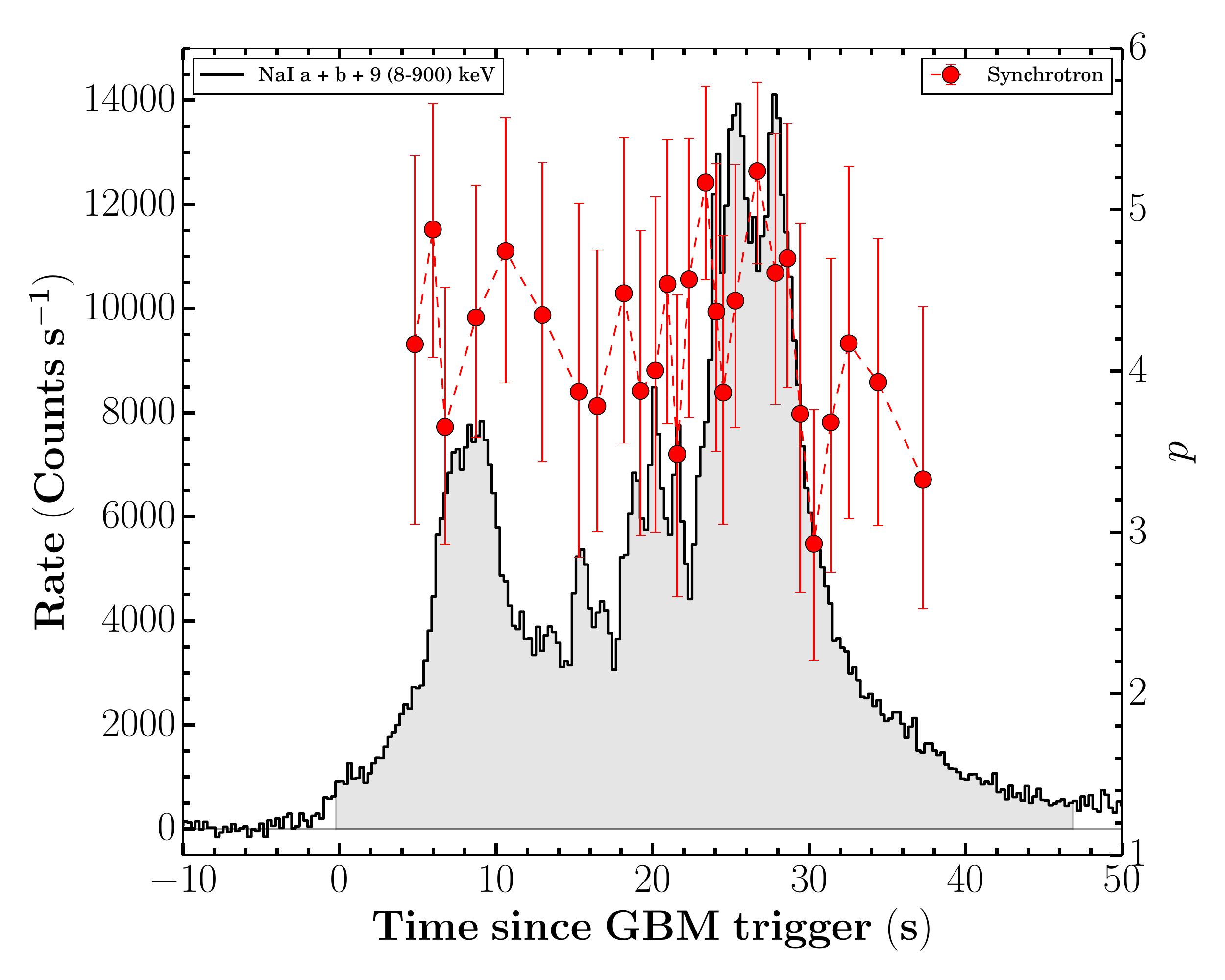}
\includegraphics[scale=0.32]{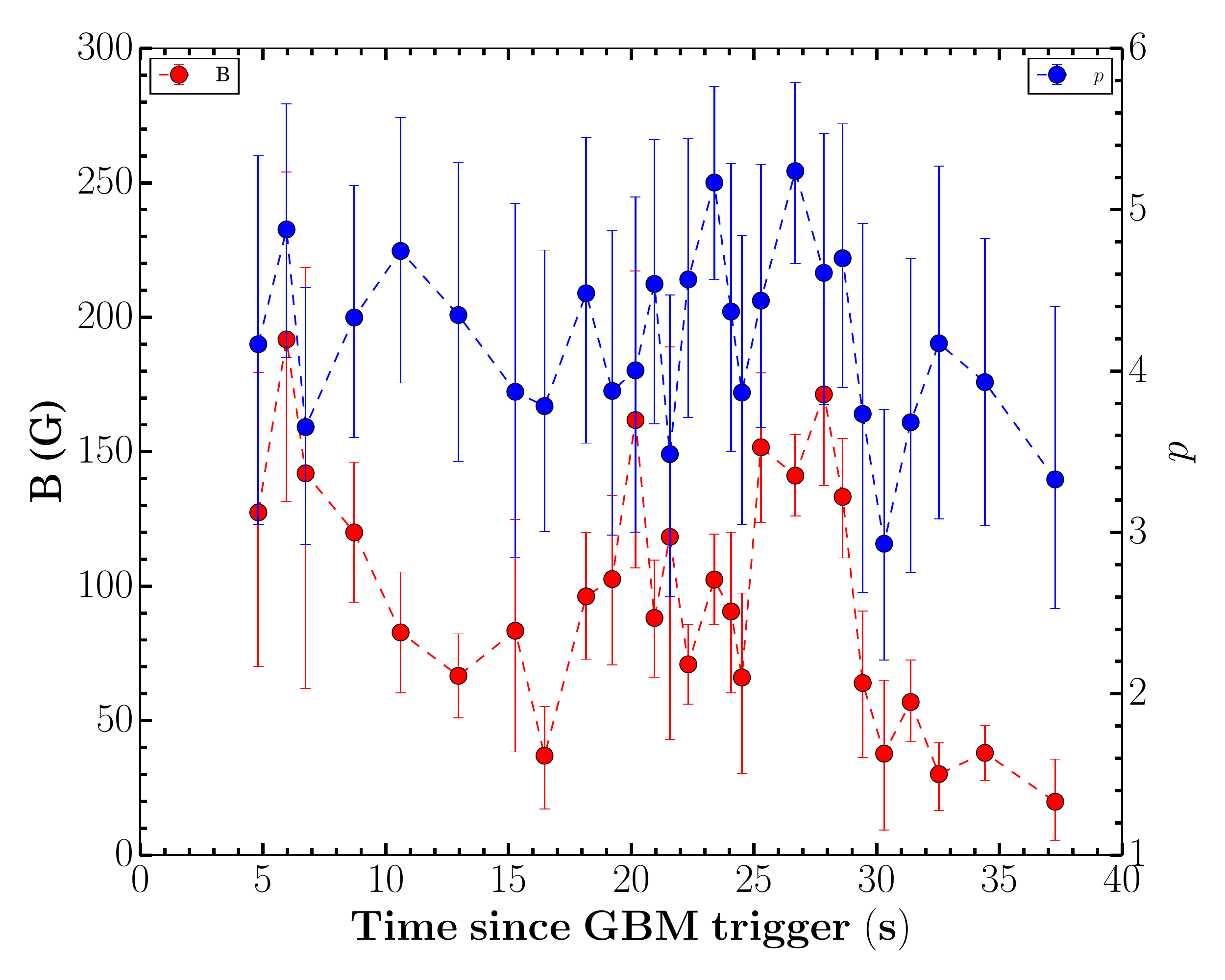}
\caption{The evolution of spectral parameters obtained from fitting of physical \sw{Synchrotron} model. Top panel: evolution of magnetic field strength B along with the count rate light curve. Middle panel: evolution of electron energy injection index $p$ along with the count rate light curve. Bottom panel: a simultaneous plot between B and $p$ to compare their evolution with time.}
\label{fig:Synchrotron_para_evolution}
\end{figure}

{\bf The evolution of spectral parameters:} The studies of spectral evolution are a very powerful tool to probe the emission process responsible for the prompt emission. The peak energy of the \sw{Band} function shows four different possible types of spectral evolution. 1. Hard to soft pattern \citep{1986ApJ...301..213N}, 2. Soft to hard pattern \citep{1994ApJ...422..260K}, 3. flux tracking pattern \citep{1983Natur.306..451G} and 4. Chaotic pattern. On the other hand, the low energy spectral index of the \sw{Band} function also changes with time, however, it does not show any particular trend. There are some recent studies suggesting a flux tracking pattern of $\alpha_{\rm pt}$, supporting the double tracking trend \citep{2019ApJ...884..109L}. However, most of the spectral evolution studies are performed for single pulsed bursts. The prompt light curve of \thisgrbb shows a more complex multi-pulsed structure, and the evolution of empirical and physical spectral parameters are very interesting. Figure \ref{fig:Band_para_evolution} shows the evolution of the spectral parameters (\Ep, and $\alpha_{\rm pt}$) of \thisgrbb obtained using empirical \sw{Band} function. The evolution of \Ep of \thisgrbb shows a flux tracking trend, i.e., \Ep increases and decrease as flux increases and decrease in respective bins. We noticed that the $\alpha_{\rm pt}$ values are changing with time and do not exceed the expected values of spectral indices of synchrotron fast and slow cooling cases. We have also shown the evolution of $\alpha_{\rm pt}$ and \Ep together in Figure \ref{fig:Band_para_evolution}, and we can see that the evolution patterns are quite similar throughout the emission. Next, we study the spectral evolution of parameters obtained from the physical synchrotron modeling. Figure \ref{fig:Synchrotron_para_evolution} shows spectral evolution of the magnetic field strength (B), and spectral index of electrons ($p$). We noticed that the magnetic field strength is also following the intensity of the burst. We could not confirm the evolution trend of $p$ due to large associated errors. In light of the above, we suggest that multi-pulsed \thisgrbb have flux tracking characteristics. We further investigated the correlation among these spectral parameters.

{\bf Spectral parameters correlations:} Studying the correlations between the different spectral parameters obtained using the time-resolved spectral analysis using empirical and physical modeling gives an important clue about the physics of GRBs and jet composition. We calculated the correlation between \sw{Band} and \sw{Band+Blackbody} spectral parameters: log (Flux) and log (\Ep), log (Flux) and $\alpha_{\rm pt}$, log (\Ep) and $\alpha_{\rm pt}$, and log (Flux) and k$T$. We found a high degree (the correlation coefficient ranges from 0.60-0.79) of correlation between the \sw{Band} spectral parameters, i.e., log (Flux) and log (\Ep), log (Flux) and $\alpha_{\rm pt}$, and log (\Ep) and $\alpha_{\rm pt}$. We also found a high degree of correlation between log (Flux) and k$T$ for the \sw{Band+ Blackbody} model. The correlation results for the \sw{Band} and \sw{Band+ Blackbody} models (Pearson correlation) are listed in Table \ref{tab:correlation} and Figure \ref{fig:correlation_band} of the appendix. We also studied the correlation between physical spectral parameters (log (Flux)-log (B), log (Flux)-$p$, and log (B)-$p$) obtained using \sw{Synchrotron} model. We found a very high degree (the correlation coefficient is $\geq$ 0.80) of correlation between log (Flux)-log (B) and medium correlation (the correlation coefficient ranges from 0.40-0.59) between log (B)-$p$, however, we found a low degree (the correlation coefficient ranges from 0.20-0.39) of correlation between log (Flux)-$p$. The correlation results obtained using physical modeling are shown in Figure \ref{fig:correlation_sync} and in Table \ref{tab:correlation} of the appendix.

In addition, we also studied the correlation between empirical and physical models parameters: log (B)-log (\Ep), log (B)-$\alpha_{\rm pt}$, and $p$-$\alpha_{\rm pt}$. The correlation results are shown in Figure \ref{fig:correlation_sync_band} and Table \ref{tab:correlation} of the appendix. We found a very high degree of correlation between log (B) of physical synchrotron model and log (\Ep) of empirical \sw{Band} function. We also found a medium degree of correlation between log (B) and $\alpha_{\rm pt}$, however, there is a low degree of correlation between $p$ and $\alpha_{\rm pt}$.

\subsection{Afterglow emission of \thisgrba and \thisgrbb:}
\begin{figure}
\centering
{\hspace{-2em}
\includegraphics[scale=0.32]{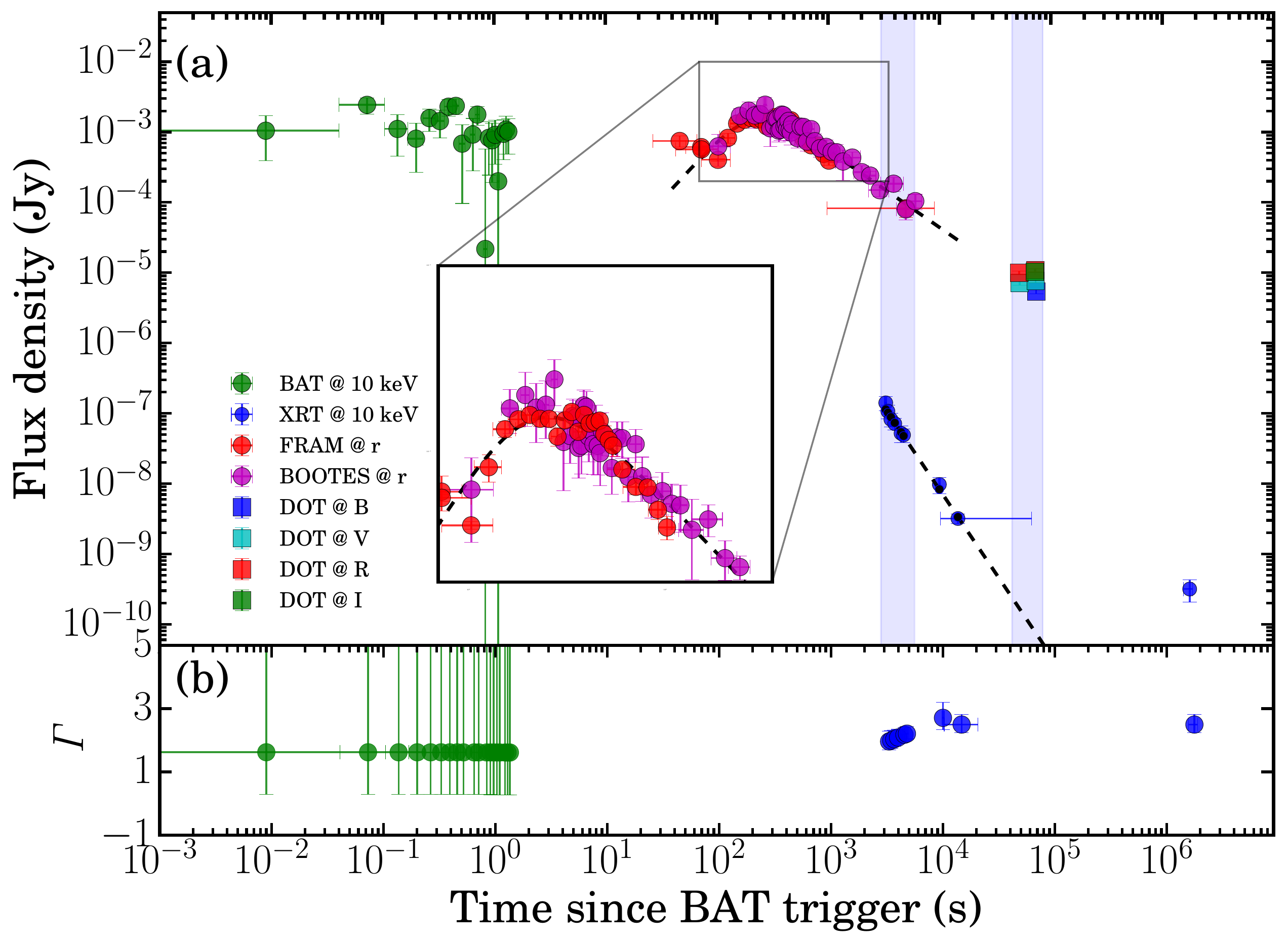}}
\includegraphics[scale=0.32]{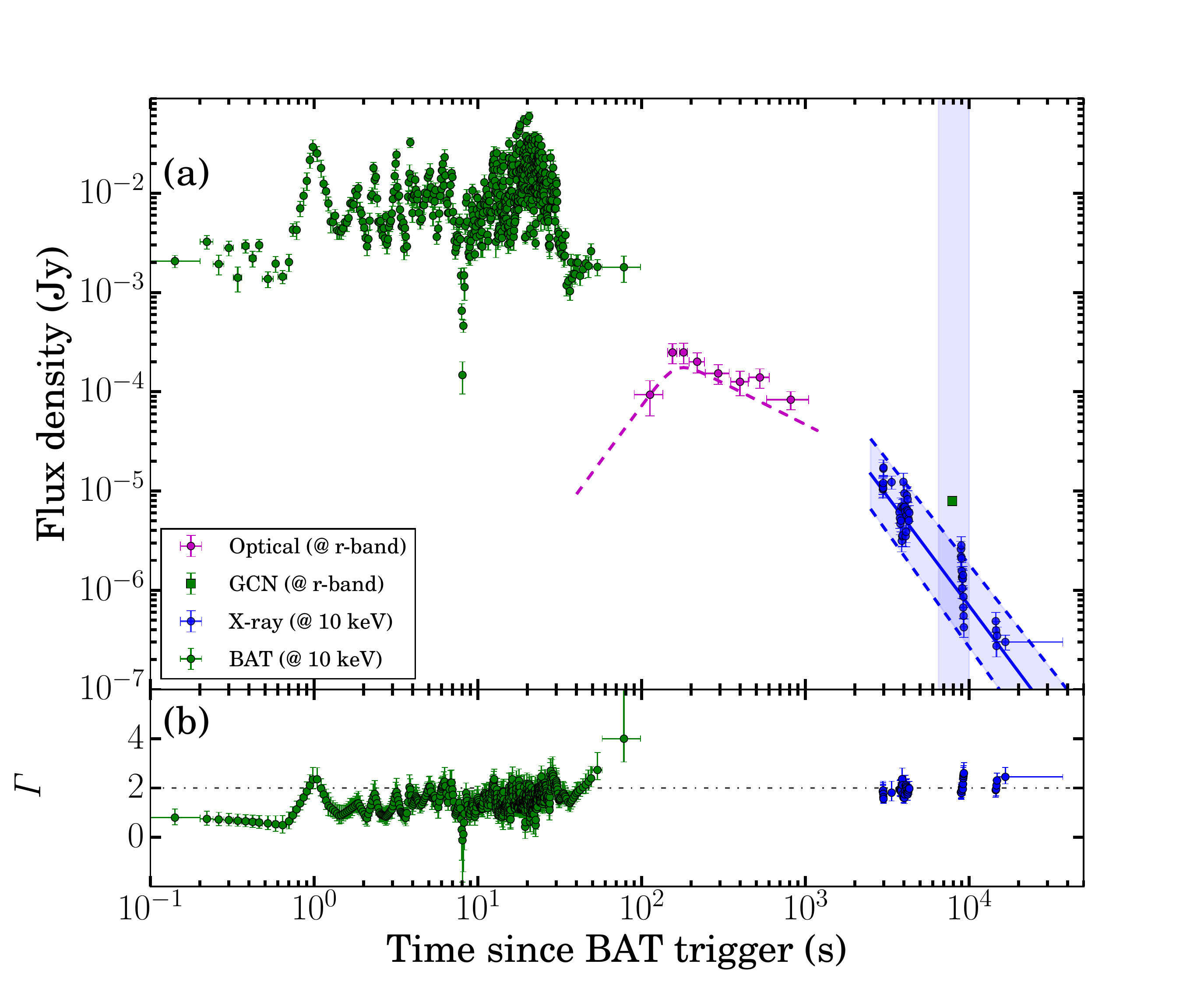}
\caption{Multi-wavelength light curves of \thisgrba (top) and \thisgrbb (bottom). (a) Represent the \swift-BAT (@ 10 \keV, green), \swift-XRT (@ 10 \keV, blue) and optical (@ r band, red/magenta) flux density light curve. The inset represents the zoomed part of the optical bump. The vertical blue shaded regions indicate the time ranges used for the SEDs analysis. (b) Represents the evolution of the spectral photon index in the \swift-BAT and \swift-XRT energy channels.}
\label{multiwavelength_LC}
\end{figure}

In the present section, we study the results of multi-wavelength afterglow of \thisgrba and \thisgrbb, detected by \swift-XRT (X-ray), FRAM-ORM, BOOTES and 3.6m DOT (optical). Multi-wavelength light curves of \thisgrba and \thisgrbb are shown in Figure \ref{multiwavelength_LC}, whereas SEDs are discussed in Figure \ref{fig:SED}.

\subsubsection{X-ray afterglows}
The X-ray afterglow light curves of \thisgrba and \thisgrbb do not show any flare, bump, break, or plateau-like activities (see Figure \ref{multiwavelength_LC}).\\
In the case of \thisgrba, we initially tried to fit the X-ray light curve using a simple power-law model (temporal index = -2.36$^{+0.17}_{-0.26}$). The calculated XRT temporal index is consistent with the temporal index reported on the XRT catalogue page using count rate light curve fitting\footnote{\url{https://www.swift.ac.uk/xrt_live_cat/01000452/}}. However, the model shows a significant deviation from the last observed data point (chi/dof =1.34). We noted that the last two data points are considered unreliable because no centroid could be determined. We calculated the XRT temporal decay index = -2.27$^{+0.43}_{-0.58}$ excluding these points. We found that both the calculated XRT decay indices are consistent, considering the large associated errors. We also used the broken power-law function to fit the light curve and obtained chi/dof = 0.27, indicating that the model is over fitting the data. Further, we used \sw{F-test} to find the best fit model among the two empirical models. The \sw{F-test} suggests that the simple power-law model is better fitting the X-ray afterglow light curve of \thisgrba, consistent with those reported on the XRT catalogue page. The X-ray light curve and the best fit model are shown in Figure \ref{multiwavelength_LC}. For the XRT spectral analysis, we fitted the late time spectrum (\swiftT+4000 to \swiftT+22019 s) to constrain the intrinsic hydrogen column density ($\rm NH_{\rm z}$) of \thisgrba and calculated $\rm NH_{\rm z}$ = $5.56 \times 10^{21}$ cm$^{-2}$. The time-averaged X-ray spectrum (\swiftT+3300 to \swiftT+1800000 s) of \thisgrba is described using an absorbed power-law model with a spectral index = 1.10$^{+0.25}_{-0.25}$. Further, we divided the XRT light curve into three segments based on available observations. For the first (\swiftT+3300 to \swiftT+4800 s), second (\swiftT+10000 to \swiftT+15000 s), third (\swiftT+10000 to \swiftT+1800000 s) time bins, we obtained the spectral indices = 1.07$^{+0.31}_{-0.30}$, 1.10$^{+0.53}_{-0.55}$, and 1.38$^{+0.49}_{-0.47}$, respectively. Our analysis does not find any noticeable evolution among the spectral indices within the observed duration considering large values of associated errors. \\
The X-ray light curve of \thisgrbb is shown in Figure \ref{multiwavelength_LC}. The light curve (@ 10 \keV) decay as a power-law with a decay index of $\alpha_x$ = $-2.21_{-0.11}^{+0.10}$ with (chi/dof =3.35). For the XRT spectral analysis, we calculated the $\rm NH_{\rm z}$ of the host using the late time spectral fitting (\swiftT+4001 to \swiftT+37346 s) and found $\rm NH_{\rm z}$ = $2.71\times 10^{22}$ cm$^{-2}$. The joint PC and WT mode time-averaged X-ray spectrum (\swiftT+2900 to \swiftT+17000 s) of \thisgrbb is described using an absorbed power-law model with a spectral index = 0.97$^{+0.05}_{-0.05}$. Further, we created the spectrum for individually segmented WT and PC mode observations using the Swift Science Data Centre web-page\footnote{\url{https://www.swift.ac.uk/xrt_spectra/addspec.php?targ=01013243&origin=GRB}} and performed the spectral fitting. For the WT (\swiftT+2900 to \swiftT+3000 s), and PC (\swiftT+3300 to \swiftT+17000 s) time bins, we obtained the spectral indices = 0.87$^{+0.22}_{-0.22}$, and 0.90$^{+0.10}_{-0.09}$, respectively. Our analysis indicates that the spectral index is not changing with time (see Figure \ref{multiwavelength_LC}).

\subsubsection{Optical afterglows}
\label{onset_afterglow}
For both these VHE detected bursts, we can examine the early afterglow behavior using the earliest optical observations from the 0.25m robotic FRAM-ORM and BOOTES telescopes \citep{2020GCN.28664....1J, 2020GCN.29070....1J, 2020GCN.28645....1H}.

The optical light curve of \thisgrba is highly rich in features. Following an early decay, the light curve has a smooth optical bump, which may be either due to reverse shock emission or the onset of afterglow in the forward shock. We calculated the decay rate for the very early optical emission $\alpha_{\rm o1}=- 0.68 \pm 0.15$. To characterize the nature of the early bump, we fitted the optical bump using a smoothly broken power-law function, given in equation \ref{eqn_3}. 

\begin{equation} 
\label{eqn_3}
F(t)=F_0\left[\left(\frac{t}{t_b}\right)^{k\alpha_r}+ \left(\frac{t}{t_b}\right)^{k\alpha_d}\right] ^{{-1}/{k}}
\end{equation}

Where $\alpha_r$ and $\alpha_d$ are the rise and decay temporal indices, respectively. $F_{0}$ is the normalization constant, and $t_{b}$ is the break time. The break time is related to the observed peak time ($t_{\rm p}$) by the following equation: 

\begin{equation} \label{eqn_4}
t_{p} = t_b\left(-\frac{\alpha_r}{\alpha_d}\right)^{[{-1}/{k({\alpha_d - \alpha_r})]}}
\end{equation}

In the case of \thisgrba, the smoothly broken power-law fit (smoothness parameter $k=1$) shows that initially, the optical afterglow light curve rises with a temporal index of $\alpha_r$=$-$1.74$_{-0.23}^{+0.19}$ and then decays with an index of $\alpha_d$=1.10$_{-0.06}^{+0.06}$, with the break time $t_{\rm b}$=217.04$_{-18}^{+19}$ s post burst. Further, we obtained the peak time $t_{\rm p}$= 184.64$_{-17}^{+17}$ s post burst using equation \ref{eqn_4}.

In the case of \thisgrbb, our early optical observations with FRAM-ORM telescope reveals a smooth bump in the optical afterglow light curve. We fitted the optical bump using the smoothly broken power-law function (see equation \ref{eqn_3}) and calculated the following temporal parameters: rising temporal index $\alpha_r$=$-0.83_{-0.41}^{+0.45}$, decay temporal index $\alpha_d$=$2.24_{-2.27}^{+2.14}$, and the break time $t_{\rm b}$= $161.60_{-42}^{+50}$ s post burst, respectively. Further, we determine the peak time $t_{\rm p}$= 179.90$_{-50}^{+50}$ s post burst using the equation \ref{eqn_4} (smoothness parameter $k$ = 3).

{\bf Reverse shock origin:} According to the fireball model, the forward (moving towards the external/surrounding medium), and the reverse (propagating into the blast wave) shocks are originated due to the results of external shock between the blast-wave and ambient medium. The observed early optical peak in the afterglow light curve might be created due to the reverse shock \citep{2003ApJ...582L..75K}. We used the expected temporal indices (for ISM and wind mediums in the thin and thick shells reverse shock cases) to determine the origin of early optical bump for both bursts (see Table 1 of \cite{2015ApJ...810..160G}). We note that the observed temporal indices values during the rising and decaying part of the bump of both the bursts are inconsistent with the expected values from the reverse shock decay in ISM or wind mediums. Therefore, the observed early bump in the optical light curves could not be due to the reverse shock. 

{\bf Onset of optical afterglow:} 
The early peak in the afterglow light curve can be produced by the onset of the afterglow and it can be used to calculate the the bulk Lorentz factor of the fireball \citep{1999ApJ...520..641S}. \cite{2018AA...609A.112G} examined the early temporal coverage of GRBs with a measured redshift and calculated the bulk Lorentz factor for 67 bursts using the peak of the afterglow. In addition, they also summarize the methods used by various authors to estimate the bulk Lorentz factor.
In the case of thin shell regime (\tninty is less than deceleration time), the peak time (in the rest frame) of the light curve provides a direct measurement of the deceleration time $t_{\rm dec}$. At the deceleration time, the Lorentz factor diminishes by a factor of two from its initial value ($\Gamma_{0}$) and enters into the self-similar deceleration phase. For the homogeneous medium surrounding the burst, the thin shell deceleration time is related to deceleration radius $R_{\rm dec}$ and bulk Lorentz factor with the following relation \citep{1999ApJ...520..641S, 2007AA...469L..13M}: 

\begin{equation} 
\label{eqn_1}
t_{\rm dec} = R_{\rm dec}/2c \Gamma_{0}^{2}
\end{equation}

Further,  \cite{2018AA...609A.112G} generalized the relation for both types of the surrounding medium (see equation \ref{eqn_2}).

\begin{equation} \label{eqn_2}
\Gamma_0 \sim{ K\left[\frac{E_{\rm \gamma, iso}}{ n m_{p} c^{5-s} \eta t_{p,z}^{3-s}}\right]^{\frac{1}{8-2s}}}
\end{equation}
Where s = 0 and K = 1.702 for ISM, and s = 2 and K = 1.543 for the wind-like ambient medium. $E_{\rm \gamma, iso}$ is the isotropic equivalent $\gamma$-ray energy, $m_{\rm p}$ is the proton mass, $n$ is the circumburst medium density, and $\eta$ is the radiative efficiency of the fireball. 

The optical light curve of both \thisgrba and \thisgrbb has a smooth bump that is well separated from the prompt emission, implying that the emission is in the thin shell regime. For the present analysis, we consider $\eta$ equal to 0.2 for both types of ambient medium. Further, we assume the value of $n$ = 1 cm$^{-3}$ for ISM medium. The wind medium circumburst profile is governed by the mass loss rate $\dot{M}$ and wind velocity $v_w$. Therefore, we used $n$ = $10^{35}\dot{M}_{-5}/v_{w,3}$ cm$^{-3}$ for wind medium \citep{2018AA...609A.112G}. To determine the  bulk Lorentz factor using the early bump, we have used the methodology described by (\citealt{2007AA...469L..13M},  hereafter \textbf{M2007}).

\begin{figure}
\centering
\includegraphics[scale=0.5]{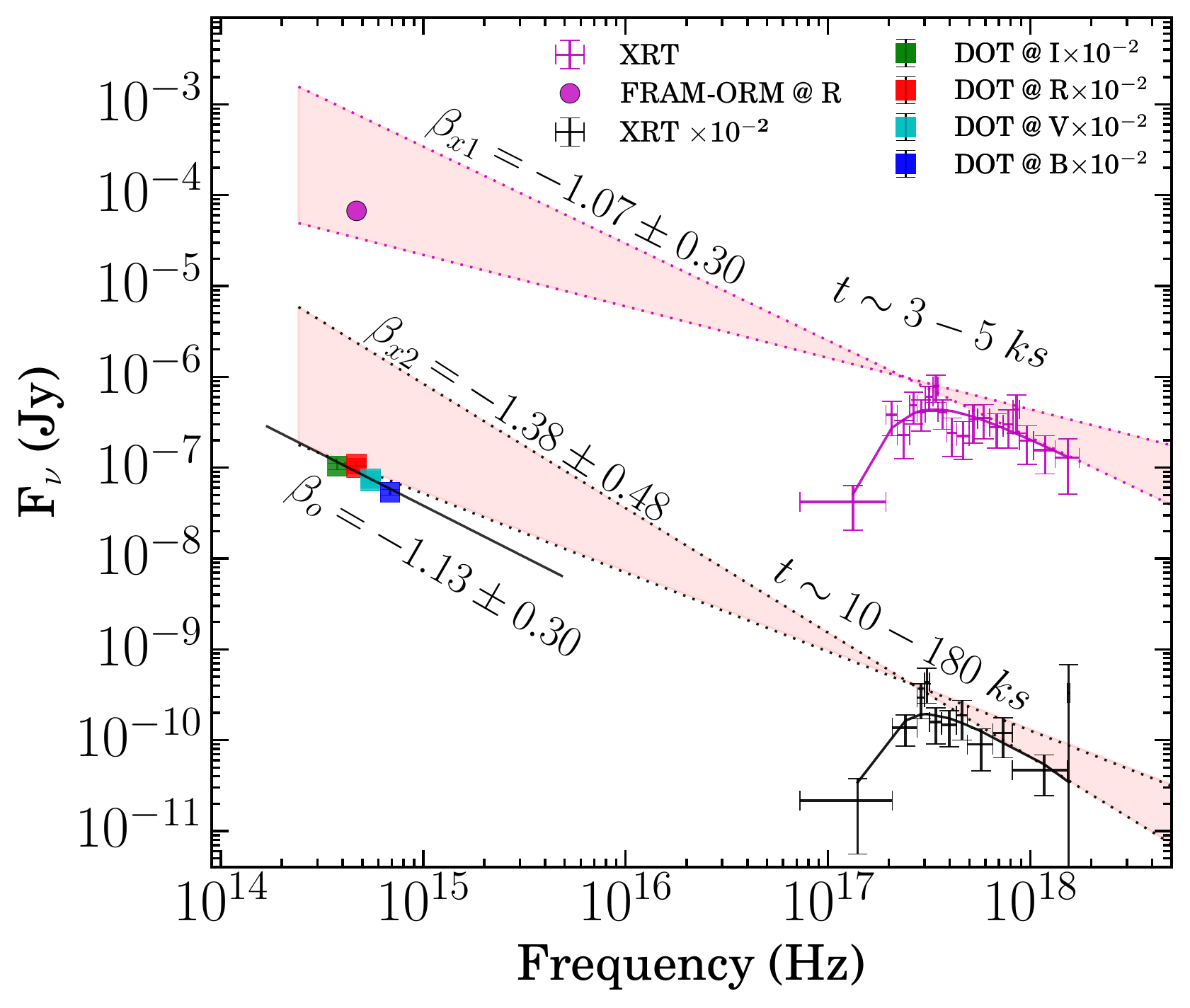}
\includegraphics[scale=0.5]{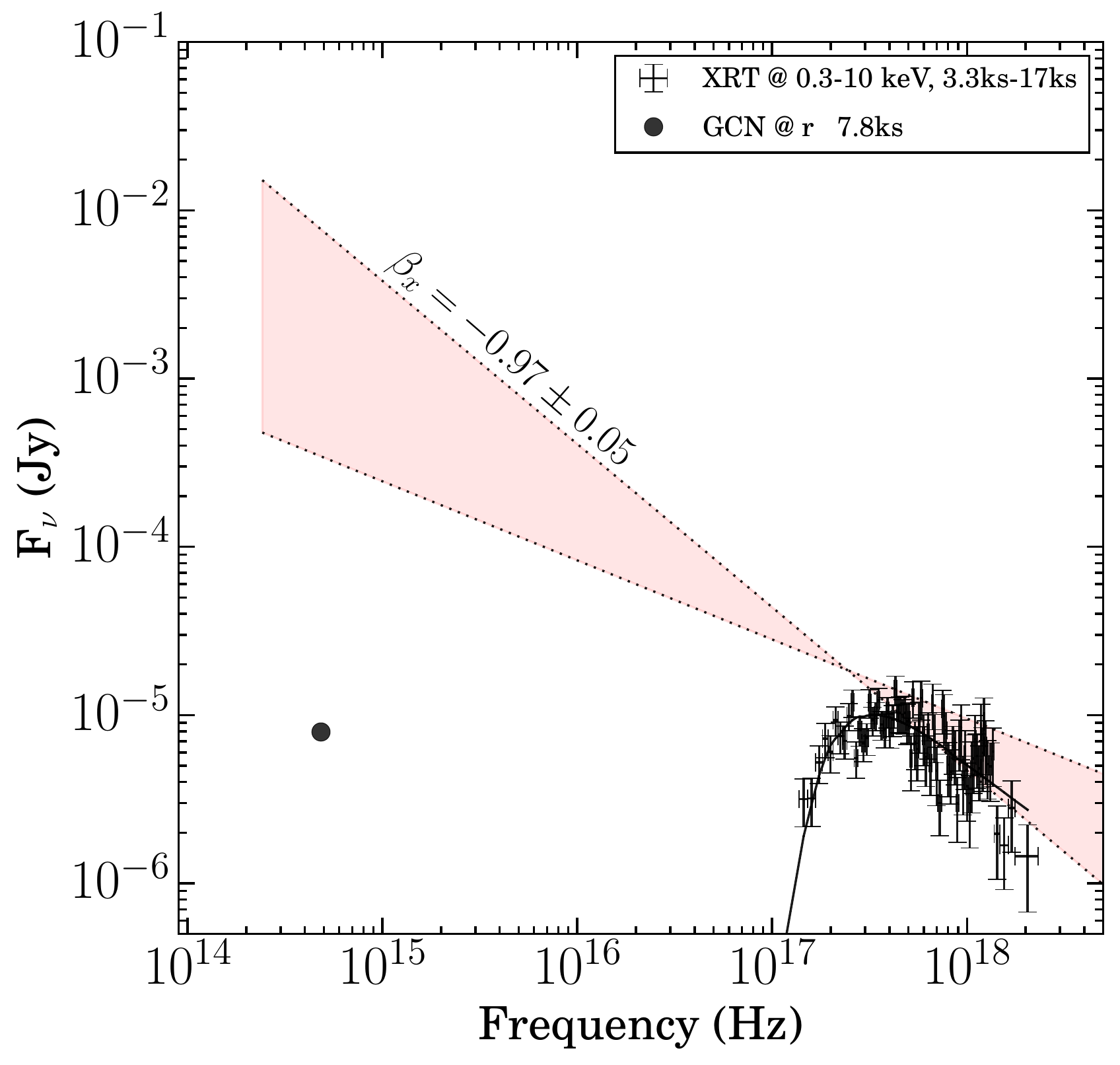}
\caption{Optical to X-ray SEDs for \thisgrba and \thisgrbb. Upper panel: represents the SEDs for \thisgrba at two different epochs; @ 3-5 ks and @ 10-180 ks. The optical data are shown with a filled circle and filled squares at 5ks, and 52-74ks, respectively, are taken from FRAM-ORM and DOT observations. Data with errors represents the \swift-XRT spectrum (@ 0.3-10 \keV). Flux densities at the second epoch are multiplied by $10^{-2}$. Lower panel: represents the SEDs for \thisgrbb at the 2.9-17 ks from \swift-BAT trigger time. The filled circle represents the r-band data point taken from GCN circular no. 29066 \citep{2020GCN.29066....1I} and the data with errors represents the \swift-XRT spectrum (@ 0.3-10 \keV). Pink shaded region represent the range of predicted synchrotron spectrum $-(p-1)$/2 = $\beta_x$+0.5 and $-p$/2 = $\beta_x$. Optical observations are corrected for Galactic extinction.}
\label{fig:SED}
\end{figure}

\subsubsection{Spectral energy distributions and closure relation} 
\label{sec:SED}
We created the SEDs using joint X-ray and optical data for both the bursts to constrain the break frequencies of the broadband synchrotron model at two epochs for \thisgrba and one epoch for \thisgrbb, respectively. We followed the methodology given in \cite{2021A&A...646A..50H, 2022arXiv220507790C, 2022JApA...43...11G} to create these SEDs. For \thisgrba, the optical temporal index ($\alpha_{\rm o}$=$- 0.92_{-0.09}^{+0.08}$) follows a simple power-law decay post optical bump and satisfies the relation $\alpha= (3p-3)/4$ in the slow cooling regime of the ISM medium, which indicates that the optical emission always remains in the $\nu_m$ $<$ $\nu_o$ $<$ $\nu_c$ spectral regime. The details of epochs and temporal/spectral indices are given in Table \ref{tab:closure_relations} of the appendix. 
For \thisgrba, we performed follow-up observations with 3.6m DOT in the BVRI filters. Our 3.6m DOT optical observations covering the temporal window from 52 ks–74 ks post burst are shown in Figure \ref{fig:SED}. We created an optical SED to calculate the optical spectral index $\beta_{o2}$ and constrain the spectral regime at the epoch of DOT observations. The magnitudes are corrected for galactic extinction. In addition, we used negligible host extinction from \cite{2022arXiv220512750G}, and fitted a power-law function to determine the optical spectral index $\beta_{o2}$. We obtained a value of $\beta_{o2}$ =-1.13$\pm$0.30. The calculated value of $\beta_{o2}$ using DOT observations is shallower than ($\sim$0.3) the X-ray spectral index ($\beta_{x2}$=-1.38$_{-0.49}^{+0.47}$) at this epoch. The measured change in the spectral indices of optical and X-ray data indicates that the cooling frequency lies in between the optical and X-ray spectral regime, although large associated errors due to limited data points and it is hard to discard other possibilities as $\beta_{o2}$ and $\beta_{x2}$ are consistent within 1 sigma. We noted that an ISM-type surrounding medium and $\nu_m$ $<$ $\nu_o$ $<$ $\nu_c$ $<$ $\nu_x$ spectral regime is also favored by the afterglow modeling of \thisgrba by \cite{2022arXiv220512750G} at this epoch, although, they mainly use the preliminary optical magnitudes reported in various circulars. Using the calculated value of $\beta_{o2}$, we determined the power-law index of electron distribution $p$ = (2 $\times$ $\beta_{o2}$ + 1) = 3.26 $\pm$ 0.60 at this epoch. 

For \thisgrba, we create optical-X-ray SED at two different epochs (as shown in Figure \ref{fig:SED}, upper panel). At the first epoch, the X-ray spectral index $\beta_{x1}$=-1.07$_{-0.30}^{+0.31}$ is shallower than ($\sim$0.3) the X-ray spectral index ($\beta_{x2}$=-1.38$_{-0.49}^{+0.47}$) at the second epoch. The evolution of X-ray spectral index $\beta_{x}$ also remain consistent within 1 sigma error. We noted that $\nu_c$ is just below or within the XRT band spectral regime, also favored by the afterglow modeling \citep{2022arXiv220512750G} at this epoch. Considering no spectral break between optical and X-ray emissions, we extrapolate the X-ray spectral index $\beta_{x1}$ toward the optical band to estimate the intrinsic optical flux. By comparing optical R-band flux with that obtained from the extrapolation, we determine an amount of extinction in the R-band $A_R$ = 2.2 mag. At the second epoch, the cooling frequency $\nu_c$ has crossed the X-ray bands at $\sim$ 52 ks.

For \thisgrbb, the late-time optical light curve seems to follow the normal decay with a power-law index, $\alpha_o$ = $-1.05_{-0.10}^{+0.11}$. Additionally, the X-ray temporal index $\alpha_x$ = $-2.21_{+0.10}^{-0.11}$ and the X-ray spectral index $\beta_{x}$ = $-0.97_{-0.05}^{+0.05}$ remain almost constant throughout the emission. The X-ray/optical temporal/spectral indices satisfied the closure relation for wind medium with emission $\nu_m$ $<$ $\nu_o$ $<$ $\nu_c$ $<$ $\nu_x$. A wind-type surrounding medium is also favored by the afterglow modeling of \thisgrbb by \cite{2022MNRAS.513.1895R}.

For \thisgrbb, similarly, we created a optical-X-ray SED from optical to X-ray frequencies and extrapolated our X-ray spectral index (@$\sim{2.5h}$) to the lower frequency, see Figure \ref{fig:SED}, lower panel. The Galactic corrected r band magnitude lies much below the extrapolated X-ray power-law slope. Considering no spectral break and a spectral break between X-ray and optical frequencies in the SED, we estimated the intrinsic optical flux by extrapolating the X-ray spectral index at r-band frequency. By comparing the estimated optical flux to the observed VLT optical flux at 2.19 h \cite{2020GCN.29066....1I}, we determine the host extinction in the r-band $A_r \sim{8-5}$ mag, supporting the dark nature of the burst, consistent with the result of \cite{2022MNRAS.513.1895R}.

\section{Discussion}
\label{discussion}

\subsection{Bulk Lorentz factor and characteristic fireball radius of VHE detected GRBs} 
\label{gamma_0} 

There are different ways to derive/put constraints on the bulk Lorentz factor of GRBs \citep{2018AA...609A.112G}. In this section, we have derived the values of bulk Lorentz factor for VHE detected GRBs using three different methods using onset of the afterglow, using Liang correlation, and using prompt $\Gamma_{\rm min}$-$t_{\rm mvts}$ relation.  

{\bf Using onset of the afterglow:} We have used equation \ref{eqn_2} to calculate the bulk Lorentz factors for \thisgrba, \thisgrbb, and the other four well-known VHE detected GRBs, GRB 160821B (this GRB has no firm detection, but an evidence of signal at VHE at the level of 3 $\sigma$), GRB 180720B, GRB 190114C, and GRB 190829A. We noted that for the last four VHE GRBs, there is no early bump/peak observed in their optical afterglow light curves. For such cases, we assumed the earliest optical observations as upper limits for the peak time for the afterglow and constrained the lower limit on the bulk Lorentz factor for these bursts. The calculated values of $\Gamma_{0}$ using equation \ref{eqn_2}, and corresponding peak times are listed in the first column of Table \ref{tab:bulk_gamma}, for both types of surrounding media. In addition, we also calculated the deceleration radius of the fireball at the peak time using the following equation taken from \textbf{M2007}: 

\begin{equation} 
\label{eqn_5}
R_{\rm dec}=2ct_{\rm p}\Gamma_{0}^{2}/(1+z)
\end{equation}

The calculated values of deceleration radius are also listed in the first column of Table \ref{tab:bulk_gamma}. Further, we also verified these parameters using various correlations between the bulk Lorentz factor and prompt emission properties, such as $\Gamma_{0}$- $E_{\rm \gamma, iso}$, and $\Gamma_{0}$-$L_{\rm \gamma, iso}$ and noted the the values are consistent for the sub-sample of VHE detected GRBs (see Figure \ref{fig:onset_corelation2}).

{\bf Using Liang correlation:} \citealt{2010ApJ...725.2209L} (hereafter \textbf{L2010}) extensively searched for the onset of the afterglow in the X-ray and optical afterglow light curves using the published literature and \swift-XRT catalog. They found that twenty bursts in the optical and twelve bursts in the X-ray bands displayed the onset features in their corresponding afterglow light curves. In addition, \textbf{L2010} also discovered tight correlations between the isotropic equivalent gamma-ray energy $E_{\rm \gamma, iso}$ and the initial bulk Lorentz factor of the fireball. The correlation can be given as: $\Gamma_0 \sim {182 (E_{\rm \gamma, iso}/10^{52}\rm  erg)^{(0.25\pm0.03)}}$.\\ 
For all the seven known VHE detected GRBs, we derive the bulk Lorentz factor using $\Gamma_{0}$ - $E_{\rm \gamma, iso}$ correlation, and calculated values along with \tninty are listed in the second column of Table \ref{tab:bulk_gamma}. We also calculated the fireball radius at the end of the prompt emission using the following relation: R$_{\rm em}=2ct\Gamma_{0}^{2}/(1+z)$ using t $\sim$ \tninty in the equation. The calculated values of fireball radius given are also listed in the second column of Table \ref{tab:bulk_gamma}.\\

Additionally, \textbf{L2010} also studied the correlations between the characteristic bump/onset parameters, such as peak time, FWHM, isotropic gamma-ray energy, etc., and they found that most of these parameters are strongly correlated with each other. We also compare the properties of the observed bump in the early optical light curves of \thisgrba and \thisgrbb with the known correlations. We noted that most of the characteristic parameters of the observed bump in the early optical light curves of \thisgrba and \thisgrbb are consistent with the correlations studied by \textbf{L2010}, confirming the nature of the bumps due to the onset of forward shocks in the external ambient medium. However, \thisgrbb does not follow the correlation between $E_{\rm \gamma, iso}$ and optical peak luminosity ($L_{\rm p, opt}$), supporting the dark behavior of \thisgrbb \citep{2022MNRAS.513.1895R}. The correlations between the different parameters of the optical bumps of \thisgrba and \thisgrbb, along with data taken from \textbf{L2010} are shown in Figure \ref{fig:onset_corelation} of the appendix.\\
{\bf Using  Prompt \boldmath $\Gamma_{\rm min}$-$t_{\rm mvts}$ relation:} In addition to the above methods, we also used the prompt $\Gamma_{\rm min}$-$t_{\rm mvts}$ (minimum variability time scales) relation to calculate the lower bound on the bulk Lorentz factor ($\Gamma_{\rm min}$) and to calculate emission radius during the prompt emission phase. We used the following relations between $\Gamma_{\rm min}$-$t_{\rm mvts}$ and $R_{\rm c}$-$t_{\rm mvts}$ derived by \citealt{2015ApJ...811...93G} (hereafter \textbf{G2015}):
\begin{equation} 
\label{eqn_7}
\Gamma_{\rm min} \geq 100 \left(\frac{L_{\rm \gamma, iso}}{10^{51} \rm erg/ s}\frac{1+z}{t_{\rm mvts}/0.1s}\right)^{1/5}
\end{equation}

\begin{equation} 
\label{eqn_8}
R_{c} \sim{7.3 \times 10^{13} \left(\frac{L_{\rm \gamma, iso}}{10^{51} \rm erg/ s}\right)^{2/5} \left(\frac{t_{\rm mvts}/0.1s}{1+z}\right)^{3/5}} \rm cm
\end{equation}

We calculated the minimum variability time scales for the VHE detected GRBs using the Bayesian block method on the prompt emission light curve in the energy range of 8-900 \keV for \fermi-GBM and 15-350 \keV for \swift-BAT, respectively. The Bayesian blocks utilize the statistically significant changes to bin the prompt emission light curves of these bursts. We determine the minimum bin size, and the minimal variability time scales are equal to half of the width of the smallest bin of Bayesian blocks \citep{2018ApJ...864..163V}. The calculated values of $t_{\rm mvts}$, $\Gamma_{\rm min}$,  and $R_{\rm c}$ are listed in the third column of Table \ref{tab:bulk_gamma}. We noted that the emission radius lies in the range of $2\times 10^{13}$ cm - $4\times 10^{14}$ cm for all the seven known VHE detected GRBs, consistent with the results of \textbf{G2015}. The calculated emission radius for these VHE detected GRBs is much larger than the typical emission radius of the photosphere, suggesting that the emission took place in an optically thin region away from the central engine \citep{2022arXiv220507790C, 2020NatAs...4..174B}.

\subsection{Lorentz factor evolution and the possible jet composition}
The evolution of the bulk Lorentz factor can provide insight into the jet composition, the prompt emission location, and the radiation physics of GRBs. The jet of GRBs can be matter dominated (originated due to photosphere) or Poynting flux dominated (originated due magnetic reconnection). However, a third hybrid jet composition (quasi-thermal component together with a non-thermal component) formed in the internal shocks, is possible. In such hybrid case, it is expected that the Lorentz factor remains almost constant during the prompt emission phase and decreases during the onset of afterglow \citep{2019ApJ...883..187L}. On the other hand, in the case of Poynting flux dominated jet (a part of the magnetic field energy dissipates to accelerate the GRB jet), the Lorentz factor measured during the onset of afterglow is expected to be larger than that measured during the prompt phase \citep{2014ApJ...782...92Z}.\\
\begin{table*}
\centering
\caption{In this table, we summarize the evaluation of the Lorentz factors at various characteristic times and radii for seven VHE detected GRBs.}
\label{tab:bulk_gamma}
\begin{center}
 \begin{tabular}{|p{1.2cm}|ccc|ccc|ccc|c|}\hline
  VHE detected &	\multicolumn{3}{c|}{\textbf{\textbf{M2007}}} & \multicolumn{3}{c|}{\textbf{L2010}}&	\multicolumn{3}{c|}{\textbf{\textbf{G2015}}}\\[1ex]	
GRBs&	 $t_{\rm p}$ (s) & $\Gamma_0$  & $R_{\rm dec}$ (cm) & \tninty (s) &	 $\Gamma_0$&	 $R_{\rm em}$ (cm) &	 $t_{\rm mvts}$  (s)& $\Gamma_{\rm 0,min}$ & $R_{\rm c}$ (cm)\\\hline
  160821B &	     -   &	      - &	   -        &	0.5&	 69 $\pm$ 8  &	1.18$\times$10$^{14}$&	0.068&	 $>$88.74  &	2.78$\times$10$^{13}$\\
  180720B &	$<$ 73  &	 $>$576 &	 $<$2.19$\times$10$^{17}$  &	49&	 506 $\pm$ 66 &	4.55$\times$10$^{17}$&	0.024&	 $>$457.20 &	1.82$\times$10$^{14}$\\
  190114C &	$<$33.2 &	 $>$341 &	 $<$4.10$\times$10$^{16}$  &	25&	 407 $\pm$ 41 &	1.74$\times$10$^{17}$&	0.016&	 $>$472.86 &	1.53$\times$10$^{14}$\\
  190829A &	     -   &	      - &	   -        &	63&	 76 $\pm$ 8  &	2.02$\times$10$^{16}$&	0.210&	 $>$47.74  &	2.67$\times$10$^{13}$\\
  201015A &	184.64&	 204 &	 8.16$\times$10$^{16}$  &	9.78&	 143 $\pm$ 4 &	8.41$\times$10$^{15}$&	       $\sim{0.1}$ &	 $>$92.07  &	$\sim$~4.81$\times$10$^{13}$\\
  201216C &	179.9&	 310 &	 1.23$\times$10$^{17}$ &	29.9&	 513 $\pm$ 68 &	2.24$\times$10$^{17}$&	0.152&	 $>$286.99 &	3.60$\times$10$^{14}$\\ \hline
 221009A & $<$179 & $>$440 & $<$4.50$\times$10$^{17}$ & 327 &  $>$757 & $<$9.7$\times$10$^{18}$ & $\sim$~0.001 & $>$450.01 & $\sim$~1.1$\times$10$^{13}$ \\ \hline
 \end{tabular}
\end{center}
\end{table*}
In \S~\ref{gamma_0}, we have calculated the bulk Lorentz factor of the fireball for \thisgrba and \thisgrbb at different epochs to examine the evolution of Lorentz factor and constrain the possible jet composition. Initially, we calculated the the bulk Lorentz factor values during the prompt emission phase using two different methods: the first one using the relation between the lower limit on bulk Lorentz factor $\Gamma_{\rm 0,min}$ and the minimum variability time scale $t_{\rm mvts}$ (\textbf{G2015}). In the second method, we have used the tight correlation between the $\Gamma_{0}$ and total isotropic gamma-ray energy released during the prompt emission to calculate the bulk Lorentz factor (\textbf{L2010}). Finally, we calculated the bulk Lorentz factor values (for the different types of ambient medium) during the onset of forward shock emission using the observed early peak in the optical afterglow light curves \citep{1999ApJ...520..641S, 2007AA...469L..13M, 2018AA...609A.112G}.  

In the case of \thisgrba, the measured values of $\Gamma_{0}$ at different epochs are following:
$>$92.07 using \textbf{G2015}, 143 using \textbf{L2010}, and 204 for ISM-like medium using the onset of forward shock emission. The lower limit on the measured value of $\Gamma_{0}$ (using \textbf{G2015}) is consistent with that measured using the tight correlation between $\Gamma_{0}$ and $E_{\rm \gamma, iso}$. For \thisgrba, the closure relations support a homogeneous medium (see \S~\ref{sec:SED}), also consistent with \cite{2022arXiv220512750G}. The value of the bulk Lorentz factor during the onset of the afterglow increases for an ISM-like ambient medium, support a Poynting flux dominated jet composition for \thisgrba, although there is not a very large difference between the bulk Lorentz factor values at different epochs. In the case of \thisgrbb, the measured values of $\Gamma_{0}$ at different epochs are following: $>$286 using \textbf{G2015}, 513 using \textbf{L2010}, and 310 for wind-like medium using the onset of forward shock emission.
For this GRB also, the lower limit on the $\Gamma_{0}$ (using \textbf{G2015}) is lower than that measured using tight correlation between the $\Gamma_{0}$ and $E_{\rm \gamma, iso}$. For \thisgrbb, the closure relations support a wind-like medium (see \S~\ref{sec:SED}), also consistent with \cite{2022MNRAS.513.1895R}. The value of the bulk Lorentz factor during the onset of the afterglow decreases for a wind-like ambient medium, support a matter dominated jet composition for \thisgrbb.

\subsection{Progenitors of \thisgrba and \thisgrbb: Collapsar origin?}

The recent discoveries of short GRBs (GRB 200826A, GRB 211227A) from the collapse of massive stars \citep{ahumada2021discovery, 2022ApJ...931L..23L} and long GRB (GRB 211211A) from binary mergers \citep{2022arXiv220410864R} challenge our understanding about progenitor systems of GRBs. In this section, we examine the possible progenitors of \thisgrba and \thisgrbb. There are two possible models (collapsar and binary merger) for the progenitor of GRBs. According to the Collapsar model, the central engine (black hole or magnetar) forms after the death of a massive stellar object emits a jet that successfully penetrates the preexisting envelope around the progenitor star. If the jet does not have sufficient energy to breakout, it deposits all of its energy into the surrounding envelope to form a mildly relativistic cocoon around it. This process is known as shock breakout and it gives emission in gamma-rays, with a luminosity 2-3 orders less than typical long GRBs \citep{2011ApJ...739L..55B}. Such subclass of GRBs with low isotropic gamma-ray luminosity (order of $10^{46}-10^{49}$ erg $\rm s^{-1}$) emitted during the prompt emission phase is assumed to have a different origin than normal long GRBs. In the VHE detected GRBs sample, GRB 190829A is a peculiar low-luminosity GRB with no shock breakout origin \cite{2020ApJ...898...42C}. In addition, \thisgrba also belongs to the low-luminosity family of GRBs with a supernova bump in the late optical light curve, which motivate us to test whether it has a collapsar or shock breakout origin. 

In the case of collapsar origin, it is expected that the observed duration (\tninty) of the burst should be greater than the jet breakout time ($T_{\rm break}$) from the surrounding envelope. \cite{2011ApJ...739L..55B} suggested that if the ratio \tninty/$T_{\rm break}$ $<$ 1 and the jet failed to cross the envelope, in such case, the burst is expected to originate from the shock breakout. On the other hand, if \tninty/$T_{\rm break}$ $>$ 1 and the jet successfully crosses the envelope, the burst is expected to originate from the collapsar. We calculated the jet breakout time for \thisgrba the following equation given by \cite{2011ApJ...739L..55B}: $T_{\rm break} (s) \sim{15 \epsilon_{\gamma}^{1/3} L_{\rm \gamma,iso,50} \theta_{10^{\circ}}^{2/3} R_{11}^{2/3} M_{15\odot}^{1/3}}$.

The jet breakout time  depends on the isotropic equivalent luminosity, the observed opening angle, progenitor mass, and radiation efficiency. To calculate T$_{\rm break}$ for \thisgrba, we assume the typical values of M = 15$M_{\odot}$, $\theta$ = $10^{\circ}$ and $\epsilon_{\gamma}=0.2$ \citep{2011ApJ...739L..55B}. We obtained the ratio \tninty/$T_{\rm break}$ = 1.75 for \thisgrba, which supports the collapsar origin of the burst. We also calculated the \tninty/T$_{\rm break}$ ratio for \thisgrbb considering $\theta$ = $1^{\circ}$ and $9^{\circ}$ with progenitor mass 12$M_{\odot}$ and 25$M_{\odot}$ \citep{2022MNRAS.513.1895R} and found that \tninty/$T_{\rm break}$ lies in the range of 30–163, which confirms the collapsar origin of \thisgrbb. 

\subsection{Comparison of \thisgrba and \thisgrbb with other VHE detected GRBs}

In all the seven cases of VHE GRBs, VHE photons were detected during the afterglow phase, although, VHE photons may enrich the prompt emission of GRBs also. In the present section, we compared the afterglow results obtained for \thisgrba and \thisgrbb (see \S~\ref{sec:result}) with a sample of well-known VHE detected GRBs \citep{2019Natur.575..455M, 2019Natur.575..459M, 2021Univ....7..503N, 2022Galax..10....7N, 2019ApJ...885...29F}. In addition to afterglow comparison, we also collected the prompt emission properties of VHE detected GRBs, listed in Table \ref{tab:gcn}.

\subsubsection{Comparison of X-ray and optical afterglow light curves of VHE detected GRBs}
In the present section, we compared the X-ray and the optical (see Figure \ref{fig:XRT_opt} of the appendix) afterglow luminosity light curves of \thisgrba and \thisgrbb with other well-studied VHE detected GRBs (GRB 160821B, GRB 180720B, GRB 190114C, GRB 190829A, and GRB 221009A). In addition, we also included X-ray (if available) and optical (if available) light curves of a nearly complete sample of nearby and supernovae-connected GRBs (GRB 050525A/SN 2005nc ($z$ = 0.606), GRB 081007A/SN 2008hw ($z$ = 0.530), GRB 091127A/SN 2009nz ($z$ = 0.490), GRB 101219B/SN 2010ma ($z$ = 0.552), GRB 111209A/SN 2011kl ($z$ = 0.677), GRB 130702A/SN 2013dx ($z$ = 0.145), GRB 130831A/SN 2013fu ($z$ = 0.479), GRB 060218/SN 2006aj ($z$ = 0.033), GRB 120422A/SN 2012bz ($z$ = 0.282), GRB 130427A/SN ($z$ = 0.340), GRB 190114C/SN ($z$ = 0.425), GRB 190829A/SN 2019oyw ($z$ = 0.078), GRB 200826A/SN ($z$ = 0.748)) for the comparison as most of VHE detected bursts are nearby and connected with supernovae. The comparison of X-ray afterglow light curves indicates that VHE detected GRB 180720B and \thisgrbb have extremely bright X-ray emission (just below the brightest X-ray emission observed from GRB 130427A at early epochs). GRB 221009A and GRB 190114C also have a very bright X-ray emission but less than those of GRB 180720B and \thisgrbb at early epochs. On the other hand, GRB 160821B, GRB 190829A, and \thisgrba have a faint X-ray emission. The X-ray afterglow of \thisgrba has a nearly comparable brightness with GRB 190829A. GRB 160821B, being a short burst, has the faintest X-ray light curve (after the steep decay phase) with respect to present VHE sample.

For the optical afterglow light curve comparison, we collected the R band light curve of VHE detected GRBs (other than GRB 221009A) from the literature \citep{2019Natur.575..455M, 2021MNRAS.504.5685M, 2021RMxAC..53..113G, 2019ApJ...885...29F, 2020ApJ...892...97J}. For GRB 221009A, we collected the optical data from the GCN circulars\footnote{\url{https://gcn.gsfc.nasa.gov/other/221009A.gcn3}}. For a nearly complete sample of nearby and supernovae-connected GRBs, we obtained the optical data from \cite{2022NewA...9701889K} and references therein. We noted that GRB 180720B has the highest optical luminosity at early epochs in comparison to the present sample (similar to the X-ray light curve) and the light curve displays a smooth power-law decay across the emission period \citep{2019ApJ...885...29F}. At later phases ($\sim$ 0.2 days post burst), GRB 221009A seems to have the highest optical luminosity in comparison to the present sample. In the case of \thisgrbb, despite of the very bright X-ray emission, the optical light curve is one of the faintest with respect to present sample, typical to those observed in the case of dark GRBs. Further, we noted that \thisgrba and \thisgrbb have a comparable optical luminous light curves. GRB 190114C \citep{2021RMxAC..53..113G},  GRB 190829A \citep{2021A&A...646A..50H}, and \thisgrba \citep{2022arXiv220512750G} had a late time bump in their optical light curve associated with their underlying supernovae explosions. In addition, we noted that \thisgrba and \thisgrbb are the only the bursts with very early smooth bump (the onset of the forward shock) in their optical light curves with respect to present sample.

\subsubsection{Possible origin of VHE emission from high and low luminosity bursts}
The broadband afterglow observations of the VHE detected GRBs could not be explained by the typical external shock synchrotron model \citep{2019Natur.575..459M, 2021Sci...372.1081H}. The multi-wavelength modeling of the observed double bump SEDs of VHE detected GRBs demands an additional SSC/Inverse Compton component. VHE detected GRBs are expected to be luminous and nearby such as GRB 180720B, GRB 190114C, and GRB 221009A. However, some of the recent detection of VHE emission from low/intermediate-luminosity bursts (such as GRB 190829A, and \thisgrba), open a new question about their possible progenitors and viewing geometry. The high luminosity GRBs are typically observed on-axis with narrow viewing, on the other hand, low-luminosity bursts are typically observed off-axis with wide viewing angle. Recently, some authors \citep{2022arXiv220813987S, 2022MNRAS.513.1895R} used two different jet components (narrow and wide) model to explain the origin of the observed properties of high and low luminosity VHE detected GRBs. According to this model, the early broadband emission of the high luminosity GRBs are explained using narrow jet component with typical opening angle $\theta < 0.86^{\circ}$. On the other hand, the low luminosity GRBs are explained using wide jet component with typical opening angle $\theta > 6^{\circ}$ \citep{2022arXiv220813987S}. \cite{2022arXiv220813987S} performed the broadband modeling of nearby and low-luminosity GRB 190829A using two-component jet model and noted that the  prompt and afterglow emissions could not be described from the narrow jet component. The observed low/intermediate luminosity nature of GRB 190829A is explained by assuming that the viewing angle is greater than the opening angle of its narrow jet component (off-axis observations). \thisgrba is also a nearby intermediate luminosity GRB and might have a very similar viewing geometry to GRB 190829A (viewing angle greater than the narrow jet opening angle). In the case of GRB 180720B, and GRB 190114C, the observed high luminosity nature of these GRBs are explained from emission within the narrow jet component (on-axis observations). The same applies to \thisgrbb, given its observed high luminosity.

The observed typical sub-TeV bump are explained either using SSC or external Inverse Compton \citep{2019Natur.575..459M, 2019Natur.575..464A}. According to two component model, during the early phase, both the components (narrow and wide) of the jet have almost equal velocities. This negligible difference in the velocities helps to discard the possibility of the contribution from the external Inverse Compton during early phase \citep{2022arXiv220813987S}. Therefore, the SSC emission mechanism is expected to explain the early VHE emission, for example GRB 190114C, and GRB 190829A \cite{2019Natur.575..459M, 2021Sci...372.1081H}. In the case of GRB 180720B, VHE photons were observed several hours after the detection. Therefore, considering the two jets moving at different velocities, Inverse Compton significantly contributed to the observed VHE emission \citep{2019Natur.575..464A}.  In the case of \thisgrbb, considering the synchrotron process as a possible radiation mechanism at t$\sim$ 3$\times10^3$ - $10^4$ s post burst at X-ray frequencies, we extrapolated the X-ray spectral index $\beta_x$ towards the GeV-TeV energies (for the spectral regime $\nu_{c} ~< \nu_{x}$), and estimated the expected flux density $F_\nu \sim{2.3\times10^{-11}}$ Jy, and $F_\nu \sim{3\times10^{-13}}$ Jy at 1 GeV and 0.1 TeV, respectively. \fermi-LAT could not detect the GeV emission from \thisgrbb during the interval \fermiT + 3500 s to \fermiT + 5500 s post burst. This is in agreement with the fact that the expected flux density at 1 GeV at t$\sim$ 3$\times10^3$ - $10^4$ s post burst is below the sensitivity of \fermi-LAT instrument. Furthermore, from the observed peak in the early optical light curve of \thisgrbb, we calculated the initial Lorentz factor of the burst $\Gamma\sim$ 300. With this Lorentz factor and $z=1.1$, photons of maximum energy, $\leq$15 GeV, are allowed by synchrotron process \cite{2019ApJ...885...29F}. 

Similarly, for \thisgrba expected flux density $F_\nu \sim{3.8\times10^{-13}}$ Jy, and $F_\nu \sim{2.8\times10^{-15}}$ Jy at 1 GeV and 0.1 TeV, respectively, is obtained at 3ks-5ks. A bulk Lorentz factor $\sim$200 is obtained from the observed peak in the optical light curve. With this Lorentz factor and $z=0.426$, we estimated the maximum energy of synchrotron photons $\leq$14 GeV \cite{2019ApJ...885...29F}.  

Therefore, observed early VHE emission could not be explained using the synchrotron emission model. The previous discussion suggests that the early detection of VHE photons from \thisgrba and \thisgrbb required that the ultra-relativistic outflow boosts the energy of low-energy synchrotron photons to the VHE energies via the SSC process.

\section{Summary and Conclusion}
\label{summaryandconclusion}

In this work, we presented a detailed analysis of the prompt and afterglow emission of two VHE detected GRBs \thisgrba and \thisgrbb and their comparison with a subset of similar bursts. In spite of showing prompt emission characteristics of other typical LGRBs, \thisgrba is a low/intermediate luminosity whereas \thisgrbb is one of the high luminosity ones. Detailed time-resolved spectral analysis of \fermi observations of \thisgrbb suggests that the low energy spectral index ($\alpha_{\rm pt}$) remained within the expected values of synchrotron slow and fast cooling limits supporting the synchrotron emission as a possible emission mechanism. Searches for the additional thermal component indicate that some of the Bayesian bins have a quasi-thermal component centered around the beginning or near the peaks of the light curve.

Further, we studied the evolution of spectral parameters and find a rare feature where \Ep and $\alpha_{\rm pt}$ both showing flux tracking behavior (double tracking) throughout the duration of \thisgrbb as published recently \citep{2021MNRAS.505.4086G}, proposing the observed relation between \Ep and flux in terms of fireball cooling and expansion. In such a scenario, during the fireball expansion, the magnetic field reduces resulting in a lower intensity and \Ep. However, increased central engine activity during the bursting phase might increase the magnetic field, resulting in higher \Ep and/or intensity. If such a scenario is true, the magnetic field should be strongly correlated with \Ep \citep{2021A&A...656A.134G}. Interestingly, we found a strong correlation between \Ep (derived using empirical \sw{Band} function) and the magnetic field (B) (derived using the physical \sw{Synchrotron} model), supporting our results discussed above. On the other hand, strong correlation observed between $\alpha_{\rm pt}$-flux can be explained in terms of sub-photospheric heating in a flow of varying entropy \citep{2019MNRAS.484.1912R}.

Our earliest optical observations of the afterglow of \thisgrba using FRAM-ORM and BOOTES and \thisgrbb using FRAM-ORM robotic telescopes display a smooth bumps, consistent with the onset of afterglow in the framework of the external forward shock model \citep{1999ApJ...520..641S}. Using the observed optical peak, we determined the initial bulk Lorentz factors of \thisgrba and \thisgrbb: $\Gamma_{0}$ = 204 for the ISM-like, and $\Gamma_0$ = 310 for wind-like ambient media, respectively. Further, we studied evolution of the Lorentz factors and constrain the possible jet compositions for both the bursts. Evolution of the Lorentz factors suggests a Poynting flux-dominated jet for \thisgrba whereas for \thisgrbb, preference for an internal shock-dominated jet, consistent with time-resolved spectral analysis. Furthermore, we investigated the possible progenitors of \thisgrba and \thisgrbb by constraining the time taken by the jet to break through the surrounding envelope ($T_{\rm break}$) using the relation given by \cite{2011ApJ...739L..55B}, and taking the ratio of observed \tninty to $T_{\rm break}$. We find that for both the bursts, this ratio is greater than one, supporting the collapsar as the possible progenitors of \thisgrba and \thisgrbb. Also, late time optical follow-up observations of \thisgrba reveal an associated supernova \citep{2022arXiv220512750G}, additional evidence confirming the collapsar origin.

Finally, we compare the properties of \thisgrba and \thisgrbb with other similar VHE detected GRBs. Our findings suggest that VHE emission is common both in high and low luminosity GRBs. Our study also suggests that SSC process is needed to explain the VHE emission of these bursts. Early follow-up observations of similar sources using robotic telescopes are very crucial not only to constrain the Lorentz factors/their evolution but also to decipher other yet least explored aspects of underlying physics like radiation mechanism, jet compositions, etc. 

\section*{Acknowledgments}
Authors are thankful to the anonymous referee for his/her positive and valuable comments. RG and SBP acknowledge the financial support of ISRO under AstroSat archival Data utilization program (DS$\_$2B-13013(2)/1/2021-Sec.2). AA acknowledges funds and assistance provided by the Council of Scientific \& Industrial Research (CSIR), India with file no. 09/948(0003)/2020-EMR-I. AJCT acknowledges support from the Spanish Ministry project PID2020-118491GB-I00 and Junta de Andalucia grant P20\_010168. YDH acknowledges support under the additional funding from the RYC2019-026465-I. MCG acknowledges support from the Ram\'on y Cajal Fellowship RYC2019-026465-I (funded by the MCIN/AEI /10.13039/501100011033 and the European Social Funding). This research has used data obtained through the HEASARC Online Service, provided by the NASA-GSFC, in support of NASA High Energy Astrophysics Programs. SBP thankfully acknowledge inclusion of the photometric calibration data of GRB 201015A taken with the 4Kx4K CCD Imager acquired as a part of the present analysis and extends sincere thanks to all the observing and support staff of the 3.6m DOT to maintain and run the observational facilities at Devasthal Nainital.

\facilities{FRAM-ORM, BOOTES, 3.6m DOT (IMAGER)}

\software{DAOPHOT-II \citep{1987PASP...99..191S}, IRAF \citep{1986SPIE..627..733T, 1993ASPC...52..173T}, 3ML \citep{2015arXiv150708343V}, XSPEC \citep{1996ASPC..101...17A}, HOTPANTS \citep{2015ascl.soft04004B}, Astropy \citep{2013A&A...558A..33A},
Matplotlib \citep{2007CSE.....9...90H}}

\bibliography{GRB201216C}{}
\bibliographystyle{aasjournal}

\clearpage
\appendix

\section{Additional Figures}
\restartappendixnumbering

\begin{figure}
\centering
\includegraphics[scale=0.45]{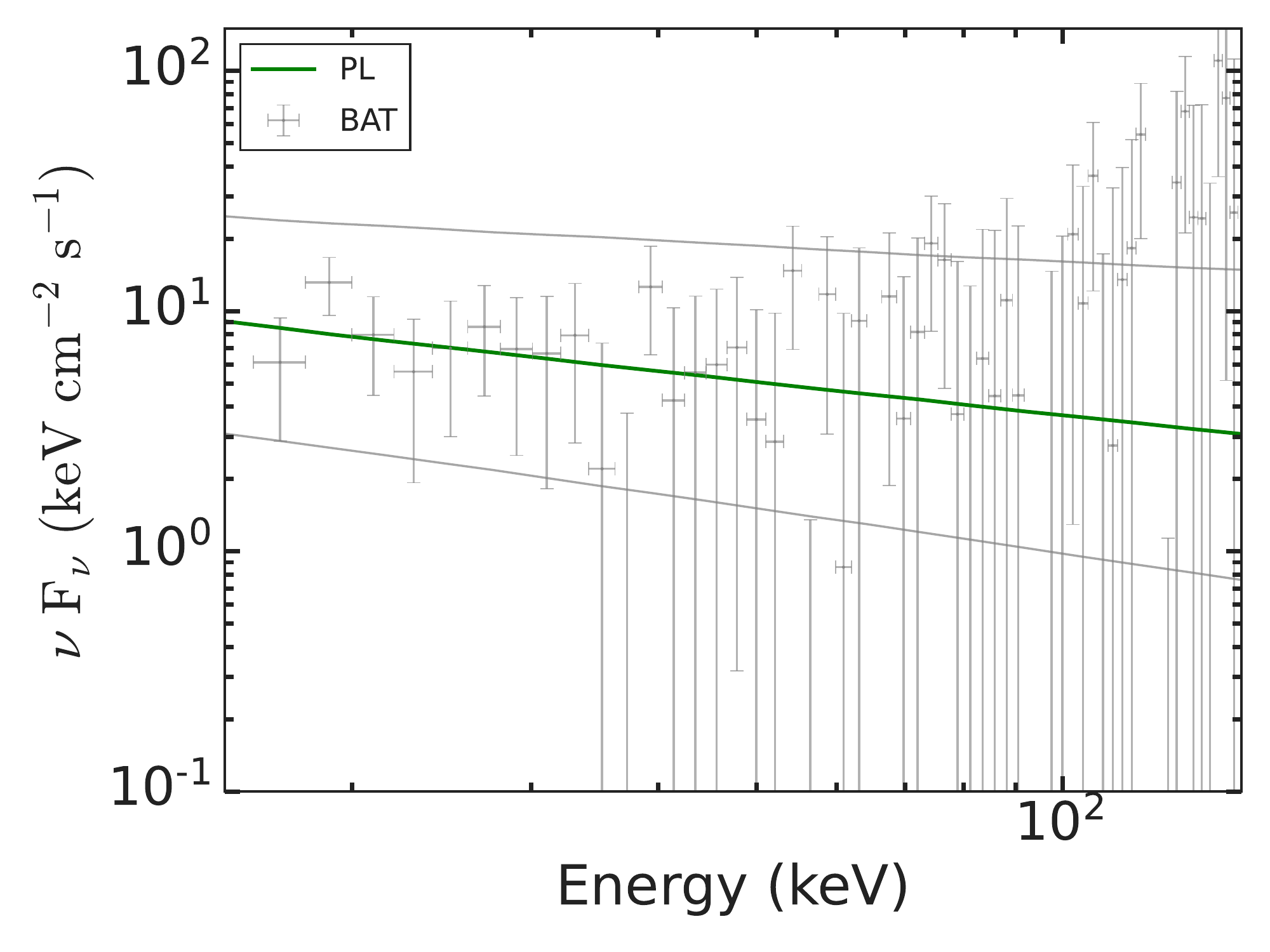}
\includegraphics[scale=0.5]{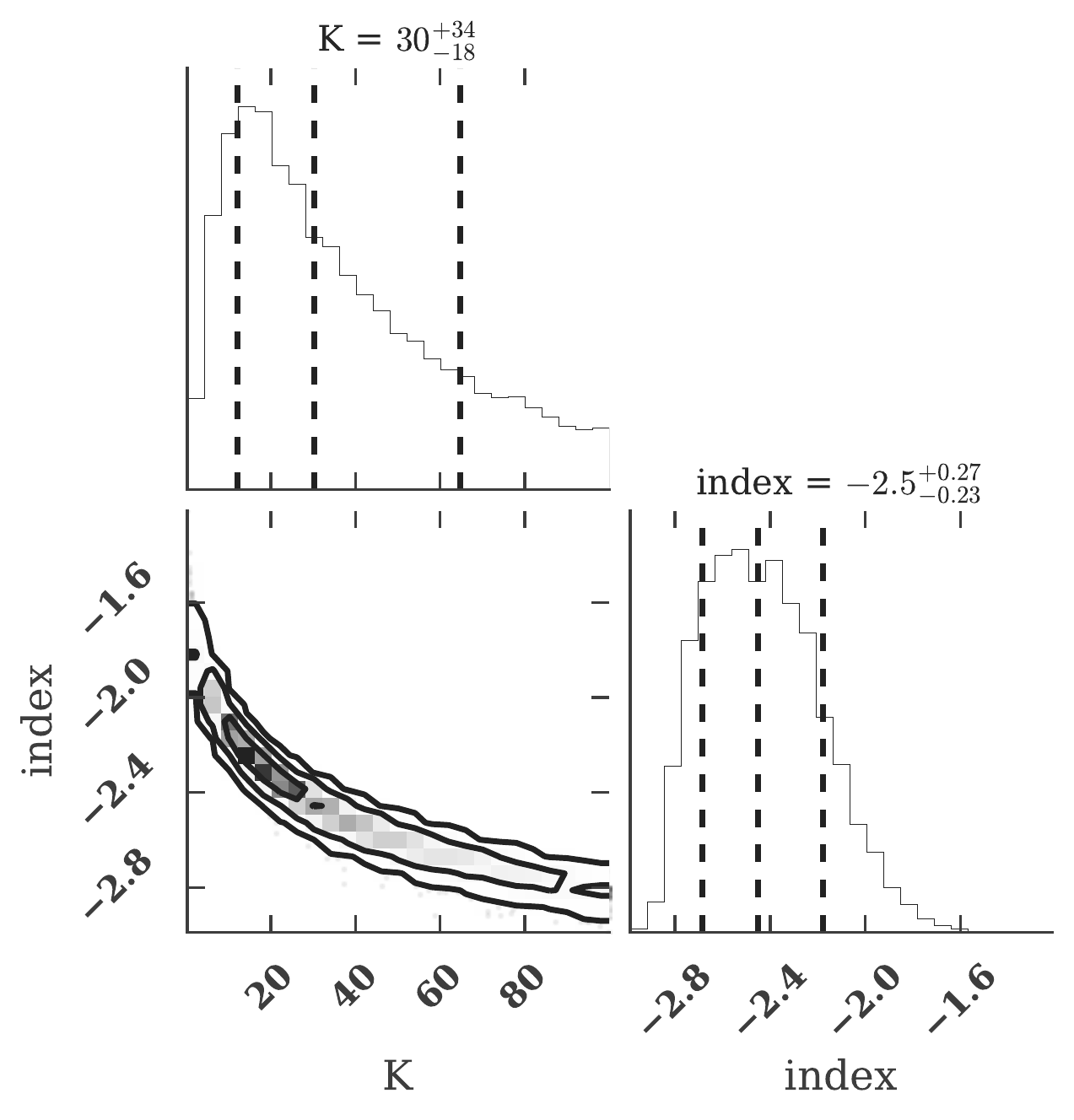}
\includegraphics[scale=0.45]{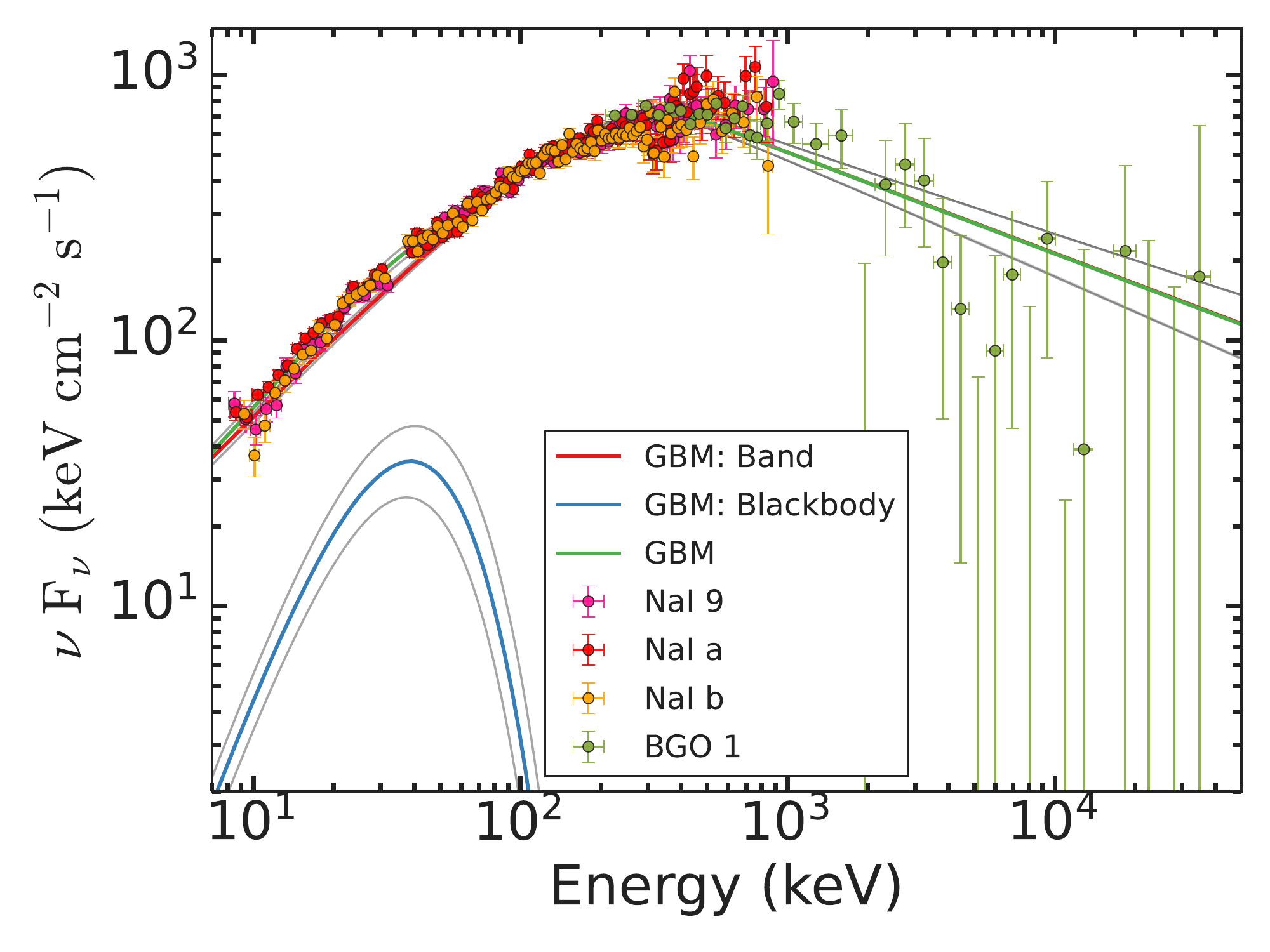}
\includegraphics[scale=0.25]{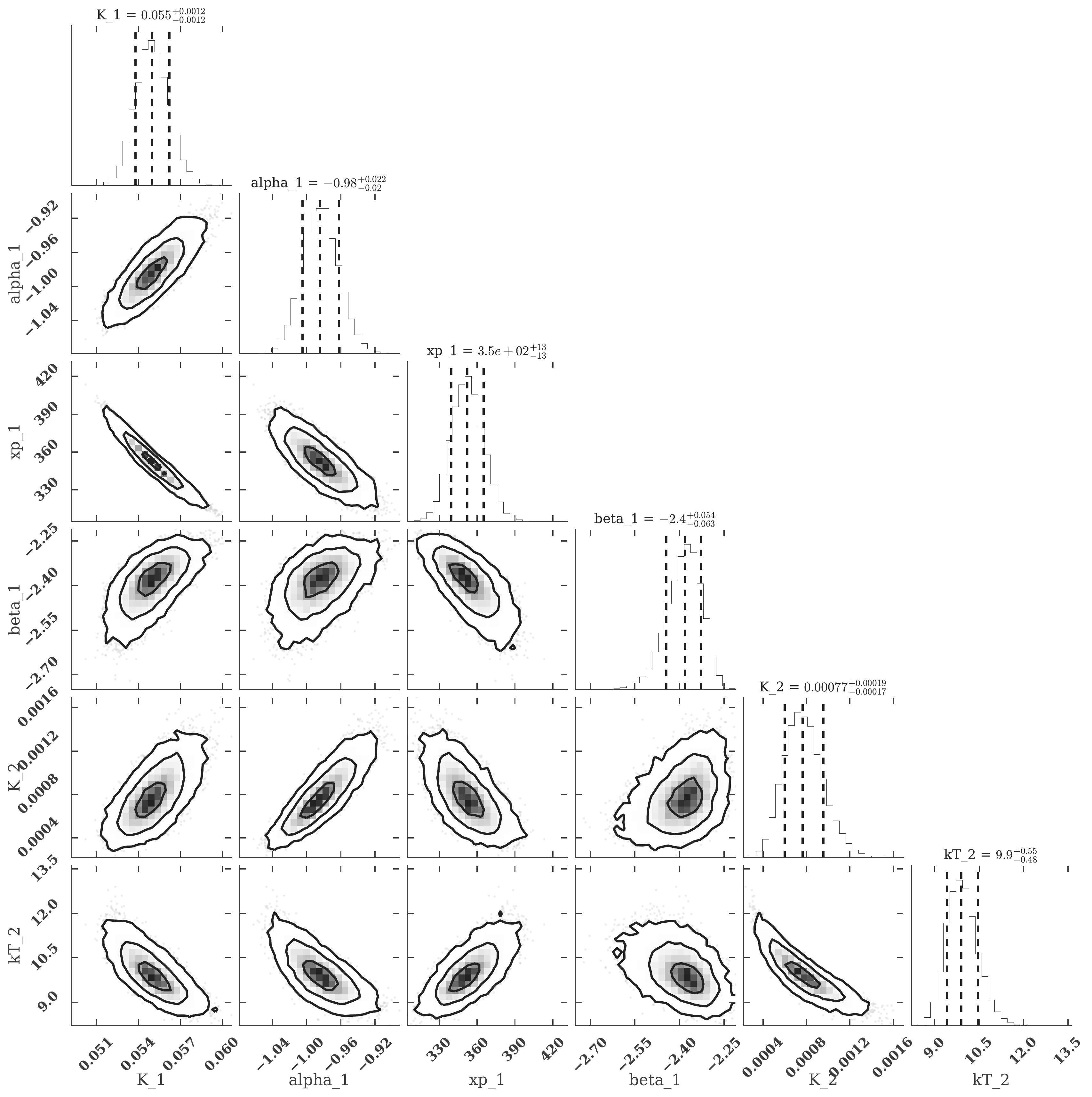}
\caption{The best fit time-averaged energy spectrum using power-law (for \thisgrba) and \sw{Band + Blackbody} (for \thisgrbb) models, respectively. The green solid lines (for \thisgrba and \thisgrbb) show the best fit energy spectrum of the models. The corresponding white shaded regions show the 95\% confidence interval. Right plots show the corresponding corner plots for the best fit models.}
\label{fig:TAS_band_bb}
\end{figure}

\begin{figure}
\centering
\includegraphics[scale=0.35]{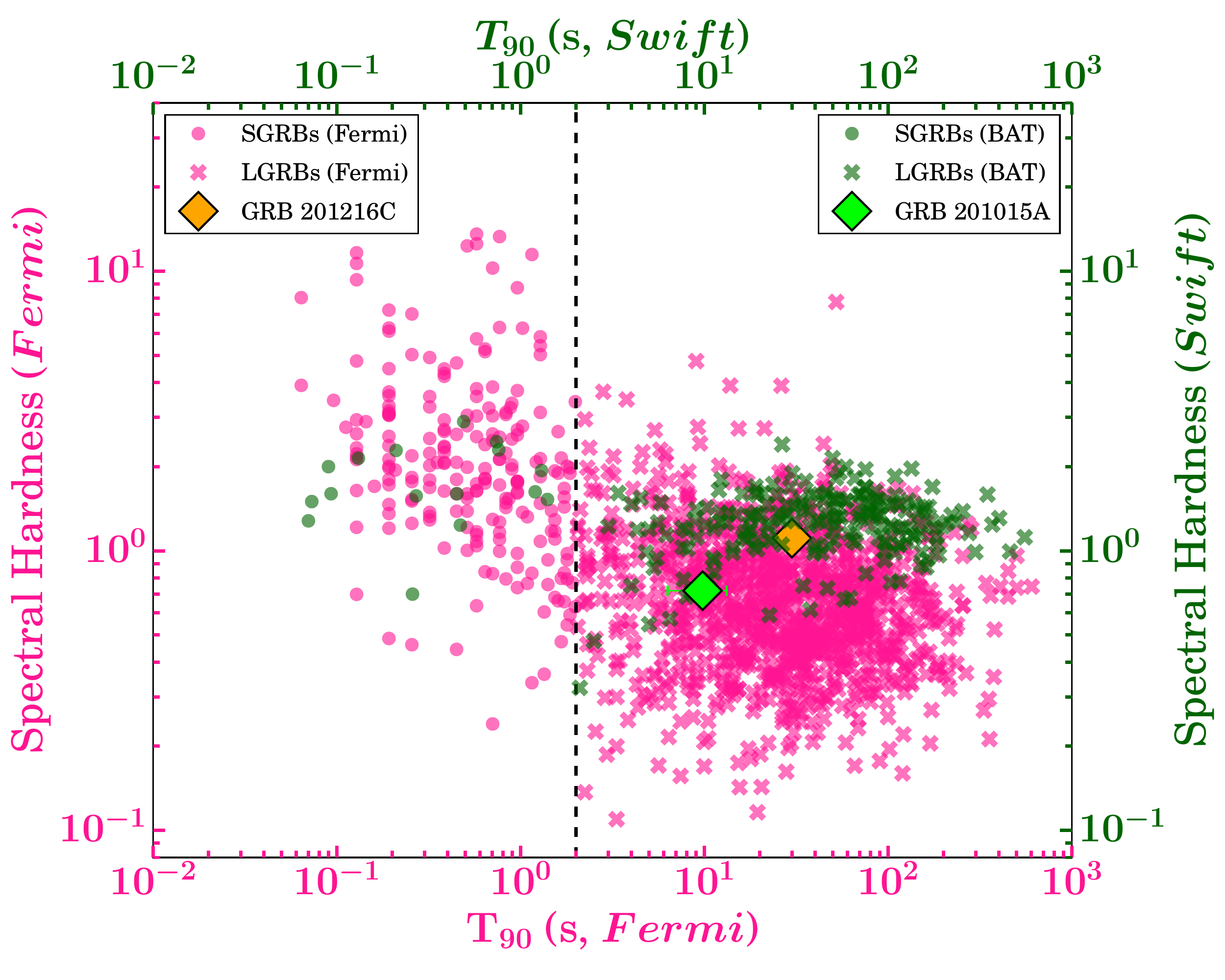}
 \includegraphics[scale=0.35]{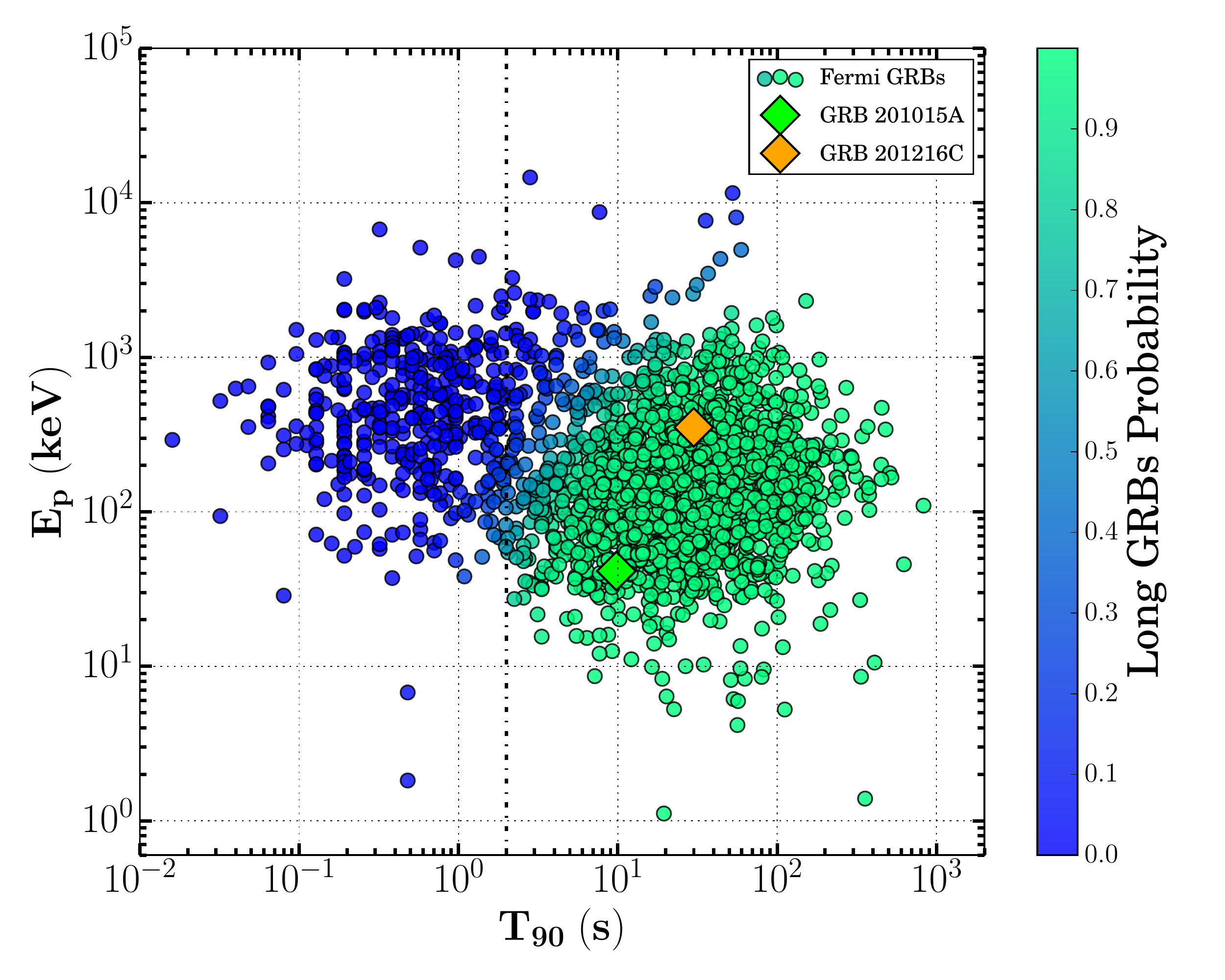}
\includegraphics[scale=0.35]{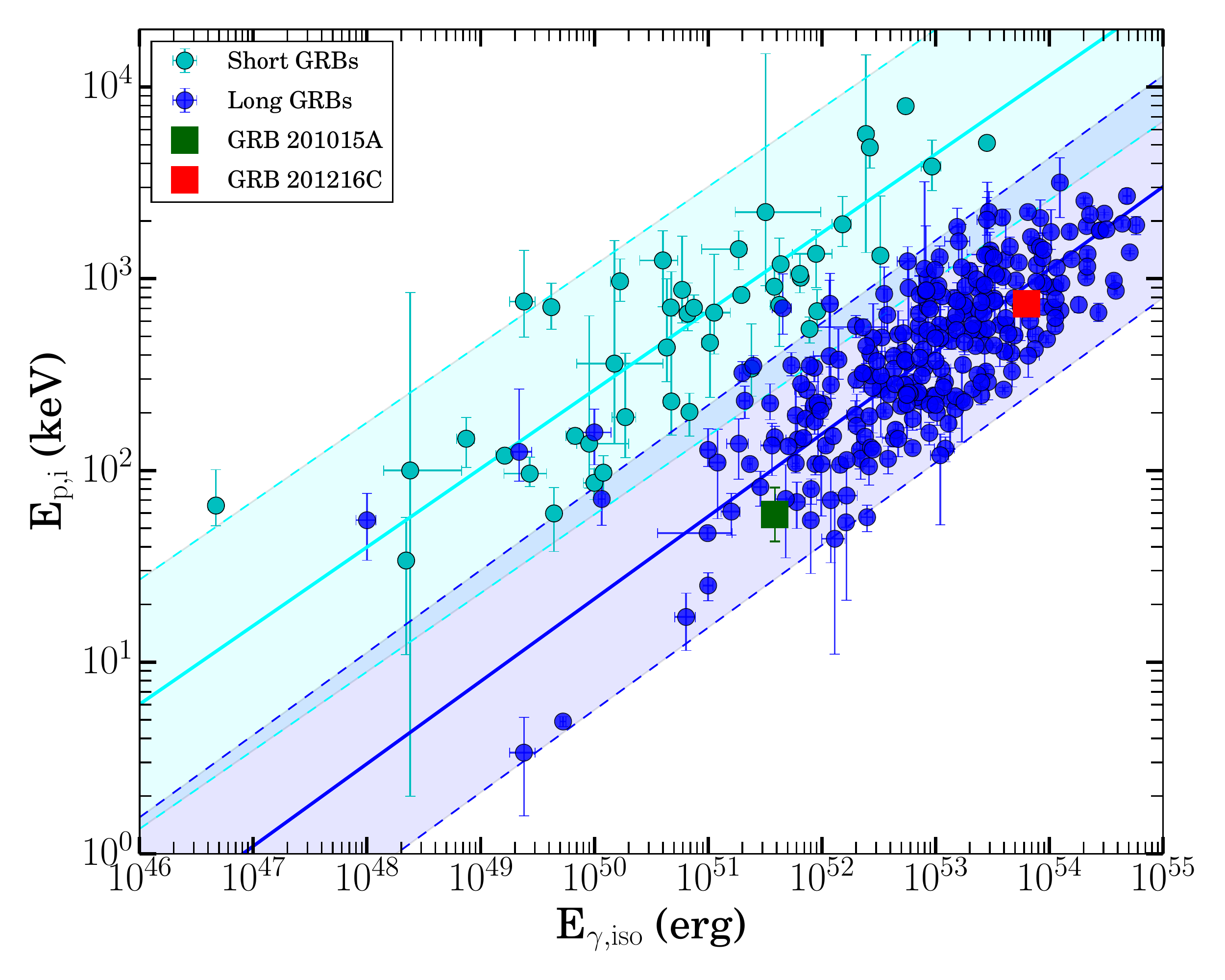}
\includegraphics[scale=0.35]{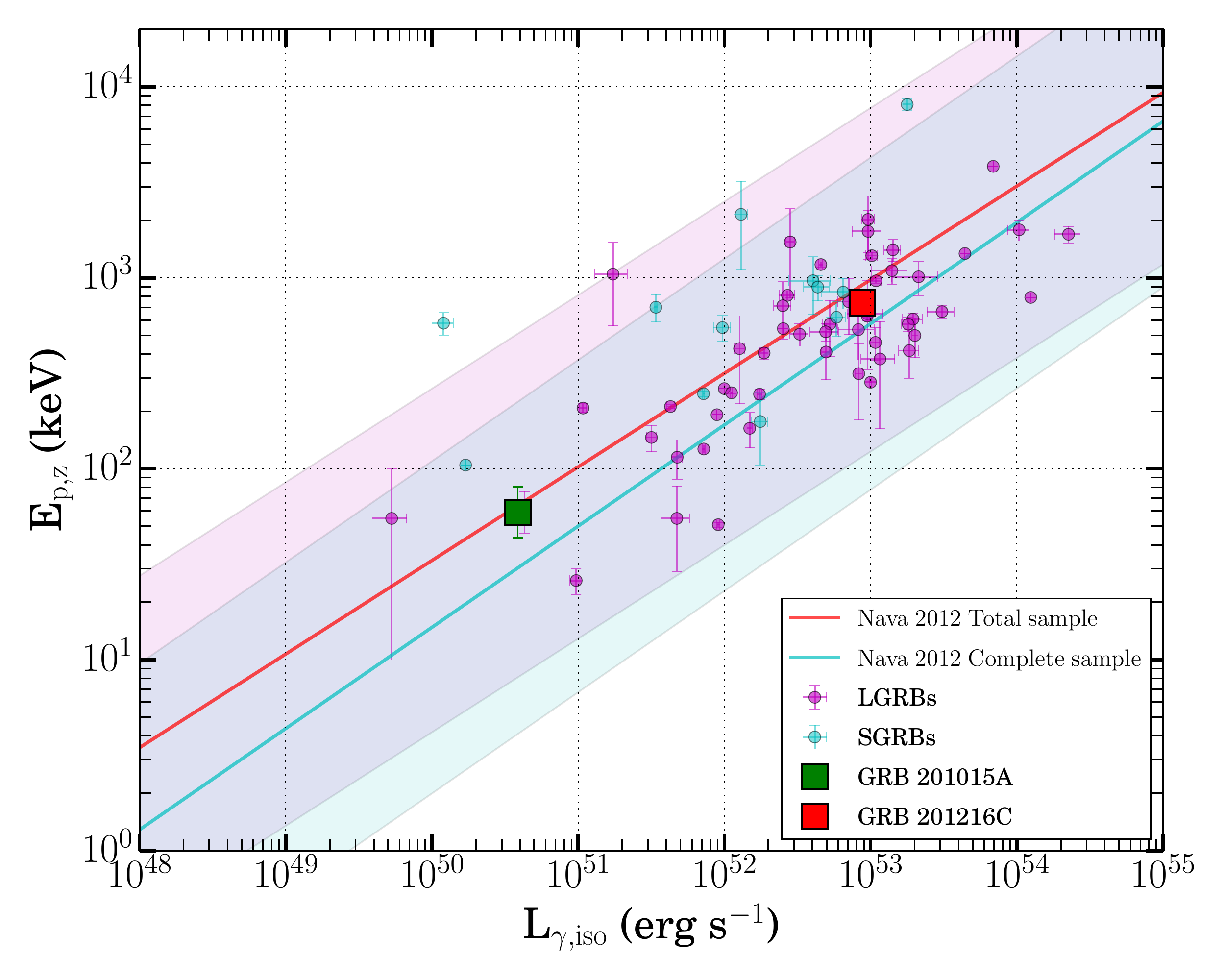}
\caption{{Top left panel:} The hardness-ratio and \tninty distribution plane for the \swift-BAT (green) and the \fermi-GBM (deep pink) detected GRBs. {Top right panel:} The \Ep-\tninty distribution plane for the \fermi-GBM detected GRBs \cite{Goldstein_2017}. The right-side color bar denotes the probability of the GRB being a long burst. The locations of \thisgrba and \thisgrbb are shown with lime and orange diamonds, respectively. The black dashed vertical line denotes the boundary between long and short bursts. Prompt emission correlations: {Bottom left panel:} Amati correlation: \thisgrba and \thisgrbb are represented by the green and red squares, respectively, while the cyan and blue dots represent the short and long GRBs collected from \cite{2020MNRAS.492.1919M}. {Bottom right panel:} Yonetoku correlation: \thisgrba and \thisgrbb are represented by the green and red squares, respectively, along with the GRBs sample reported in \cite{2012MNRAS.421.1256N}. The best-fit is shown by several solid colored lines and the shaded region represents the 3$\sigma$ scatter of the correlation.}
\label{fig:prompt_properties}
\end{figure}

\begin{figure}
\centering
\includegraphics[scale=0.5]{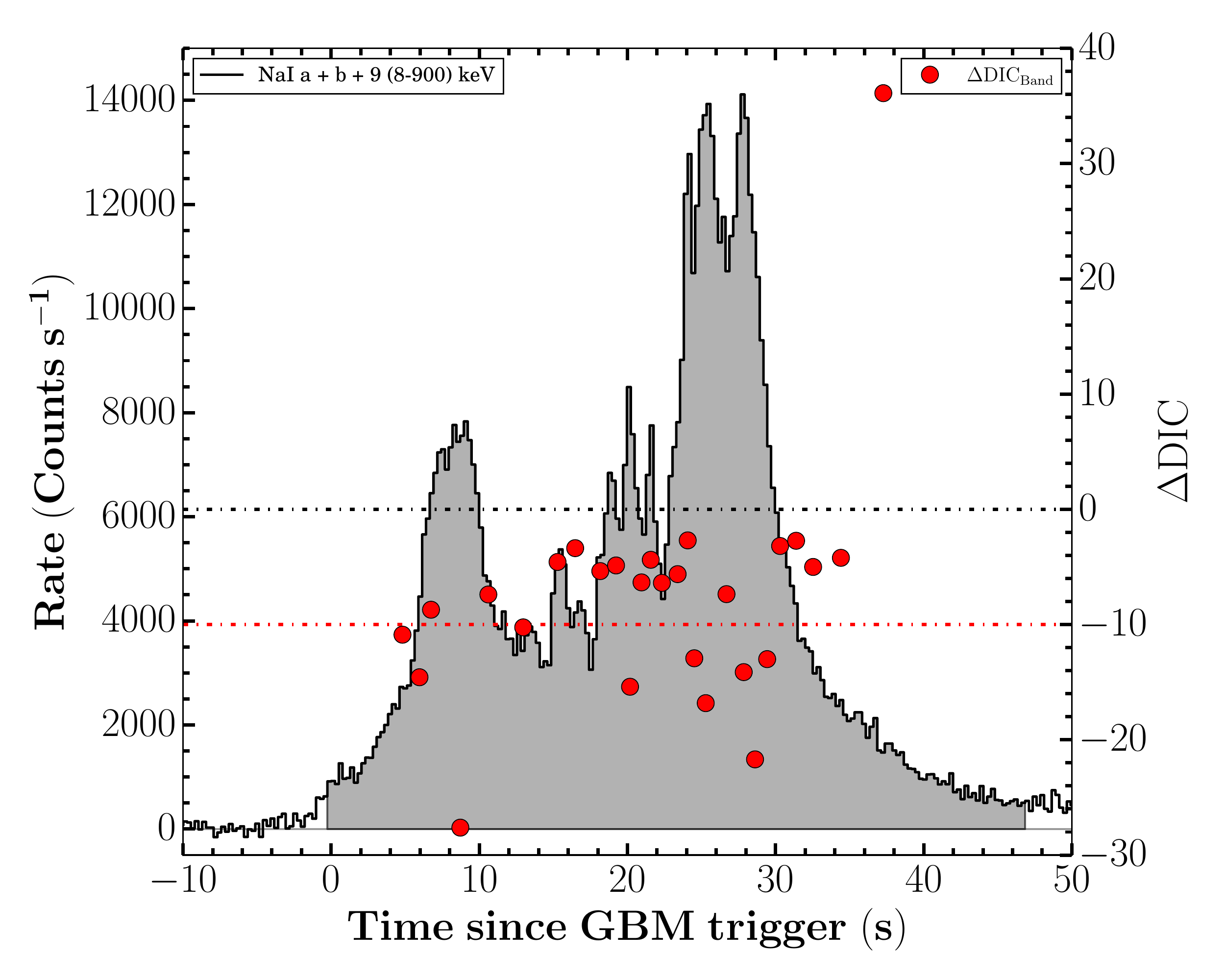}
\caption{The evolution of $\rm \Delta DIC$ within the burst interval for \sw{Band} and \sw{Band + Blackbody} models for \thisgrbb. The black dotted line is at $\rm \Delta DIC = 0$; the data points below indicate an improvement in fit after adding the \sw{Blackbody} function. The red dotted line is at $\rm \Delta DIC = -10$; the data points below indicate a significant amount of thermal components in the corresponding spectrum.}
\label{fig:DIC}
\end{figure}

\begin{figure}
\centering
\includegraphics[scale=0.35]{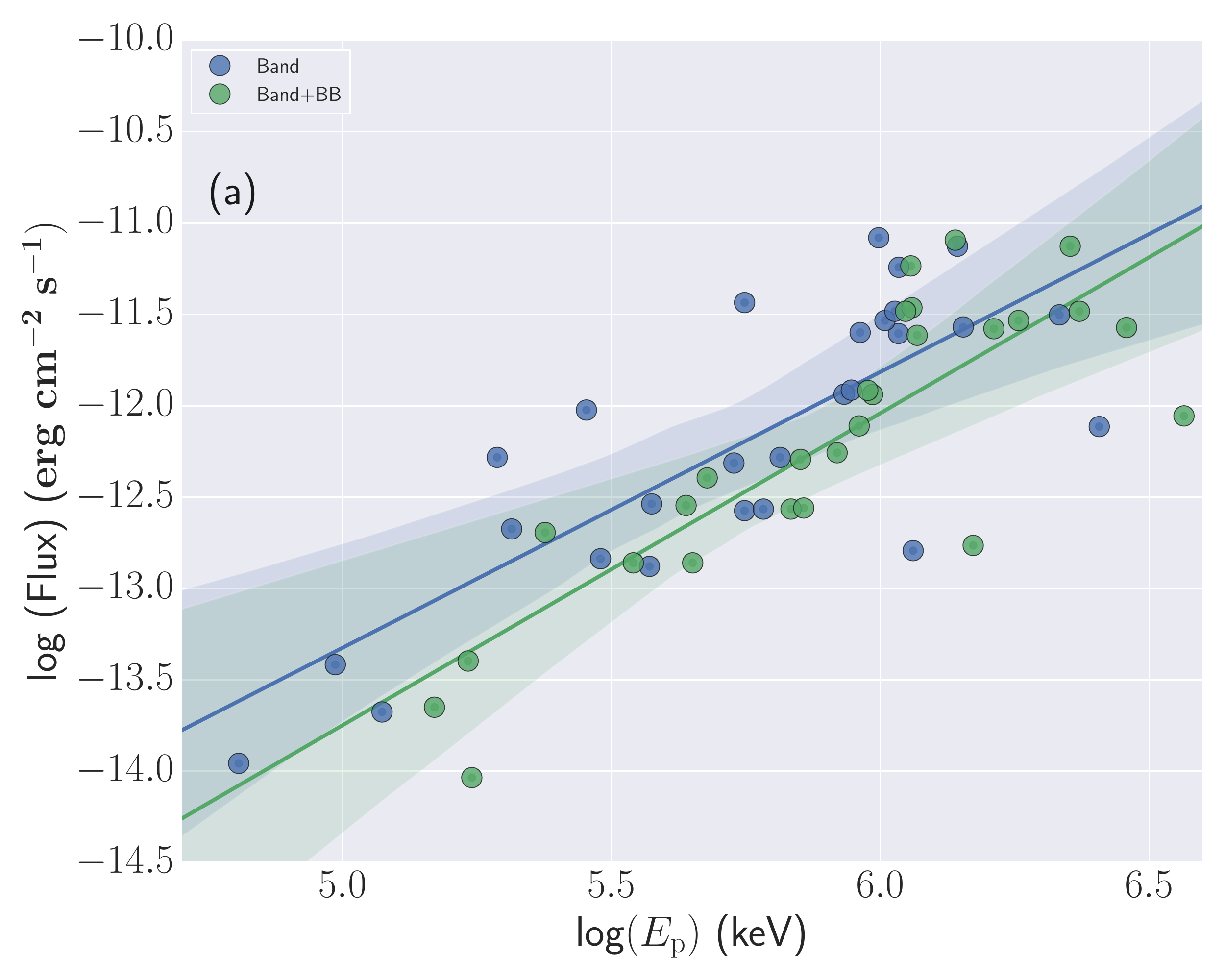}
\includegraphics[scale=0.35]{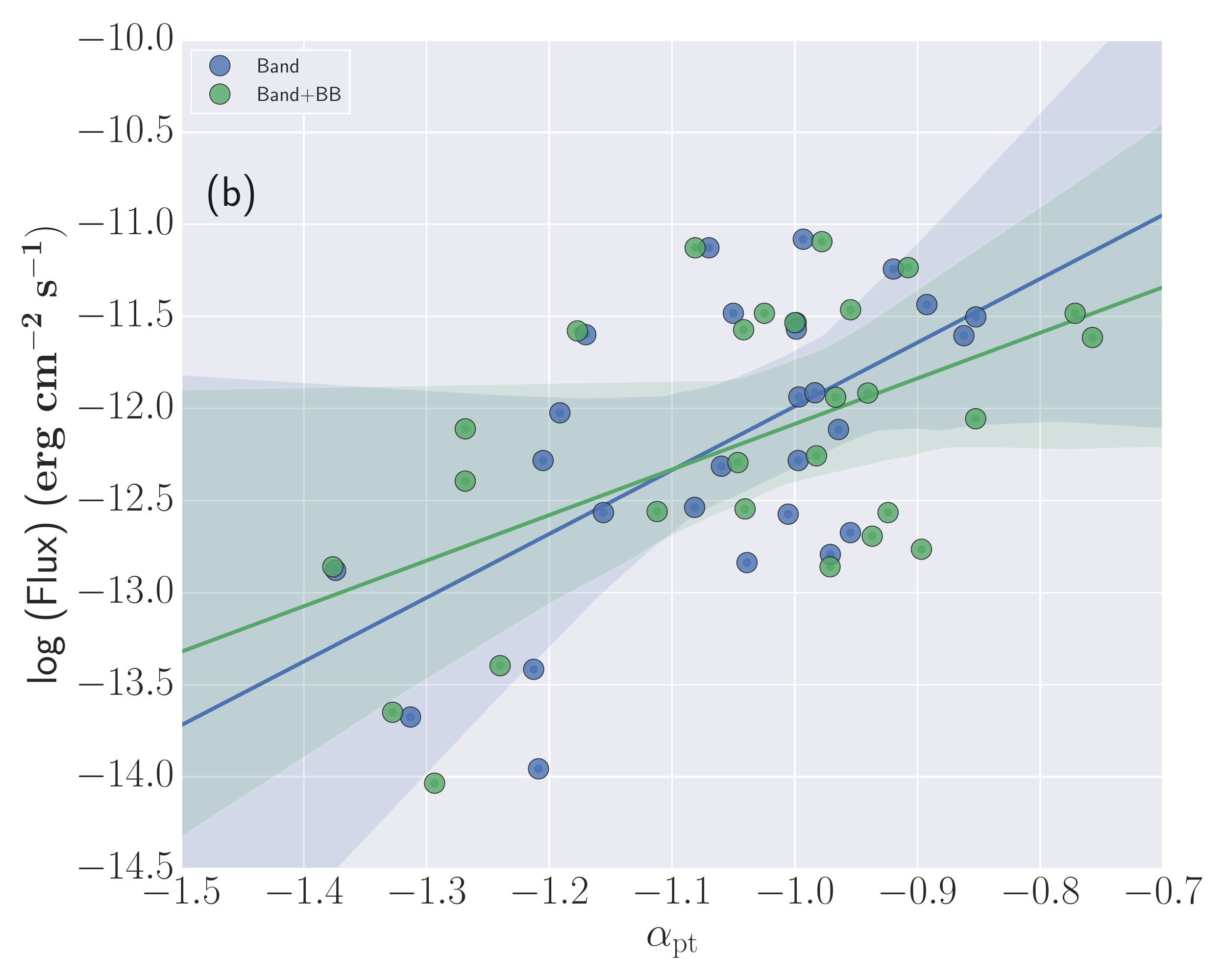}
\includegraphics[scale=0.35]{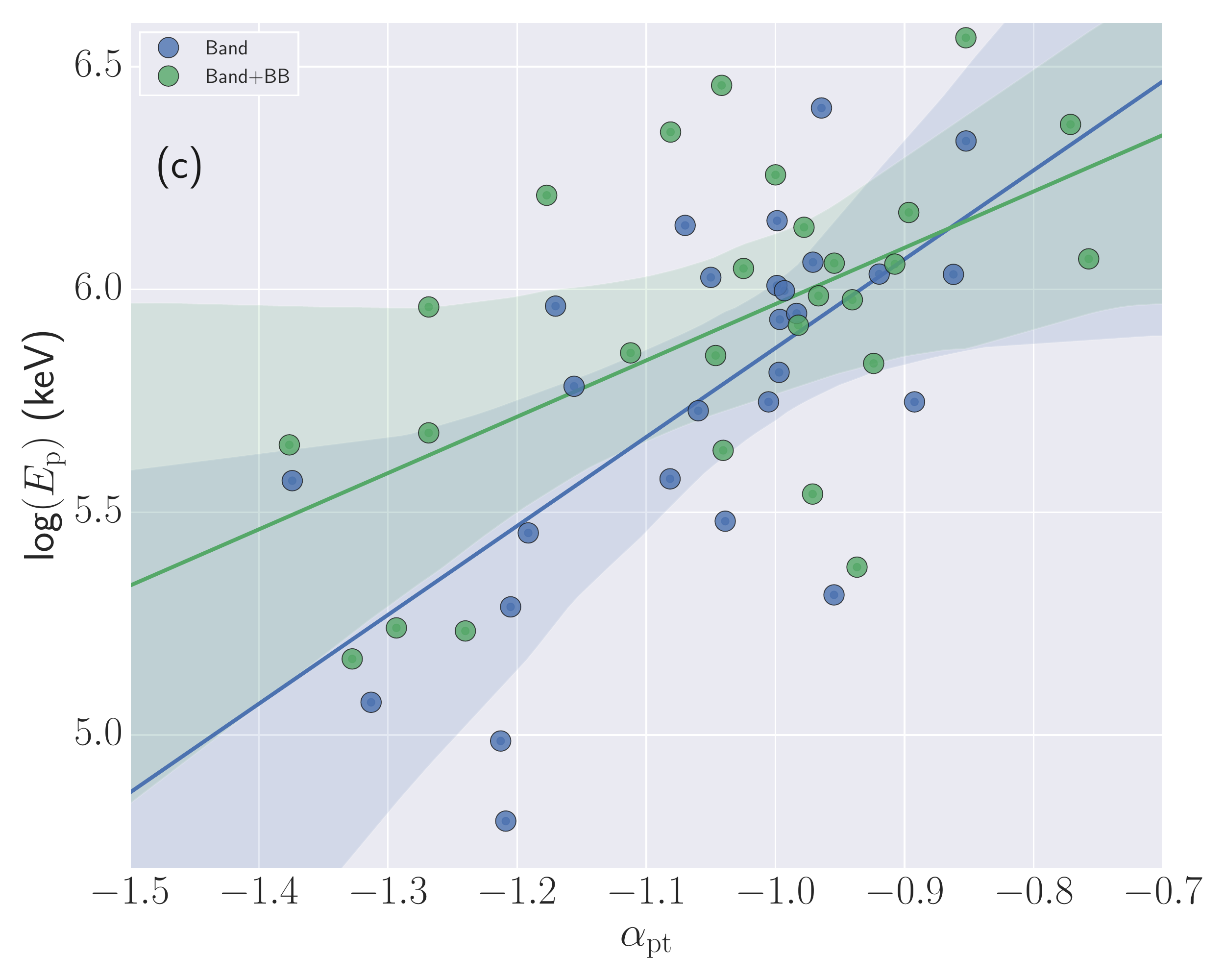}
\includegraphics[scale=0.35]{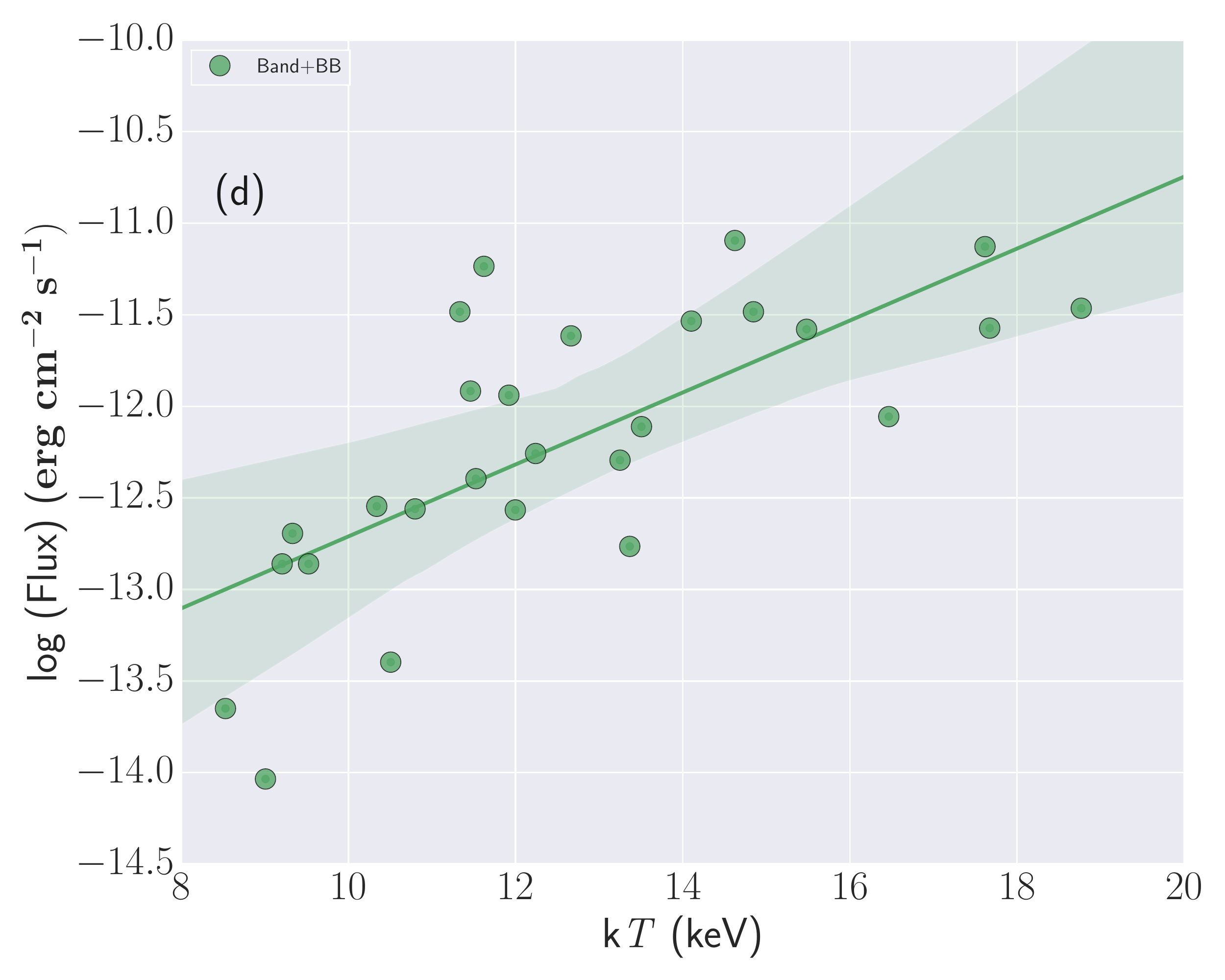}
\caption{The correlations between spectral parameters obtained from the fitting of \sw{Band} and \sw{Band + Blackbody} models to the GBM data for \thisgrbb. (a): correlation between the log(flux)-log(\Ep) obtained from \sw{Band} (blue) and \sw{Band + Blackbody} (green) models. Similarly, (b) and (c) represent the correlation between log(flux)-$\alpha_{\rm pt}$ and log(\Ep)-$\alpha_{\rm pt}$. (d) represents the correlation between log(flux)-kT obtained from \sw{Band + Blackbody} models. Solid lines represent the best fit, and the shaded region shows the 2$\sigma$ confidence interval of the correlation.}
\label{fig:correlation_band}
\end{figure}

\begin{figure}
\centering
\includegraphics[scale=0.35]{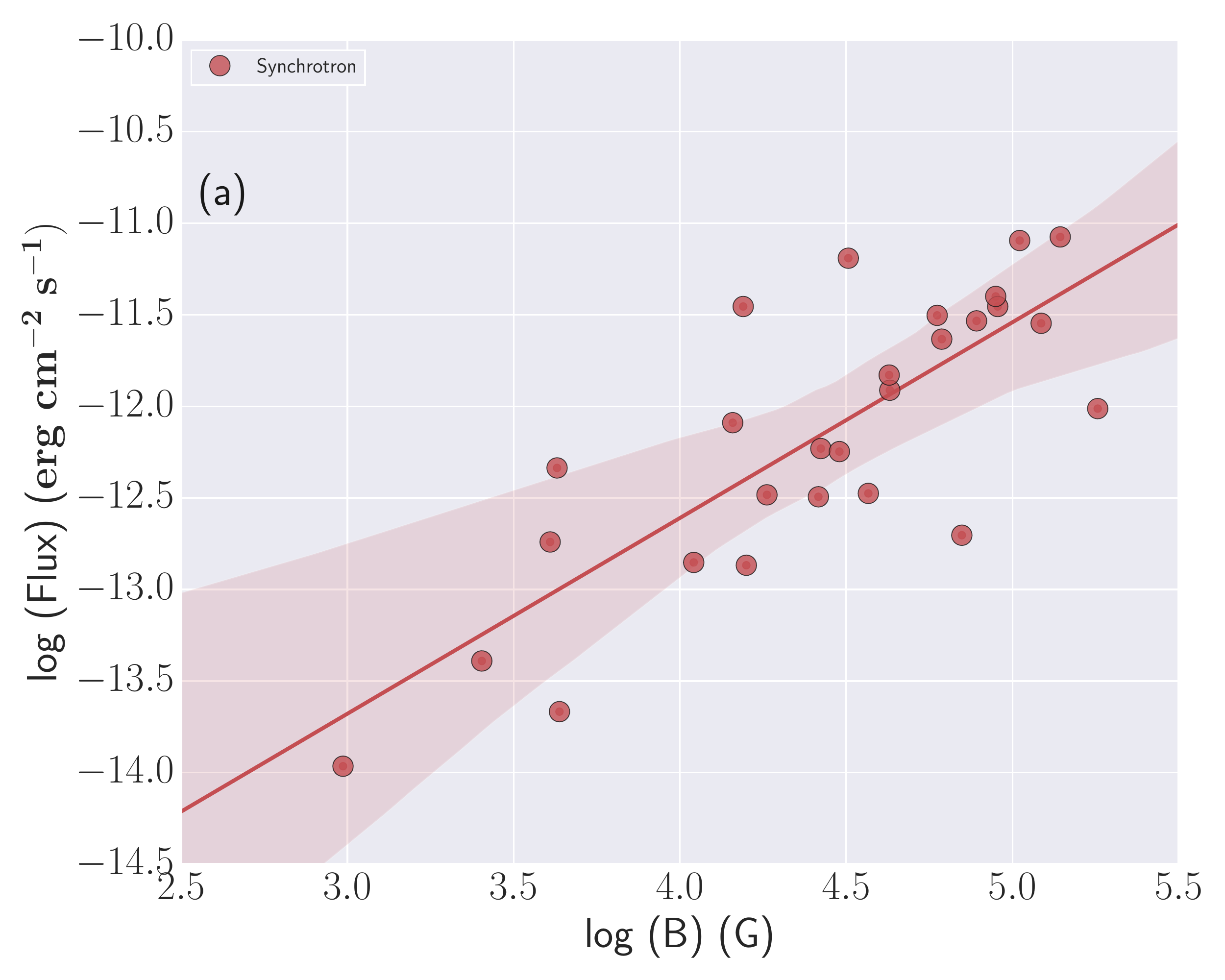}
\includegraphics[scale=0.35]{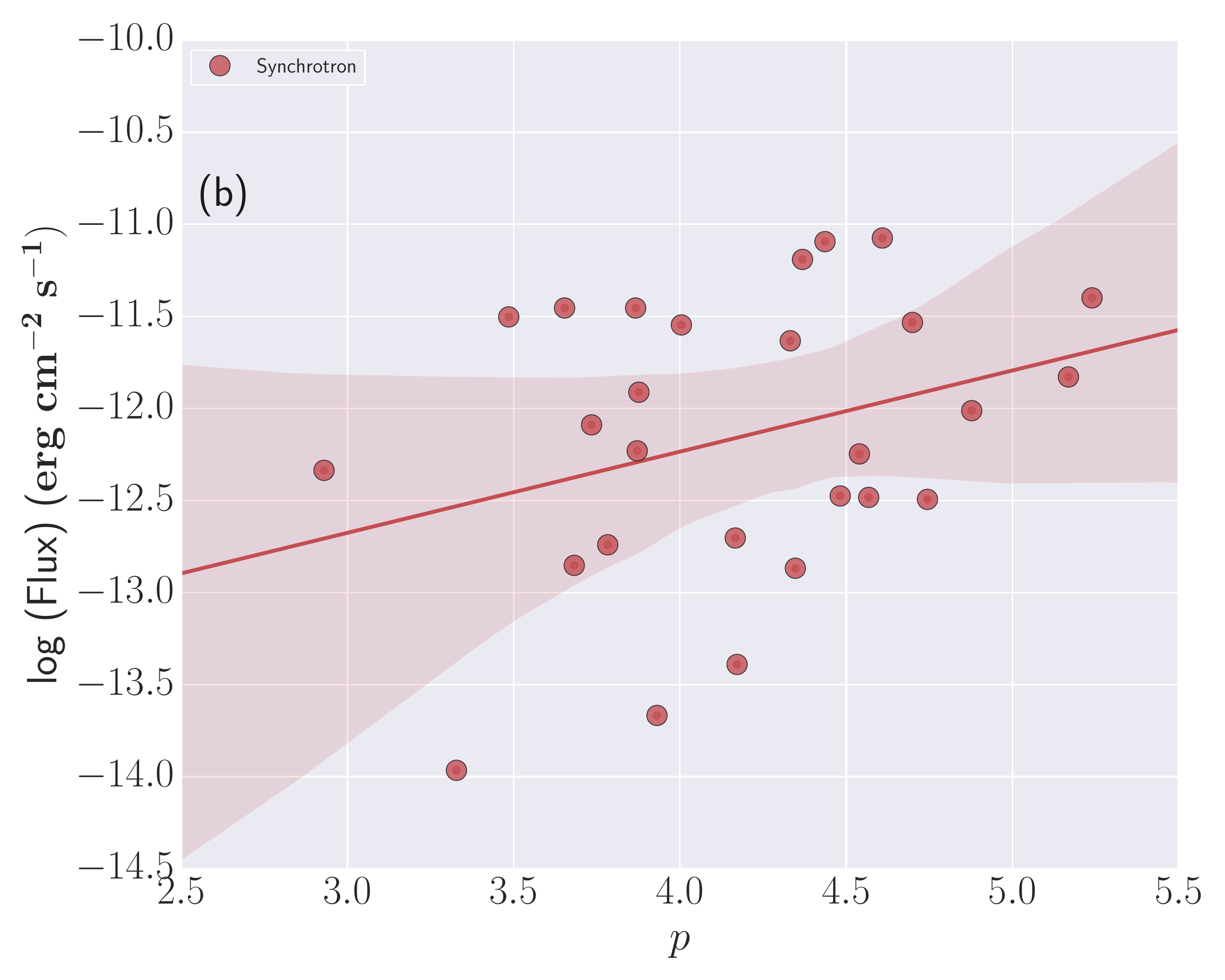}
\includegraphics[scale=0.35]{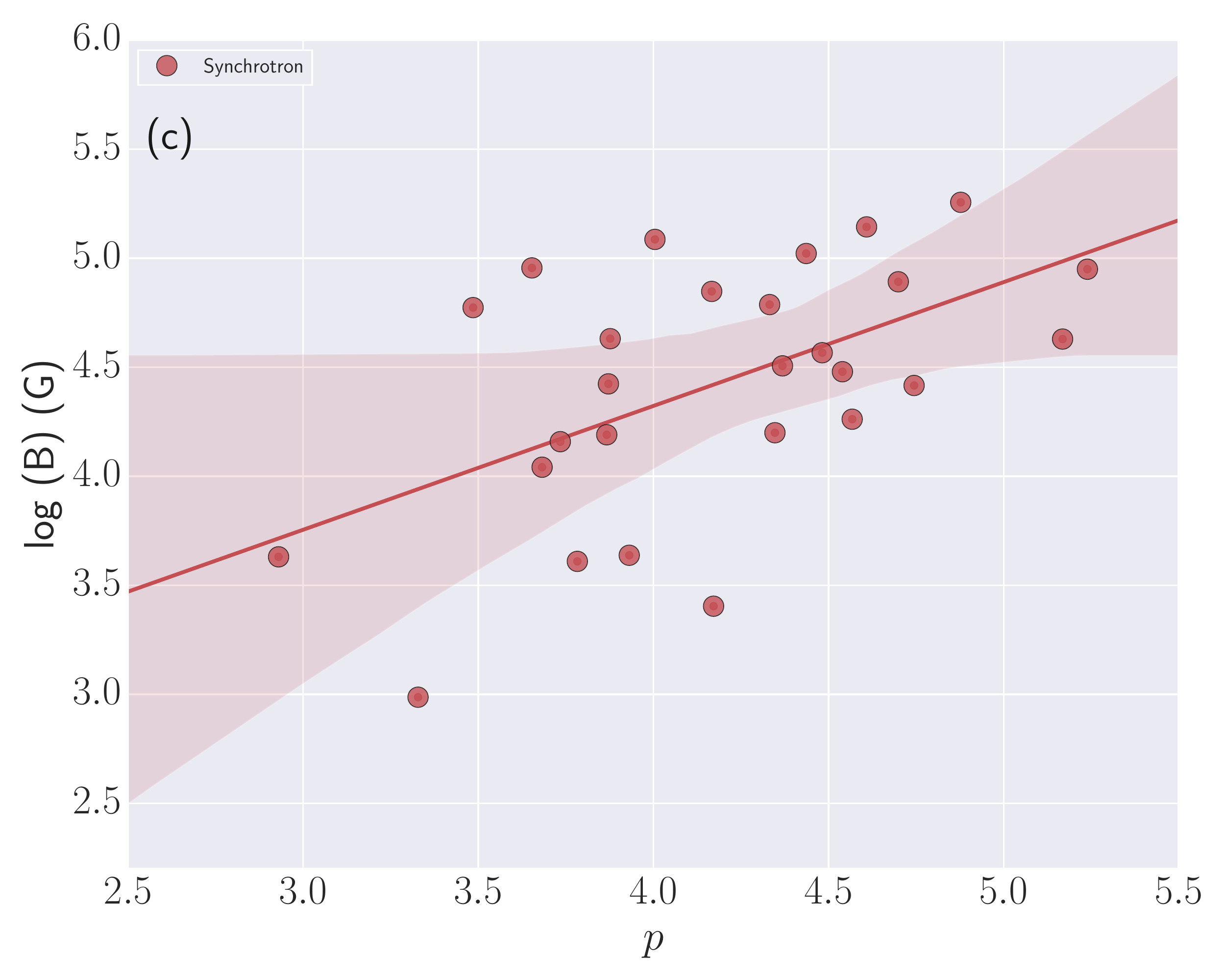}
\caption{The correlations between spectral parameters obtained from fitting the physical \sw{Synchrotron} model to the GBM data for \thisgrbb. (a) logarithm of magnetic field strength B versus the log(flux), (b) electron energy index $p$ versus log(flux) and (c) log(B) versus $p$. Solid red lines represent the best fit, and the shaded region shows the 2$\sigma$ confidence interval of the correlation.}
\label{fig:correlation_sync}
\end{figure}

\begin{figure}
\centering
\includegraphics[scale=0.32]{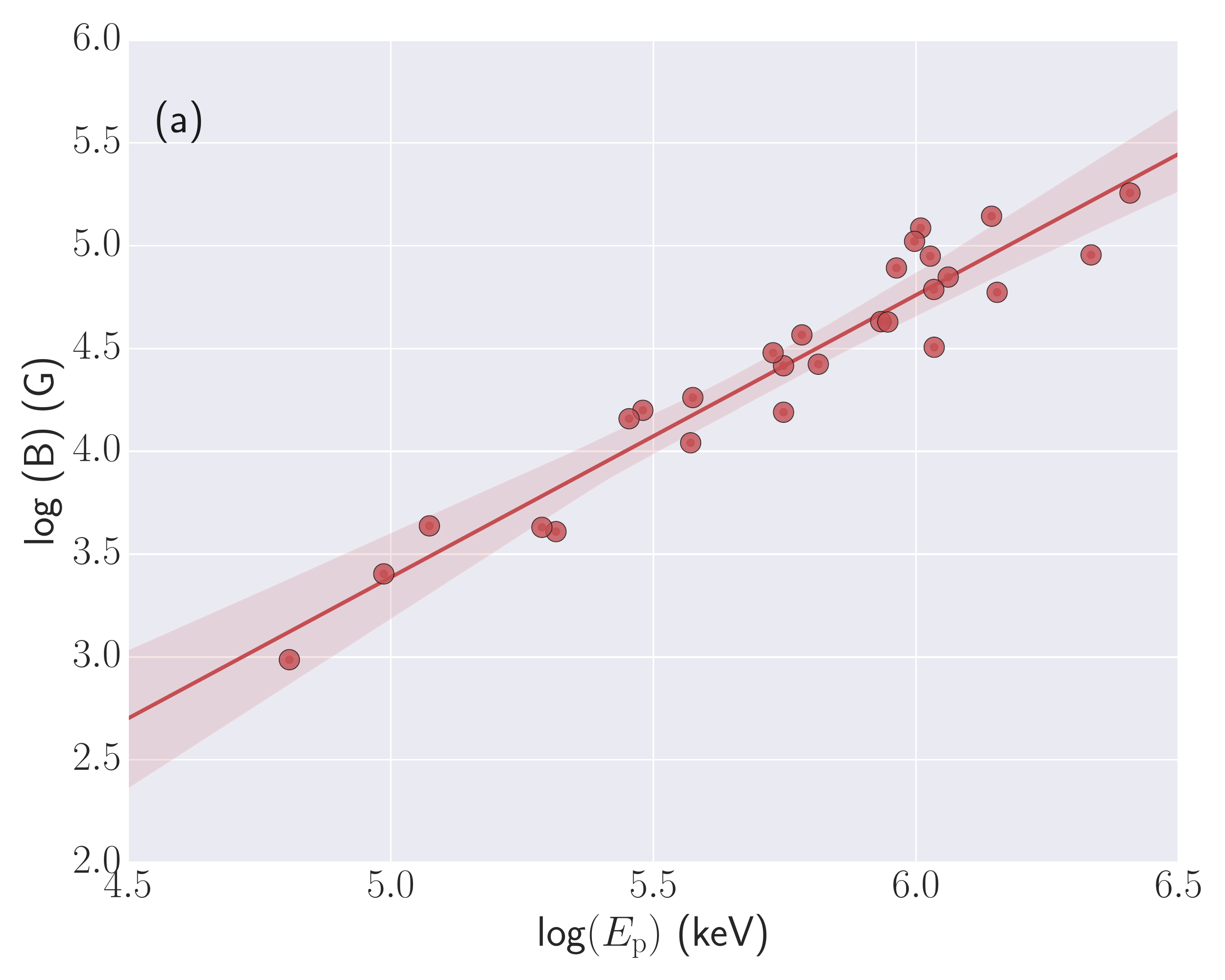}
\includegraphics[scale=0.32]{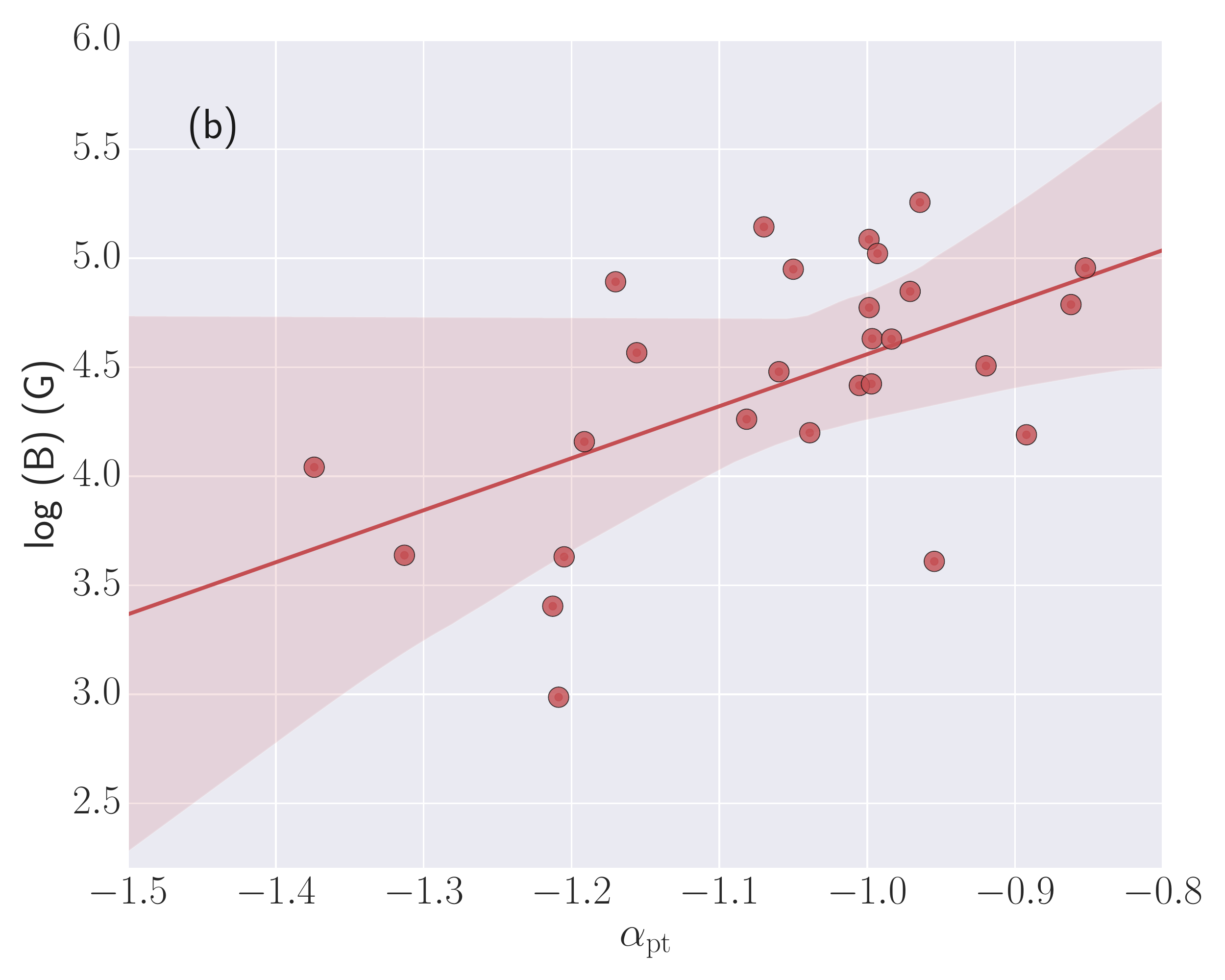}
\includegraphics[scale=0.32]{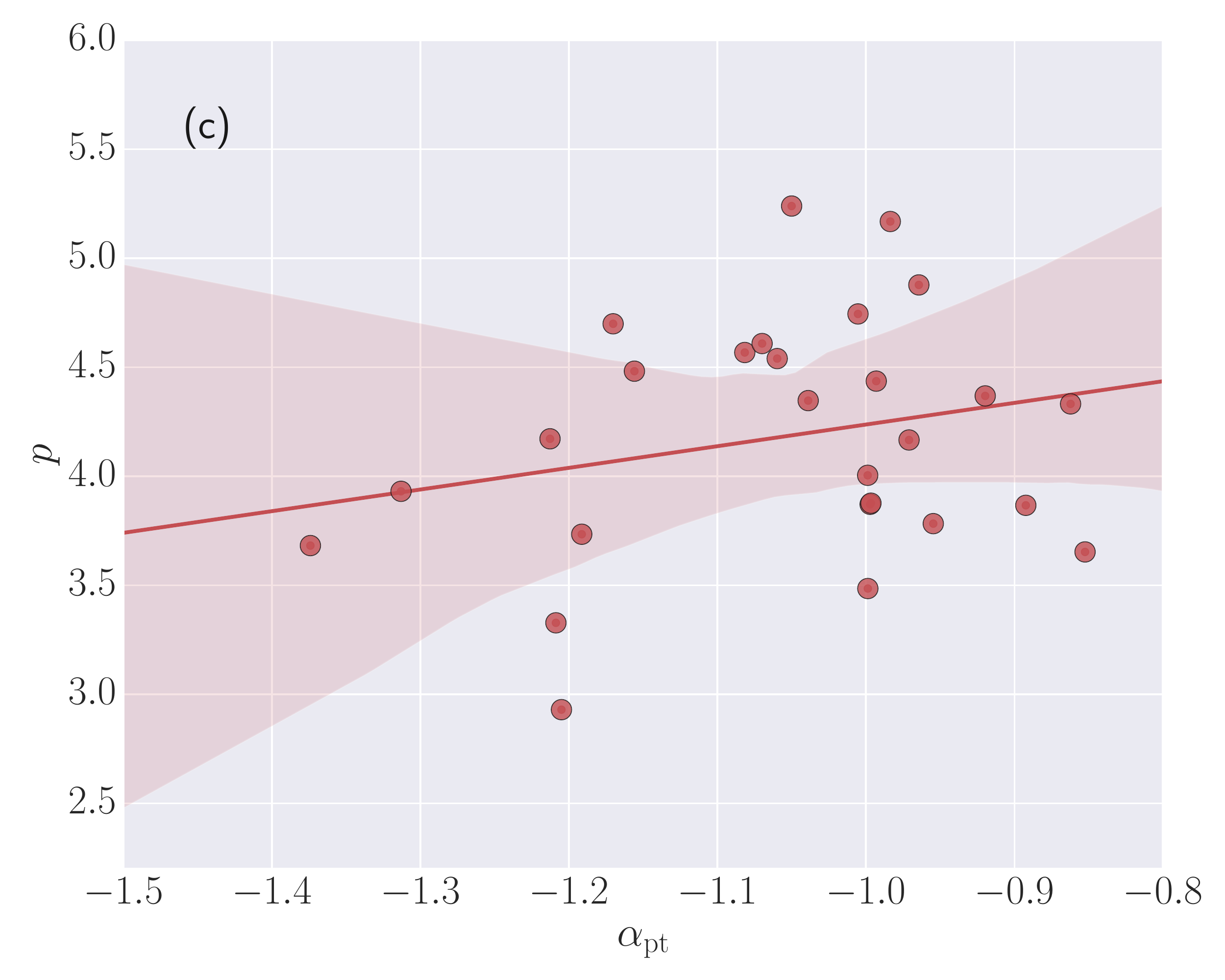}
\caption{The correlations between spectral parameters obtained from the \sw{Synchrotron} and \sw{Band} models to the GBM data for \thisgrbb. (a) logarithm of magnetic field strength B versus the log(\Ep) (b) logarithm of magnetic field strength B versus log($\alpha_{\rm pt}$), (c) electron energy index $p$ versus $\alpha_{\rm pt}$. Solid red lines represent the best fit, and the shaded region shows the 2$\sigma$ confidence interval of the correlation.}
\label{fig:correlation_sync_band}
\end{figure}

\begin{figure}
\centering
\includegraphics[scale=0.43]{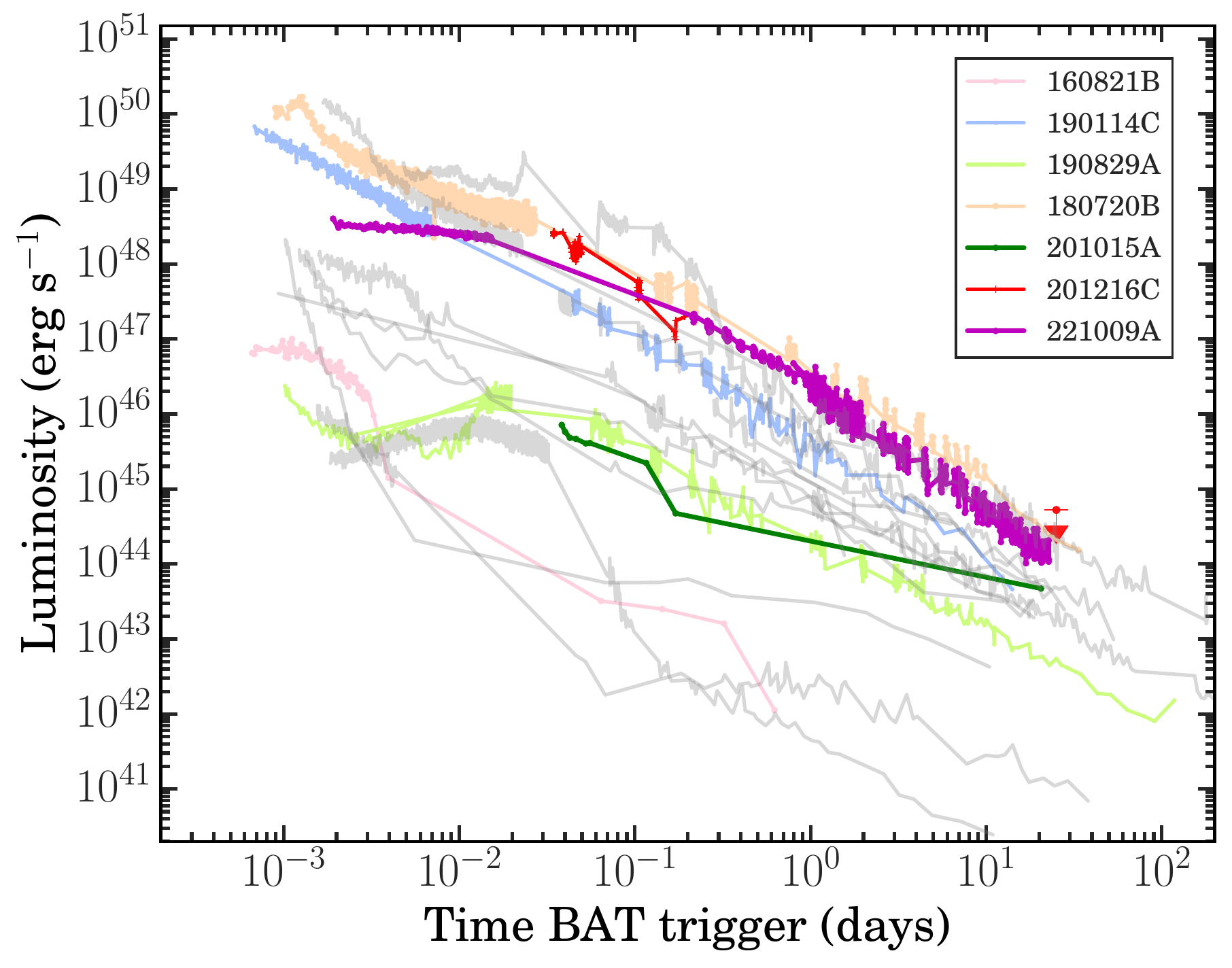}
\includegraphics[scale=0.43]{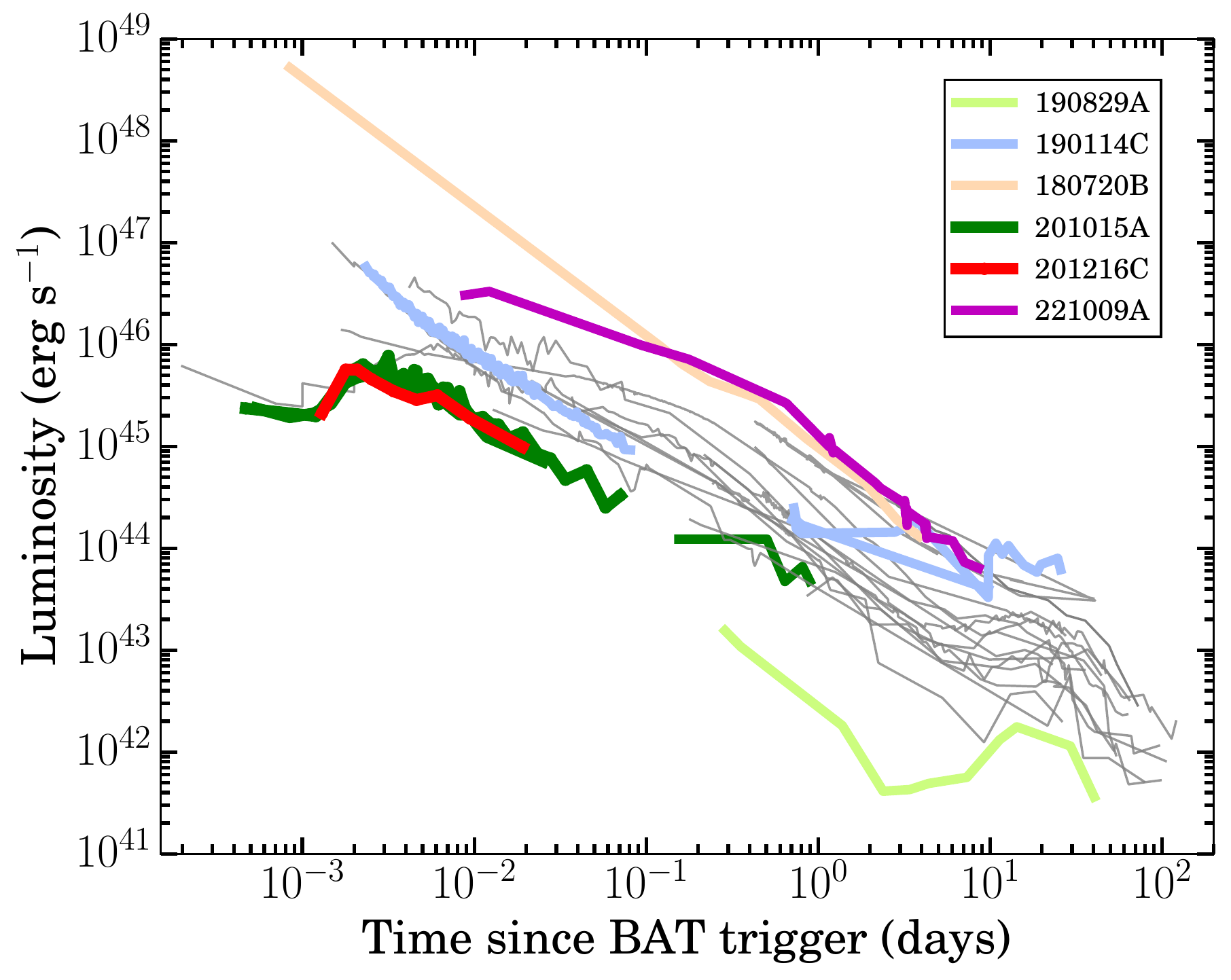}
\caption{Comparison of the afterglow light curves of VHE detected GRBs. {\it Left panel:} consists of X-ray (@ 0.3-10 \keV) light curves of VHE detected GRBs represented with various colors, as shown in the legends. Light curves shown in the background with the grey color are a sample of nearby supernova-connected GRBs taken from \cite{2022NewA...9701889K}. {\it Right panel:} the comparison among the optical light curves of VHE detected GRBs and a sample of nearby supernova-connected GRBs in the R-band.}
\label{fig:XRT_opt}
\end{figure}

\begin{figure}
\centering
\includegraphics[scale=0.5]{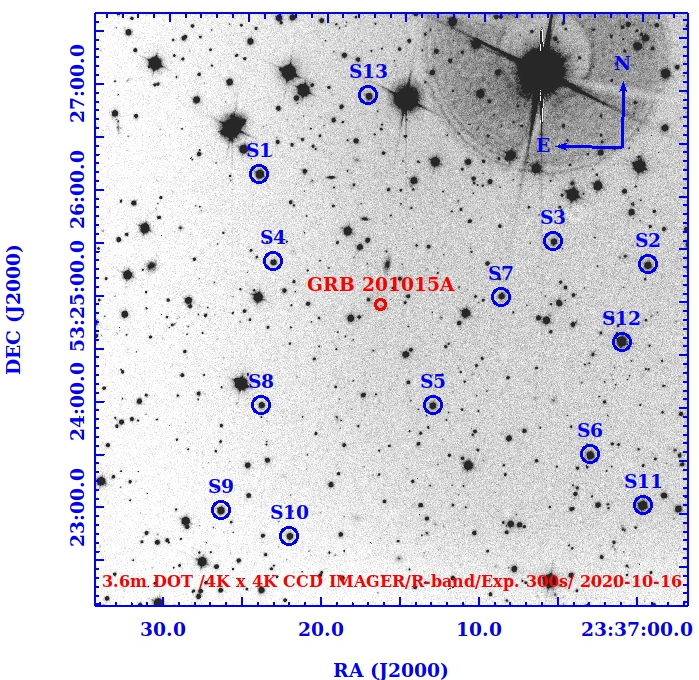}
\caption{The finding chart of \thisgrba observed in the R-band utilizing the 4K $\times$ 4K CCD Imager mounted on the 3.6m DOT \citep{2022JApA...43...27K}. The arrows in the image denote the directions (north and east). The stars marked with S1 to S13 are used to calibrate the magnitude in the standard system. The image has an FoV of 6.5$'$ $\times$ 6.5$'$.}
\label{fig:FC_DOT}
\end{figure}

\begin{figure}
\centering
\includegraphics[scale=0.45]{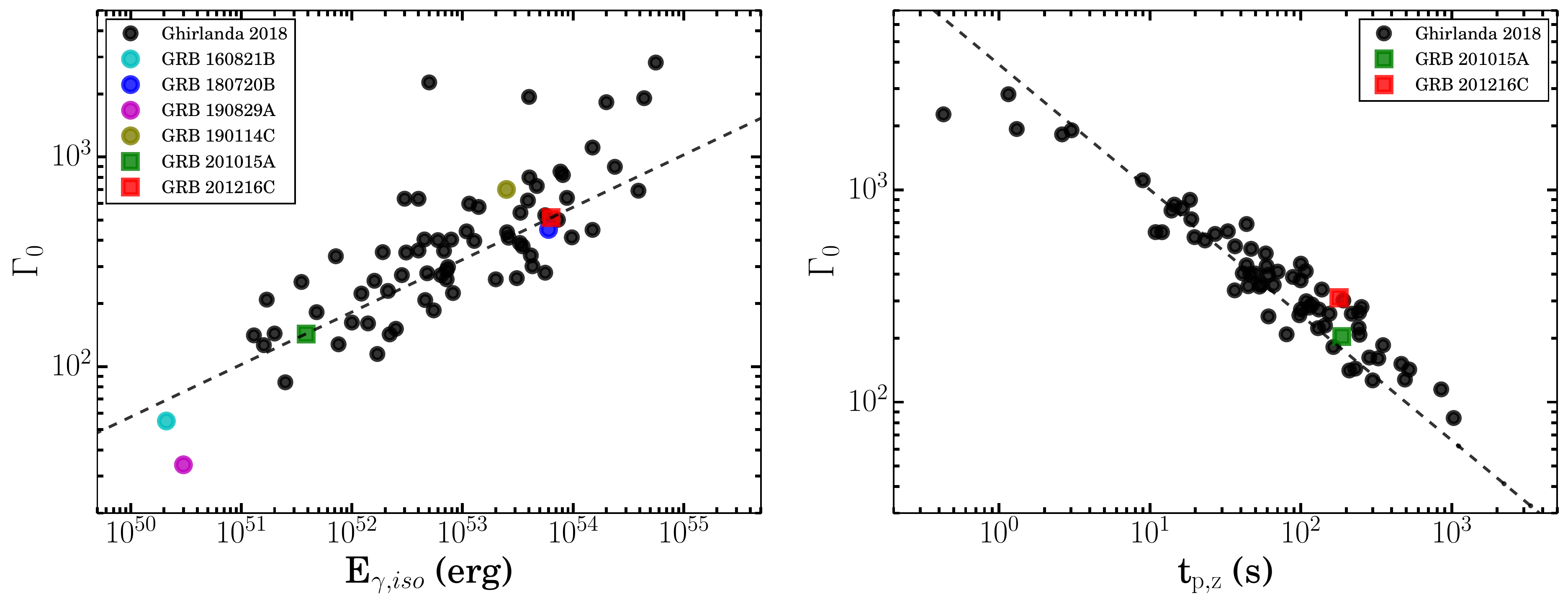}
\includegraphics[scale=0.45]{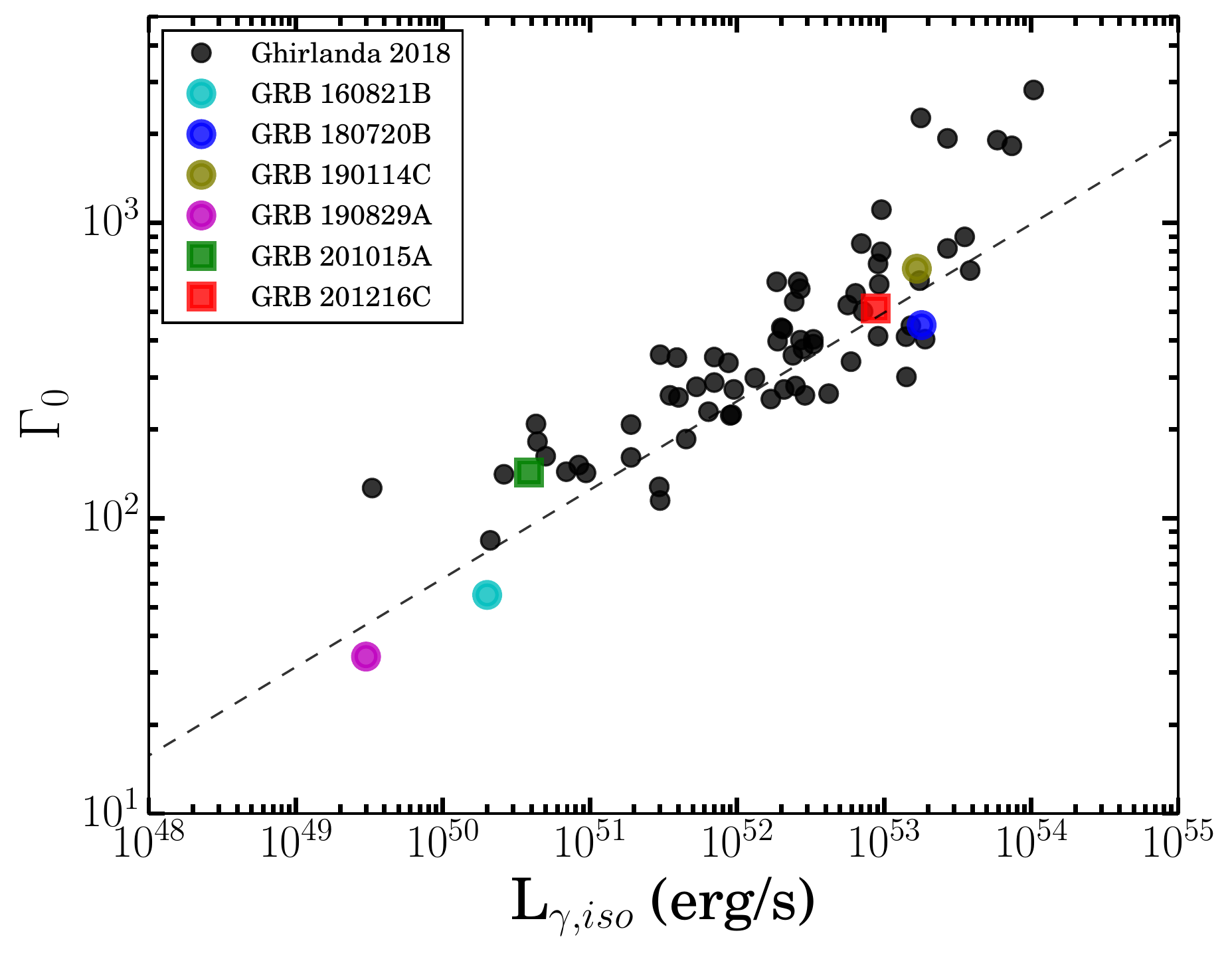}
\caption{Upper left and right panels represent the two important tight correlations $\Gamma_0- E_{\rm \gamma, iso}$ and $\Gamma_0- t_{p,z}$, respectively, discovered by the \textbf{L2010}. The lower panel represents the $\Gamma_0- L_{\rm \gamma, iso}$ correlation given by \cite{2012ApJ...751...49L}. Data points shown with black dots are taken from \cite{2018AA...609A.112G}. Green and red squares represent the \thisgrba and \thisgrbb, respectively, clearly satisfying the given correlations. Data points shown with various colored dots are taken from publications cited in Table \ref{tab:gcn}.}
\label{fig:onset_corelation2}
\end{figure}

\begin{figure}
\centering
\includegraphics[scale=0.52]{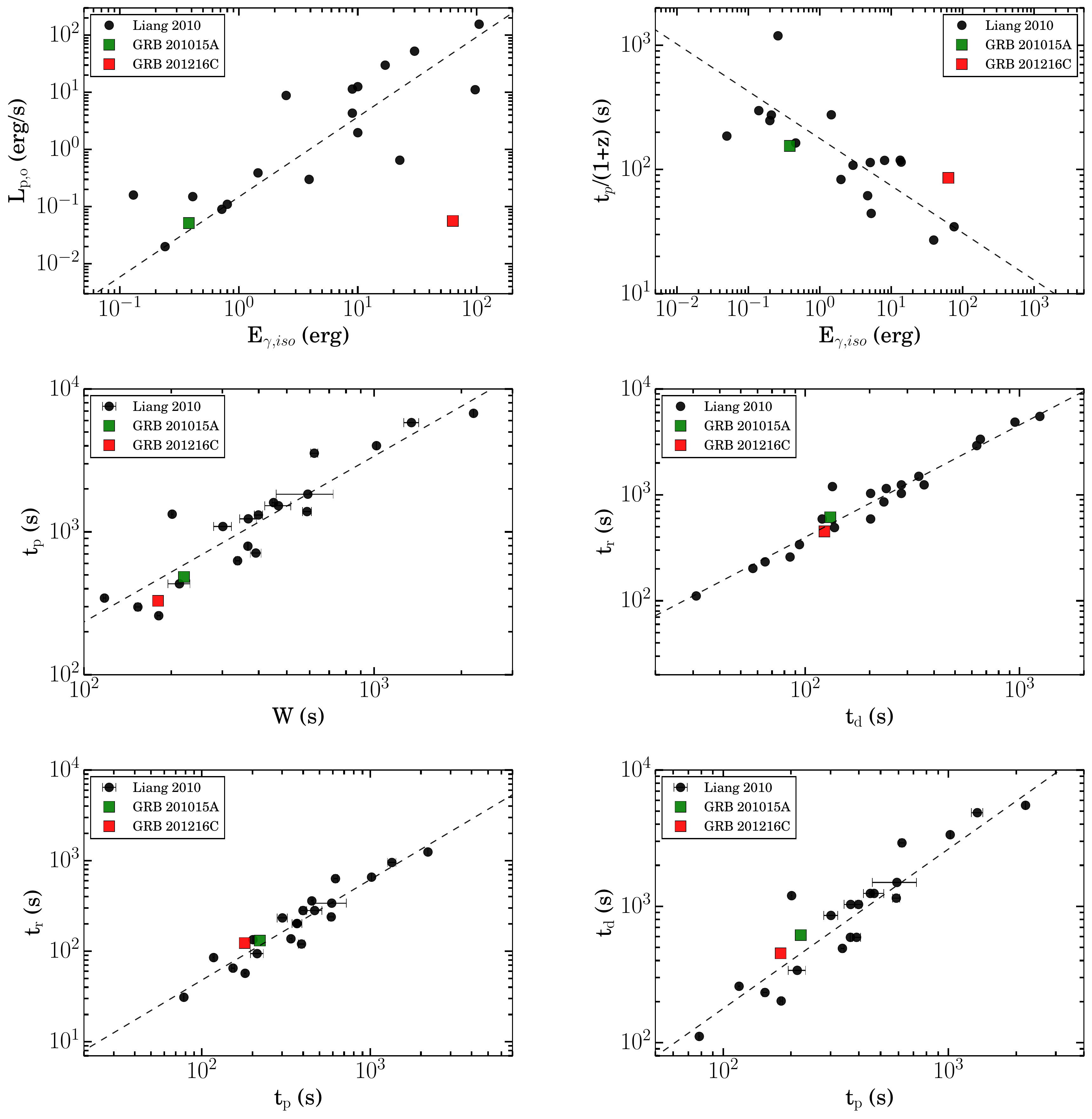}
\caption{Correlations between onset parameters studied by the \textbf{L2010}. Green and red squares represent the \thisgrba and \thisgrbb, respectively. \thisgrbb is not satisfying the L$_{p,o}$-E$_{\gamma,iso}$ correlation, indicating the dark nature of the burst.}
\label{fig:onset_corelation}
\end{figure}

\section{Additional Tables}

\FloatBarrier
\begin{table}
\caption{Results of time-average (\fermiT-0.503 to \fermiT+ 47.09 s) spectral fitting of the GBM data for \thisgrbb. Time-integrated flux has been calculated from 10 \keV to 10 MeV energy range.}
\label{tab:TAS}
\centering
\begin{tabular}{|c|c|c|c|c|c|c|c|} \hline 
\textbf{Model} & \multicolumn{5}{c|}{\textbf{Spectral parameters}} & \multicolumn{2}{c|}{\textbf{Statistics}} \\ \hline
 &\boldmath $ \alpha_{\rm \bf pt}$ & \boldmath \Ep /\boldmath $ E_{\it c}$ (\boldmath $\keV$) & \boldmath $ \beta_{\rm \bf pt}$ & $\rm \bf k{\it \bf T}_{\rm \bf BB}$ (\boldmath $\keV$) &\bf Flux \bf (erg \boldmath$\rm cm^{-2}$ $\rm s^{-1}$)& \textbf{DIC} & \textbf{$\rm \bf \Delta~ {DIC}$}\\ \hline
\sw{Band}&$-1.07_{-0.01}^{+0.01}$&$344.15_{-9.90}^{+9.86}$&$-2.35_{-0.05}^{+0.05}$&-&4.29 $\times 10^{-06}$&7609.74&-\\
\sw{Band+BB}&$-0.98_{-0.02}^{+0.02}$&$352.31_{-12.74}^{+12.77}$&$-2.39_{-0.06}^{+0.06}$&$9.94_{-0.52}^{+0.52}$&4.26 $\times 10^{-06}$ &7502.07& -107.67\\
\sw{CPL}&$-1.11_{-0.01}^{+0.01}$&$452.95_{-11.70}^{+11.52}$&-&-&3.38 $\times 10^{-06}$ &7684.57&- \\
\sw{CPL+BB}&$-1.07_{-0.01}^{+0.01}$&$464.40_{-17.64}^{+17.94}$&-&$12.63_{-0.83}^{+0.86}$&3.45 $\times 10^{-06}$ &7558.00&-126.57\\\hline
\sw{Synchrotron}&\bf B (G)&\boldmath$p$& \boldmath $\gamma_{cool}$& &\bf Flux \bf (erg \boldmath$\rm cm^{-2}$ $\rm s^{-1}$)& \textbf{DIC} & \textbf{$\rm \bf \Delta~ {DIC}$}\\ \hline
&$142.60_{-11.42}^{+11.49}$&$5.02_{-0.56}^{+0.60}$&$180875.73_{-10325.27}^{+9787.31}$&-&4.11 $\times 10^{-06}$ &7454.89&- \\\hline
\end{tabular}
\end{table}

\begin{sidewaystable*}
\addtolength{\tabcolsep}{1.5pt}
\caption{Time-resolved spectral analysis results of \thisgrbb modeled by empirical \sw{Cutoff power-law} (CPL) and \sw{CPL + Blackbody} models. The reported flux (in erg $\rm cm^{-2}$ $\rm s^{-1}$) are calculated in 10 \keV-10 MeV energy range.}
\label{TRS:cpl_band}
\begin{scriptsize}
\begin{center}
\begin{tabular}{|c|c|ccc|cccc|ccc|} \hline
&\bf S &\multicolumn{3}{c|}{\bf CPL}&\multicolumn{4}{c|}{\bf CPL+BB}& {\bf CPL} & {\bf CPL+BB} &\\
\bf t$_{\rm start}$-t$_{\rm stop}$  &  &\boldmath $\Gamma$ &\boldmath \Ep  & \bf Flux  & {\boldmath $\alpha_{\rm pt}$} & {\boldmath \Ep \boldmath} &\bf kT & \bf Flux &\bf DIC & \bf DIC & \bf $\Delta$DIC\\
\bf (s)&  &  &(\boldmath$\keV$)& &  & (\boldmath$\keV$) & (\boldmath$\keV$) & & & & \\[1ex] \hline
3.93-5.68 & 40.93 & $-1.00_{-0.05}^{0.05}$ & $483.52_{-77.87}^{78.55}$ & 2.42$\times 10^{ 6}$ & $-0.94_{-0.08}^{0.08}$ & $517.54_{-111.94}^{110.79}$ & $14.66_{-3.74}^{4.17}$ & 2.68$\times 10^{ 6}$ & 2394.02 & 2365.42 & -28.6  \\ 
5.68-6.23 & 35.83 & $-0.98_{-0.05}^{0.05}$ & $637.45_{-111.95}^{112.58}$ & 5.2$\times 10^{ 6}$ & $-0.87_{-0.08}^{0.08}$ & $681.70_{-140.22}^{139.82}$ & $16.88_{-2.84}^{2.87}$ & 5.73$\times 10^{ 6}$ & 665.66 & 637.12 & -28.54  \\ 
6.23-7.24 & 63.53 & $-0.89_{-0.03}^{0.03}$ & $583.88_{-61.01}^{61.08}$ & 8.08$\times 10^{ 6}$ & $-0.83_{-0.05}^{0.05}$ & $598.39_{-83.68}^{83.53}$ & $17.04_{-3.05}^{3.38}$ & 8.36$\times 10^{ 6}$ & 1739.16 & 1719.61 & -19.55  \\ 
7.24-10.19 & 120.81 & $-0.91_{-0.02}^{0.02}$ & $452.07_{-23.52}^{23.97}$ & 7.21$\times 10^{ 6}$ & $-0.85_{-0.04}^{0.04}$ & $460.25_{-37.54}^{37.65}$ & $15.52_{-2.06}^{2.05}$ & 7.43$\times 10^{ 6}$ & 3437.91 & 3396.15 & -41.76  \\ 
10.19-11.02 & 46.78 & $-1.02_{-0.05}^{0.05}$ & $333.80_{-46.34}^{44.99}$ & 3.21$\times 10^{ 6}$ & $-0.94_{-0.07}^{0.07}$ & $339.63_{-53.97}^{52.78}$ & $12.22_{-1.68}^{1.87}$ & 3.32$\times 10^{ 6}$ & 1256.32 & 1242.7 & -13.63  \\ 
11.02-14.90 & 77.15 & $-1.08_{-0.03}^{0.03}$ & $297.07_{-25.36}^{25.32}$ & 2.15$\times 10^{ 6}$ & $-1.01_{-0.04}^{0.04}$ & $289.91_{-28.00}^{27.71}$ & $10.13_{-1.16}^{1.12}$ & 2.19$\times 10^{ 6}$ & 3505.93 & 3491.02 & -14.91  \\ 
14.90-15.66 & 48.58 & $-1.04_{-0.05}^{0.05}$ & $401.31_{-58.67}^{57.84}$ & 3.92$\times 10^{ 6}$ & $-1.03_{-0.07}^{0.07}$ & $443.88_{-89.72}^{87.04}$ & $14.13_{-5.39}^{5.97}$ & 4.21$\times 10^{ 6}$ & 1227.59 & 1214.58 & -13.02  \\ 
15.66-17.28 & 59.96 & $-1.03_{-0.04}^{0.04}$ & $251.80_{-26.80}^{26.75}$ & 2.35$\times 10^{ 6}$ & $-1.03_{-0.07}^{0.07}$ & $277.08_{-47.17}^{43.24}$ & $12.59_{-4.46}^{5.70}$ & 2.61$\times 10^{ 6}$ & 2286.17 & 2267.03 & -19.14  \\ 
17.74-18.58 & 48.27 & $-1.17_{-0.04}^{0.04}$ & $405.31_{-62.19}^{61.81}$ & 3.22$\times 10^{ 6}$ & $-1.12_{-0.05}^{0.06}$ & $413.46_{-67.32}^{65.97}$ & $10.87_{-1.98}^{2.11}$ & 3.31$\times 10^{ 6}$ & 1236.56 & 1229.42 & -7.15  \\ 
18.58-19.87 & 73.73 & $-1.02_{-0.03}^{0.03}$ & $430.15_{-39.88}^{39.74}$ & 5.27$\times 10^{ 6}$ & $-0.99_{-0.04}^{0.04}$ & $439.19_{-49.23}^{48.92}$ & $13.13_{-3.47}^{3.99}$ & 5.44$\times 10^{ 6}$ & 2005.36 & 1995.71 & -9.65  \\ 
19.87-20.47 & 62.39 & $-1.08_{-0.03}^{0.03}$ & $587.08_{-76.91}^{76.28}$ & 7.57$\times 10^{ 6}$ & $-1.06_{-0.05}^{0.05}$ & $730.58_{-127.81}^{127.50}$ & $18.14_{-2.78}^{2.80}$ & 8.36$\times 10^{ 6}$ & 1052.87 & 1028.02 & -24.85  \\ 
20.47-21.42 & 61.86 & $-1.08_{-0.04}^{0.04}$ & $362.22_{-43.14}^{42.79}$ & 4.06$\times 10^{ 6}$ & $-1.07_{-0.05}^{0.05}$ & $408.69_{-68.74}^{69.94}$ & $14.25_{-3.31}^{3.71}$ & 4.36$\times 10^{ 6}$ & 1480.71 & 1468.67 & -12.04  \\ 
21.42-21.72 & 44.21 & $-1.02_{-0.05}^{0.05}$ & $528.13_{-83.90}^{82.39}$ & 7.55$\times 10^{ 6}$ & $-1.01_{-0.05}^{0.05}$ & $566.01_{-106.30}^{105.48}$ & $14.49_{-8.60}^{7.99}$ & 7.95$\times 10^{ 6}$ & 13.17 & 7.63 & -5.53  \\ 
21.72-22.92 & 61.99 & $-1.10_{-0.04}^{0.04}$ & $310.73_{-33.62}^{33.31}$ & 3.25$\times 10^{ 6}$ & $-1.06_{-0.05}^{0.05}$ & $313.66_{-36.88}^{36.36}$ & $10.56_{-1.58}^{1.64}$ & 3.28$\times 10^{ 6}$ & 1845.48 & 1837.52 & -7.96  \\ 
22.92-23.85 & 76.31 & $-0.99_{-0.03}^{0.03}$ & $387.04_{-32.61}^{32.54}$ & 6.35$\times 10^{ 6}$ & $-0.95_{-0.04}^{0.04}$ & $381.01_{-35.13}^{35.02}$ & $11.60_{-2.07}^{2.23}$ & 6.46$\times 10^{ 6}$ & 1581.2 & 1573.82 & -7.38  \\ 
23.85-24.28 & 76.51 & $-0.93_{-0.03}^{0.03}$ & $412.51_{-36.18}^{36.40}$ & 1.17$\times 10^{ 6}$ & $-0.92_{-0.04}^{0.04}$ & $418.79_{-41.01}^{38.80}$ & $12.71_{-6.75}^{6.72}$ & 1.19$\times 10^{ 6}$ & 638.32 & 634.18 & -4.14  \\ 
24.28-24.73 & 66.46 & $-0.97_{-0.04}^{0.04}$ & $370.73_{-36.39}^{36.53}$ & 8.69$\times 10^{ 6}$ & $-0.99_{-0.06}^{0.06}$ & $471.95_{-81.47}^{81.21}$ & $20.52_{-4.24}^{4.20}$ & 9.39$\times 10^{ 6}$ & 621.01 & 598.12 & -22.88  \\ 
24.73-25.83 & 123.46 & $-1.05_{-0.02}^{0.02}$ & $517.88_{-33.73}^{33.69}$ & 1.24$\times 10^{ 6}$ & $-1.04_{-0.03}^{0.03}$ & $601.55_{-66.48}^{66.01}$ & $18.14_{-3.08}^{3.10}$ & 1.33$\times 10^{ 6}$ & 2082.07 & 2047.75 & -34.32  \\ 
25.83-27.53 & 134.88 & $-1.06_{-0.02}^{0.02}$ & $463.91_{-26.29}^{26.31}$ & 9.58$\times 10^{ 6}$ & $-1.04_{-0.03}^{0.03}$ & $467.15_{-34.63}^{31.36}$ & $12.74_{-3.14}^{2.85}$ & 9.92$\times 10^{ 6}$ & 2664.38 & 2650.17 & -14.21  \\ 
27.53-28.15 & 97.35 & $-1.10_{-0.02}^{0.02}$ & $579.32_{-50.35}^{50.10}$ & 1.28$\times 10^{ 6}$ & $-1.09_{-0.03}^{0.03}$ & $672.64_{-82.16}^{81.84}$ & $18.28_{-3.13}^{3.16}$ & 1.36$\times 10^{ 6}$ & 1248.52 & 1229.01 & -19.51  \\ 
28.15-29.06 & 97.18 & $-1.19_{-0.02}^{0.02}$ & $519.64_{-47.69}^{48.03}$ & 8.29$\times 10^{ 6}$ & $-1.18_{-0.03}^{0.03}$ & $622.46_{-77.14}^{77.40}$ & $15.59_{-1.86}^{1.93}$ & 8.95$\times 10^{ 6}$ & 1679.26 & 1654.19 & -25.07  \\ 
29.06-29.79 & 69.43 & $-1.28_{-0.03}^{0.04}$ & $426.52_{-58.70}^{58.11}$ & 4.71$\times 10^{ 6}$ & $-1.28_{-0.04}^{0.04}$ & $558.76_{-99.10}^{98.59}$ & $13.60_{-1.12}^{1.10}$ & 5.06$\times 10^{ 6}$ & 1202.71 & 1182.16 & -20.56  \\ 
29.79-30.81 & 63.57 & $-1.31_{-0.04}^{0.04}$ & $407.51_{-61.43}^{61.76}$ & 3.2$\times 10^{ 6}$ & $-1.30_{-0.04}^{0.04}$ & $491.35_{-95.63}^{94.87}$ & $12.20_{-1.52}^{1.60}$ & 3.4$\times 10^{ 6}$ & 1659.75 & 1647.92 & -11.83  \\ 
30.81-31.95 & 50.74 & $-1.39_{-0.04}^{0.04}$ & $471.88_{-95.45}^{94.56}$ & 2.29$\times 10^{ 6}$ & $-1.39_{-0.05}^{0.05}$ & $506.05_{-116.77}^{113.31}$ & $9.75_{-5.41}^{5.23}$ & 2.35$\times 10^{ 6}$ & 1649.24 & 1645.29 & -3.95  \\ 
31.95-33.11 & 39.54 & $-1.26_{-0.06}^{0.06}$ & $217.73_{-38.05}^{37.27}$ & 1.34$\times 10^{ 6}$ & $-1.25_{-0.07}^{0.07}$ & $262.15_{-58.42}^{56.56}$ & $10.61_{-1.44}^{1.53}$ & 1.43$\times 10^{ 6}$ & 1644.47 & 1635.39 & -9.08  \\ 
33.11-35.71 & 43.68 & $-1.36_{-0.05}^{0.05}$ & $288.97_{-55.81}^{54.46}$ & 1.01$\times 10^{ 6}$ & $-1.39_{-0.08}^{0.08}$ & $358.73_{-109.39}^{108.86}$ & $11.04_{-6.40}^{7.31}$ & 1.16$\times 10^{ 6}$ & 2812.15 & 2795.55 & -16.6  \\ 
35.71-38.83 & 35.6 & $-1.36_{-0.07}^{0.07}$ & $259.59_{-61.17}^{59.37}$ & 6.66$\times 10^{ 6}$ & $-1.34_{-0.08}^{0.08}$ & $337.47_{-105.31}^{103.36}$ & $9.46_{-1.31}^{1.37}$ & 7.35$\times 10^{ 6}$ & 3034.54 & 3022.18 & -12.36 \\ \hline
\end{tabular}
\end{center}
\end{scriptsize}
\end{sidewaystable*}

\begin{sidewaystable}
\caption{Time-resolved spectral analysis results of \thisgrbb modeled by empirical \sw{Band} function and its combination using \sw{Blackbody} (BB) function. The reported flux (in erg $\rm cm^{-2}$ $\rm s^{-1}$) are calculated in 10 \keV-10 MeV energy range.} 
\label{TRS:Band}
\begin{scriptsize}
\begin{center}
\begin{tabular}{|c| c c c c |c c c c c |c c c|}\hline
&  &\bf Band & & & & &\bf Band+BB & & &\bf Band&\bf Band+BB& \\ [1ex]
\bf t$_{\rm start}$-t$_{\rm stop}$&\boldmath $ \alpha_{\rm \bf pt}$ &\boldmath $\beta_{\rm \bf pt}$ &\boldmath \Ep &\bf Flux  &\boldmath $ \alpha_{\rm \bf pt}$ &\boldmath $\beta_{\rm \bf pt}$ &\boldmath \Ep & \boldmath ${\rm k}{T}$ &\bf Flux &\bf DIC &\bf DIC &\boldmath $\Delta DIC$ \\ [1ex]
\bf (s)&  & &\bf  (\boldmath$\keV$)& & & & (\boldmath$\keV$) & (\boldmath$\keV$) & & & & \\[1ex] \hline
3.93-5.68 & $-0.97_{-0.06}^{0.06}$ & $-2.81_{-0.58}^{0.51}$ & $428.87_{-65.55}^{65.43}$ & 2.78$\times 10^{ 6}$ & $-0.90_{-0.10}^{0.10}$ & $-3.02_{-0.69}^{0.64}$ & $479.50_{-89.33}^{89.10}$ & $13.37_{-3.34}^{3.81}$ & 2.86$\times 10^{ 6}$ & 2391.24 & 2380.36 & -10.88 \\
5.68-6.23 & $-0.96_{-0.06}^{0.05}$ & $-3.30_{-0.45}^{0.44}$ & $606.17_{-93.56}^{94.31}$ & 5.48$\times 10^{ 6}$ & $-0.85_{-0.09}^{0.09}$ & $-3.63_{-0.59}^{0.57}$ & $709.63_{-120.59}^{121.34}$ & $16.46_{-2.80}^{2.91}$ & 5.81$\times 10^{ 6}$ & 665.42 & 650.85 & -14.57 \\
6.23-7.24 & $-0.85_{-0.04}^{0.04}$ & $-2.36_{-0.15}^{0.16}$ & $562.91_{-50.84}^{50.08}$ & 1.01$\times 10^{ 5}$ & $-0.77_{-0.06}^{0.06}$ & $-2.40_{-0.16}^{0.16}$ & $584.21_{-57.14}^{56.27}$ & $14.85_{-2.37}^{2.73}$ & 1.03$\times 10^{ 5}$ & 1720.89 & 1712.18 & -8.71 \\
7.24-10.19 & $-0.86_{-0.02}^{0.02}$ & $-2.39_{-0.11}^{0.11}$ & $417.29_{-23.53}^{23.51}$ & 9.12$\times 10^{ 6}$ & $-0.76_{-0.04}^{0.04}$ & $-2.44_{-0.12}^{0.12}$ & $432.09_{-27.61}^{27.71}$ & $12.66_{-1.01}^{1.00}$ & 9.03$\times 10^{ 6}$ & 3412.17 & 3384.55 & -27.62 \\
10.19-11.02 & $-1.01_{-0.05}^{0.05}$ & $-3.24_{-0.42}^{0.40}$ & $313.48_{-29.52}^{29.92}$ & 3.46$\times 10^{ 6}$ & $-0.92_{-0.07}^{0.07}$ & $-3.43_{-0.55}^{0.52}$ & $341.72_{-37.52}^{37.28}$ & $12.00_{-1.54}^{1.72}$ & 3.49$\times 10^{ 6}$ & 1254.32 & 1246.93 & -7.39 \\
11.02-14.90 & $-1.04_{-0.04}^{0.04}$ & $-2.58_{-0.31}^{0.29}$ & $239.83_{-21.76}^{21.47}$ & 2.66$\times 10^{ 6}$ & $-0.97_{-0.05}^{0.05}$ & $-2.70_{-0.34}^{0.35}$ & $254.91_{-23.81}^{23.22}$ & $9.52_{-0.98}^{0.99}$ & 2.6$\times 10^{ 6}$ & 3496.97 & 3486.73 & -10.24 \\
14.90-15.66 & $-1.00_{-0.06}^{0.06}$ & $-2.65_{-0.47}^{0.43}$ & $334.92_{-48.66}^{47.49}$ & 4.63$\times 10^{ 6}$ & $-0.98_{-0.07}^{0.07}$ & $-2.80_{-0.56}^{0.52}$ & $372.32_{-65.40}^{61.92}$ & $12.24_{-4.88}^{5.07}$ & 4.75$\times 10^{ 6}$ & 1222.70 & 1218.13 & -4.57 \\
15.66-17.28 & $-0.95_{-0.06}^{0.06}$ & $-2.46_{-0.26}^{0.27}$ & $203.28_{-22.37}^{22.21}$ & 3.13$\times 10^{ 6}$ & $-0.94_{-0.07}^{0.07}$ & $-2.56_{-0.32}^{0.33}$ & $216.35_{-27.74}^{26.31}$ & $9.33_{-3.47}^{2.82}$ & 3.07$\times 10^{ 6}$ & 2277.67 & 2274.30 & -3.37 \\
17.74-18.58 & $-1.16_{-0.05}^{0.05}$ & $-3.13_{-0.50}^{0.48}$ & $324.62_{-37.61}^{37.21}$ & 3.49$\times 10^{ 6}$ & $-1.11_{-0.06}^{0.06}$ & $-3.34_{-0.62}^{0.59}$ & $349.90_{-44.38}^{43.94}$ & $10.80_{-1.85}^{1.99}$ & 3.51$\times 10^{ 6}$ & 1234.22 & 1228.86 & -5.36 \\
18.58-19.87 & $-1.00_{-0.04}^{0.04}$ & $-2.55_{-0.30}^{0.30}$ & $377.18_{-36.28}^{36.48}$ & 6.54$\times 10^{ 6}$ & $-0.97_{-0.05}^{0.05}$ & $-2.64_{-0.34}^{0.35}$ & $397.69_{-43.41}^{42.65}$ & $11.92_{-3.25}^{3.55}$ & 6.53$\times 10^{ 6}$ & 1997.41 & 1992.54 & -4.87 \\
19.87-20.47 & $-1.00_{-0.07}^{0.07}$ & $-2.31_{-0.28}^{0.28}$ & $406.97_{-85.82}^{86.54}$ & 9.79$\times 10^{ 6}$ & $-1.04_{-0.06}^{0.06}$ & $-2.93_{-0.60}^{0.55}$ & $637.73_{-111.05}^{110.37}$ & $17.67_{-3.00}^{3.03}$ & 9.42$\times 10^{ 6}$ & 1042.53 & 1027.14 & -15.39 \\
20.47-21.42 & $-1.06_{-0.04}^{0.04}$ & $-2.95_{-0.48}^{0.44}$ & $307.31_{-29.31}^{29.47}$ & 4.49$\times 10^{ 6}$ & $-1.05_{-0.06}^{0.06}$ & $-3.18_{-0.59}^{0.55}$ & $347.84_{-44.89}^{43.44}$ & $13.25_{-3.15}^{3.62}$ & 4.58$\times 10^{ 6}$ & 1477.02 & 1470.68 & -6.34 \\
21.42-21.72 & $-1.00_{-0.06}^{0.06}$ & $-2.56_{-0.43}^{0.39}$ & $470.72_{-77.99}^{76.95}$ & 9.45$\times 10^{ 6}$ & $-1.00_{-0.06}^{0.06}$ & $-2.64_{-0.45}^{0.43}$ & $521.73_{-103.01}^{102.66}$ & $14.10_{-8.48}^{8.36}$ & 9.79$\times 10^{ 6}$ & 7.15 & 2.77 & -4.38 \\
21.72-22.92 & $-1.08_{-0.04}^{0.04}$ & $-3.04_{-0.51}^{0.46}$ & $263.76_{-22.35}^{22.36}$ & 3.59$\times 10^{ 6}$ & $-1.04_{-0.05}^{0.05}$ & $-3.28_{-0.65}^{0.61}$ & $281.16_{-26.48}^{26.37}$ & $10.34_{-1.54}^{1.64}$ & 3.56$\times 10^{ 6}$ & 1842.93 & 1836.55 & -6.38 \\
22.92-23.85 & $-0.98_{-0.03}^{0.03}$ & $-3.47_{-0.35}^{0.35}$ & $382.29_{-24.00}^{24.40}$ & 6.69$\times 10^{ 6}$ & $-0.94_{-0.04}^{0.04}$ & $-3.78_{-0.53}^{0.52}$ & $394.20_{-26.48}^{26.52}$ & $11.46_{-1.99}^{2.13}$ & 6.68$\times 10^{ 6}$ & 1580.47 & 1574.85 & -5.62 \\
23.85-24.28 & $-0.92_{-0.03}^{0.03}$ & $-2.94_{-0.39}^{0.36}$ & $417.58_{-29.90}^{30.28}$ & 1.31$\times 10^{ 5}$ & $-0.91_{-0.04}^{0.04}$ & $-3.01_{-0.43}^{0.42}$ & $427.04_{-34.54}^{33.68}$ & $11.62_{-5.97}^{5.38}$ & 1.32$\times 10^{ 5}$ & 632.94 & 630.25 & -2.69 \\
24.28-24.73 & $-0.89_{-0.06}^{0.06}$ & $-2.47_{-0.24}^{0.26}$ & $313.48_{-39.43}^{39.88}$ & 1.08$\times 10^{ 5}$ & $-0.95_{-0.08}^{0.08}$ & $-2.97_{-0.59}^{0.55}$ & $427.92_{-77.28}^{74.58}$ & $18.77_{-4.79}^{5.09}$ & 1.05$\times 10^{ 5}$ & 612.58 & 599.65 & -12.92 \\
24.73-25.83 & $-0.99_{-0.03}^{0.03}$ & $-2.36_{-0.11}^{0.11}$ & $402.27_{-29.46}^{29.66}$ & 1.54$\times 10^{ 5}$ & $-0.98_{-0.04}^{0.03}$ & $-2.50_{-0.16}^{0.17}$ & $463.80_{-43.91}^{43.59}$ & $14.62_{-1.92}^{1.97}$ & 1.52$\times 10^{ 5}$ & 2056.01 & 2039.19 & -16.82 \\
25.83-27.53 & $-1.05_{-0.02}^{0.02}$ & $-3.08_{-0.37}^{0.34}$ & $414.49_{-20.94}^{20.77}$ & 1.03$\times 10^{ 5}$ & $-1.02_{-0.03}^{0.03}$ & $-3.18_{-0.43}^{0.41}$ & $422.88_{-23.47}^{22.77}$ & $11.33_{-2.08}^{1.62}$ & 1.03$\times 10^{ 5}$ & 2658.61 & 2651.25 & -7.36 \\
27.53-28.15 & $-1.07_{-0.03}^{0.03}$ & $-2.70_{-0.35}^{0.34}$ & $465.77_{-50.77}^{48.69}$ & 1.47$\times 10^{ 5}$ & $-1.08_{-0.04}^{0.04}$ & $-3.11_{-0.52}^{0.48}$ & $574.32_{-64.06}^{64.42}$ & $17.62_{-3.22}^{3.33}$ & 1.47$\times 10^{ 5}$ & 1241.25 & 1227.12 & -14.13 \\
28.15-29.06 & $-1.17_{-0.03}^{0.03}$ & $-2.93_{-0.53}^{0.48}$ & $388.63_{-39.46}^{38.55}$ & 9.17$\times 10^{ 6}$ & $-1.18_{-0.03}^{0.03}$ & $-3.45_{-0.57}^{0.55}$ & $498.32_{-51.01}^{51.08}$ & $15.48_{-1.86}^{1.90}$ & 9.36$\times 10^{ 6}$ & 1674.67 & 1652.97 & -21.70 \\
29.06-29.79 & $-1.19_{-0.08}^{0.08}$ & $-2.41_{-0.39}^{0.34}$ & $233.62_{-52.57}^{52.59}$ & 6.00$\times 10^{ 6}$ & $-1.27_{-0.05}^{0.04}$ & $-3.18_{-0.65}^{0.63}$ & $387.90_{-63.20}^{63.30}$ & $13.51_{-1.15}^{1.15}$ & 5.5$\times 10^{ 6}$ & 1192.50 & 1179.51 & -13.00 \\
29.79-30.81 & $-1.21_{-0.09}^{0.09}$ & $-2.22_{-0.20}^{0.24}$ & $197.90_{-49.75}^{47.40}$ & 4.63$\times 10^{ 6}$ & $-1.27_{-0.06}^{0.06}$ & $-2.60_{-0.55}^{0.48}$ & $292.42_{-66.04}^{64.36}$ & $11.53_{-1.71}^{1.84}$ & 4.14$\times 10^{ 6}$ & 1645.06 & 1641.88 & -3.18 \\
30.81-31.95 & $-1.37_{-0.05}^{0.05}$ & $-2.85_{-0.59}^{0.55}$ & $262.67_{-46.22}^{46.46}$ & 2.55$\times 10^{ 6}$ & $-1.38_{-0.06}^{0.06}$ & $-3.00_{-0.69}^{0.65}$ & $284.65_{-56.64}^{56.88}$ & $9.21_{-4.97}^{4.86}$  & 2.6$\times 10^{ 6}$ & 1644.99 & 1642.27 & -2.73 \\
31.95-33.11 & $-1.21_{-0.09}^{0.08}$ & $-3.06_{-0.58}^{0.59}$ & $146.46_{-19.71}^{21.30}$ & 1.49$\times 10^{ 6}$ & $-1.24_{-0.08}^{0.08}$ & $-3.39_{-0.64}^{0.62}$ & $187.49_{-31.05}^{30.93}$ & $10.51_{-1.44}^{1.54}$ & 1.52$\times 10^{ 6}$ & 1640.42 & 1635.42 & -5.00 \\
33.11-35.71 & $-1.31_{-0.07}^{0.07}$ & $-2.85_{-0.57}^{0.51}$ & $159.75_{-24.54}^{23.99}$ & 1.15$\times 10^{ 6}$ & $-1.33_{-0.08}^{0.08}$ & $-3.05_{-0.69}^{0.64}$ & $176.09_{-32.33}^{31.61}$ & $8.53_{-4.61}^{5.32}$  & 1.18$\times 10^{ 6}$ & 2808.19 & 2803.98 & -4.21 \\
35.71-38.83 & $-1.21_{-0.16}^{0.18}$ & $-2.56_{-0.61}^{0.51}$ & $122.37_{-40.78}^{35.51}$ & 8.68$\times 10^{ 7}$ & $-1.29_{-0.10}^{0.09}$ & $-3.01_{-0.68}^{0.66}$ & $188.78_{-45.08}^{44.68}$ & $9.01_{-1.25}^{1.39}$  & 8.03$\times 10^{ 7}$ & 2984.82 & 3020.93 & 36.11 \\ \hline
\end{tabular}
\end{center}
\end{scriptsize}
\end{sidewaystable}

\begin{table}
\addtolength{\tabcolsep}{8pt}
\caption{Time-resolved spectral analysis results of \thisgrbb modeled by physical synchrotron model. The reported flux (in erg $\rm cm^{-2}$ $\rm s^{-1}$) column are measured in 10 \keV-10 MeV energy channels.} 
\label{TRS:Synchrotron}
\begin{scriptsize}
\begin{center} 
\begin{tabular}{|c|c|cccc|c|}\hline
\bf t$_{\rm start}$-\bf t$_{\rm stop}$  & \bf S & \multicolumn{4}{c|}{\bf Synchrotron}& \\ [1ex]
\bf (s) &  & \bf B (G) & \boldmath $p$ &\boldmath $\gamma_{cool}$ $\times$ $10^4$ ($\mathbf{\keV}$) &\bf Flux  &\bf DIC \\ [1ex] \hline
3.93-5.68 & 40.93 & $127.44_{-57.34}^{+52.09}$ & $4.17_{-1.12}^{+1.17}$ & $71.28_{-50.33}^{+14.32}$ & 3.04 $\times 10^{6}$ & 1061.27 \\
5.68-6.23 & 35.83 & $191.76_{-60.38}^{+62.20}$ & $4.88_{-0.79}^{+0.78}$ & $41.95_{-21.68}^{+10.09}$ & 6.07 $\times 10^{6}$ & 495.69 \\
6.23-7.24 & 63.52 & $141.95_{-80.01}^{+76.50}$ & $3.65_{-0.73}^{+0.86}$ & $145.89_{-115.55}^{+43.00}$ & 1.06 $\times 10^{5}$ & 268.10 \\
7.24-10.19 & 120.80 & $119.98_{-26.00}^{+26.07}$ & $4.33_{-0.75}^{+0.82}$ & $43.89_{-9.60}^{+8.73}$ & 8.87 $\times 10^{6}$ & 3220.77 \\
10.19-11.02 & 46.78 & $82.79_{-22.41}^{+22.44}$ & $4.74_{-0.82}^{+0.83}$ & $39.92_{-15.12}^{+11.87}$ & 3.75 $\times 10^{6}$ & 1135.13 \\
11.02-14.90 & 77.15 & $66.66_{-15.68}^{+15.62}$ & $4.35_{-0.91}^{+0.95}$ & $3162_{-7.43}^{+6.43}$ & 2.58 $\times 10^{6}$ & 3337.24 \\
14.90-15.66 & 48.57 & $83.41_{-45.05}^{+41.46}$ & $3.87_{-1.03}^{+1.17}$ & $95.85_{-73.26}^{+22.89}$ & 4.88 $\times 10^{6}$ & -1025.49 \\
15.66-17.28 & 59.95 & $36.96_{-19.85}^{+18.32}$ & $3.78_{-0.78}^{+0.97}$ & $163.60_{-126.22}^{+85.52}$ & 2.93 $\times 10^{6}$ & 976.25 \\
17.74-18.58 & 48.27 & $96.21_{-23.41}^{+23.76}$ & $4.48_{-0.93}^{+0.96}$ & $21.09_{-5.95}^{+4.98}$ & 3.82 $\times 10^{6}$ & 1114.54 \\
18.58-19.87 & 73.72 & $102.62_{-31.92}^{+31.12}$ & $3.88_{-0.89}^{+0.99}$ & $34.08_{-10.35}^{+8.43}$ & 6.71 $\times 10^{6}$ & 1662.13 \\
19.87-20.47 & 62.38 & $161.77_{-54.98}^{+55.49}$ & $4.00_{-1.00}^{+1.08}$ & $28.86_{-13}^{+6.06}$ & 9.67 $\times 10^{6}$ & -617.38 \\
20.47-21.42 & 61.86 & $88.19_{-22.10}^{+21.46}$ & $4.54_{-0.87}^{+0.89}$ & $28.82_{-8.04}^{+6.80}$ & 4.8 $\times 10^{6}$ & 1357.23 \\
21.42-21.72 & 44.21 & $118.33_{-75.31}^{+70.63}$ & $3.49_{-0.88}^{+0.99}$ & $89.76_{-69.89}^{+30.89}$ & 1.01 $\times 10^{5}$ & -1935.17 \\
21.72-22.92 & 61.99 & $70.93_{-14.81}^{+14.80}$ & $4.57_{-0.85}^{+0.87}$ & $29.05_{-6.89}^{+6.44}$ & 3.79 $\times 10^{6}$ & 1789.03 \\
22.92-23.85 & 76.31 & $102.43_{-16.82}^{+16.92}$ & $5.17_{-0.60}^{+0.60}$ & $36.06_{-7.11}^{+6.88}$ & 7.29 $\times 10^{6}$ & 1566.58 \\
23.85-24.28 & 76.51 & $90.59_{-30.21}^{+29.41}$ & $4.37_{-0.87}^{+0.92}$ & $89.68_{-53.08}^{+22.42}$ & 1.38 $\times 10^{5}$ & 262.16 \\
24.28-24.73 & 66.46 & $66.04_{-35.67}^{+31.36}$ & $3.87_{-0.82}^{+0.97}$ & $129.61_{-94.28}^{+62.92}$ & 1.06 $\times 10^{5}$ & -545.52 \\
24.73-25.83 & 123.46 & $151.66_{-27.87}^{+27.59}$ & $4.44_{-0.79}^{+0.84}$ & $23.00_{-3.38}^{+3.35}$ & 1.52 $\times 10^{5}$ & 1984.83 \\
25.83-27.53 & 134.88 & $141.10_{-15.02}^{+15.30}$ & $5.24_{-0.57}^{+0.55}$ & $22.36_{-2.35}^{+2.31}$ & 1.12 $\times 10^{5}$ & 2643.94 \\
27.53-28.15 & 97.34 & $171.33_{-33.88}^{+33.96}$ & $4.61_{-0.82}^{+0.86}$ & $18.96_{-3.13}^{+3.10}$ & 1.55 $\times 10^{5}$ & 1192.78 \\
28.15-29.06 & 97.18 & $133.21_{-22.68}^{+21.71}$ & $4.70_{-0.80}^{+0.83}$ & $15.87_{-2.31}^{+2.28}$ & 9.80 $\times 10^{6}$ & 1629.70 \\
29.06-29.79 & 69.43 & $63.99_{-27.73}^{+26.77}$ & $3.73_{-1.11}^{+1.18}$ & $39.21_{-25.50}^{+7.60}$ & 5.62 $\times 10^{6}$ & -6121.04 \\
29.79-30.81 & 63.56 & $37.73_{-28.41}^{+27.24}$ & $2.93_{-0.72}^{+0.83}$ & $90.80_{-75.68}^{+42.48}$ & 4.39 $\times 10^{6}$ & -11527.37 \\
30.81-31.95 & 50.73 & $56.91_{-14.74}^{+15.59}$ & $3.68_{-0.93}^{+1.02}$ & $13.96_{-3.34}^{+2.77}$ & 2.62 $\times 10^{6}$ & 1527.57 \\
31.95-33.11 & 39.54 & $30.09_{-13.44}^{+11.61}$ & $4.17_{-1.09}^{+1.10}$ & $87.37_{-67.43}^{+17.70}$ & 1.53 $\times 10^{6}$ & -310.89 \\
33.11-35.71 & 43.67 & $38.01_{-10.31}^{+10.34}$ & $3.93_{-0.89}^{+0.89}$ & $20.75_{-8.47}^{+5.92}$ & 1.16 $\times 10^{6}$ & 2521.27 \\
35.71-38.83 & 35.60 & $19.82_{-14.46}^{+15.81}$ & $3.33_{-0.80}^{+1.07}$ & $189.08_{-172.28}^{+135.21}$ & 8.61 $\times 10^{7}$ & -3405.36 \\[1ex]\hline
\end{tabular}
\end{center}
\end{scriptsize}
\end{table}

\begin{table*}
\addtolength{\tabcolsep}{4.5pt}
\begin{center}
\caption{The mean and standard deviation of spectral parameters of several fitting models obtained from the time-resolved spectral analysis of \thisgrbb.}
\label{tab:mean_std}
\begin{tabular}{|c|c|c|c|c|c|} \hline 
\textbf{Model} & \multicolumn{5}{c|}{\textbf{Spectral parameters}}  \\ \hline
 &\boldmath $ \alpha_{\rm \bf pt}$ & \boldmath \Ep /$\bf E_{\it c}$ (\boldmath$\keV$) &\boldmath $ \beta_{\rm \bf pt}$ & $\rm \bf k{\it \bf T}$ (\boldmath$\keV$) &\bf Flux (erg \boldmath$\rm cm^{-2}$ $\rm s^{-1}$)\\ \hline
\sw{Band}&$-1.06 \pm 0.13$&$339.43 \pm 119.39$&$-2.75 \pm 0.33$&-&(6.50 $\pm$ 4.08) $\times 10^{-06}$\\
\sw{Band+BB}&$-1.04 \pm 0.16$&$394.54 \pm 135.69$&$-3.01 \pm 0.36$&$12.66 \pm 2.76$&(6.48 $\pm$ 4.07) $\times 10^{-06}$ \\
\sw{CPL}&$-1.10 \pm 0.14$&$421.77 \pm 109.31$&-&-&(5.50 $\pm$ 3.45) $\times 10^{-06}$ \\
\sw{CPL+BB}&$-1.07 \pm 0.16$&$467.78 \pm 126.97$&-&$13.73 \pm 2.92$&(5.81 $\pm$ 3.63) $\times 10^{-06}$\\\hline
\sw{Synchrotron}&\bf B (G)&\boldmath$p$&\boldmath $\gamma_{cool}$& &\bf Flux (erg \boldmath$\rm cm^{-2}$ $\rm s^{-1}$)\\ \hline
&$96.00 \pm 45.82$&$4.18 \pm 0.54$&$608341.26 \pm 480884.31$&-&(6.71 $\pm$ 4.21) $\times 10^{-06}$ \\\hline
\end{tabular}
\end{center}
\end{table*}

\begin{table*}
\addtolength{\tabcolsep}{5pt}
\caption{Spectral parameter correlation results obtained using Pearson correlation. The r and P denote the Pearson correlation coefficient and the likelihood of a null hypothesis, respectively.}
\label{tab:correlation}
\centering
\begin{tabular}{|c|c|c|c|c|c|c|c|c|} \hline 
\textbf{Model} & \multicolumn{2}{c|}{\textbf{log (Flux)-log (\boldmath \Ep)}} & \multicolumn{2}{c|}{\textbf{log (Flux)-\boldmath $ \alpha_{\rm pt}$}} & \multicolumn{2}{c|}{\textbf{log (\boldmath \Ep)-\boldmath$\alpha_{\rm pt}$}}& \multicolumn{2}{c|}{\textbf{log (Flux)-k$T$}}  \\ [1ex]\hline
 & \bf r & \bf p &  \bf r & \bf p &  \bf r & \bf p & \bf r & \bf p \\ \hline
\sw{Band}& 0.79 & 7.19 $\times 10^{-7}$& 0.60 & 9.67 $\times 10^{-4}$& 0.65 & 2.11 $\times 10^{-4}$ & -&- \\ [1ex] \hline
\sw{Band+BB}& 0.83 & 7.75 $\times 10^{-8}$& 0.53 & 4.13 $\times 10^{-3}$ & 0.56 & 2.39 $\times 10^{-3}$ & 0.72 & 2.51 $\times 10^{-5}$\\ \hline
& \multicolumn{3}{c|}{\textbf{log (Flux)-log (B)}} & \multicolumn{3}{c|}{\textbf{log (Flux)-\boldmath$p$}} & \multicolumn{2}{c|}{\textbf{log (B)-\boldmath$p$}} \\ [1ex]\hline
 & \bf r & \multicolumn{2}{c|}{\textbf{p}} &  \bf r & \multicolumn{2}{c|}{\textbf{p}} & \bf r & \bf p  \\\hline
\sw{Synchrotron}& 0.80 & \multicolumn{2}{c|}{\textbf{5.43 $\times 10^{-7}$}} & 0.31 & \multicolumn{2}{c|}{\textbf{1.10 $\times 10^{-1}$}} & 0.54 & 3.53 $\times 10^{-3}$  \\\hline
& \multicolumn{3}{c|}{\textbf{log (\boldmath \Ep)-log (B)}} & \multicolumn{3}{c|}{\textbf{log (B)-\boldmath $\alpha_{\rm pt}$}} & \multicolumn{2}{c|}{\textbf{\boldmath $\alpha_{\rm pt}$-$p$}} \\ \hline
 & \bf r & \multicolumn{2}{c|}{\textbf{p}} &  \bf r & \multicolumn{2}{c|}{\textbf{p}} & \bf r & \bf p  \\ \hline
\sw{Band-Synchrotron}& 0.96 & \multicolumn{2}{c|}{\textbf{5.14 $\times 10^{-15}$}} & 0.55 & \multicolumn{2}{c|}{\textbf{3.23 $\times 10^{-3}$}} & 0.24 & 2.31 $\times 10^{-1}$  \\\hline
\end{tabular}
\end{table*}
\begin{table}
\centering
\addtolength{\tabcolsep}{14pt}
\caption{The magnitudes of 13 secondary standard stars in the vicinity of \thisgrba, calibrated against the Landolt standard field PG 0231. Observations were obtained using the 4K $\times$ 4K CCD Imager \citep{2022JApA...43...27K}, the first light instrument on the axial port of 3.6m DOT.}
 \begin{tabular}{|cccccc|}
 \hline
 \textbf{ID}&\textbf{U}&\textbf{B}&\textbf{V}&\textbf{R}&\textbf{I}\\
 \hline 
 \hline	
  1 &  17.75$\pm$0.14 &  16.81$\pm$0.02 &  15.63$\pm$0.00 &  14.94$\pm$0.01 &  14.29$\pm$0.04 \\
  2 &  17.58$\pm$0.14 &  17.19$\pm$0.02 &  16.34$\pm$0.01 &  15.82$\pm$0.01 &  15.35$\pm$0.04 \\
  3 &  17.95$\pm$0.14 &  17.72$\pm$0.02 &  17.00$\pm$0.01 &  16.55$\pm$0.01 &  16.14$\pm$0.04 \\
  4 &  18.46$\pm$0.14 &  18.06$\pm$0.02 &  17.19$\pm$0.01 &  16.66$\pm$0.01 &  16.14$\pm$0.04 \\
  5 &  17.50$\pm$0.14 &  17.28$\pm$0.02 &  16.59$\pm$0.01 &  16.15$\pm$0.01 &  15.75$\pm$0.04 \\
  6 &  17.16$\pm$0.14 &  16.79$\pm$0.02 &  16.00$\pm$0.01 &  15.53$\pm$0.01 &  15.10$\pm$0.04 \\
  7 &  18.11$\pm$0.14 &  17.86$\pm$0.02 &  17.06$\pm$0.01 &  16.55$\pm$0.01 &  16.11$\pm$0.04 \\
  8 &  18.30$\pm$0.14 &  17.76$\pm$0.02 &  16.89$\pm$0.01 &  16.34$\pm$0.01 &  15.86$\pm$0.04 \\
  9 &  17.18$\pm$0.14 &  16.77$\pm$0.02 &  15.94$\pm$0.01 &  15.43$\pm$0.01 &  14.98$\pm$0.04 \\
 10 &  17.79$\pm$0.14 &  17.40$\pm$0.02 &  16.54$\pm$0.01 &  16.02$\pm$0.01 &  15.55$\pm$0.04 \\
 11 &  15.90$\pm$0.14 &  15.69$\pm$0.02 &  15.07$\pm$0.00 &  14.69$\pm$0.01 &  14.36$\pm$0.04 \\
 12 &  18.25$\pm$0.14 &  16.60$\pm$0.02 &  15.14$\pm$0.00 &  14.33$\pm$0.01 &  13.54$\pm$0.04 \\
 13 &  17.83$\pm$0.14 &  17.49$\pm$0.02 &  16.63$\pm$0.01 &  16.04$\pm$0.01 &  15.49$\pm$0.04 \\ \hline
 \end{tabular}
 \label{tab:secondary_stars_aa}
\end{table}

\FloatBarrier
\addtolength{\tabcolsep}{8pt}
\begin{longtable}{|c|c|c|c|c|c|c|}
\caption{Optical afterglow observations of \thisgrba. The quoted magnitude values are in the AB system and are not corrected for foreground extinction.} 
\label{table:15A_optical}
\endfirsthead
\caption* {(Continued)} \\
\endhead
\endfoot
\endlastfoot \hline
\bf time (s) & \bf Exp time (s) &\bf filter &\bf Telescope &\bf Magnitude &\bf Reference\\ [1ex] \hline
46.65	&	{20}	&		C	&	FRAM-ORM	&	{17.49	$\pm$	0.19}	&	Present	work		\\
72.57	&	{20}	&		C	&	FRAM-ORM	&	{17.71	$\pm$	0.22}	&	Present	work		\\
100.22	&	{20}	&		C	&	FRAM-ORM	&	{17.61	$\pm$	0.20}	&	Present	work		\\
126.14	&	{20}	&		C	&	FRAM-ORM	&	{17.37	$\pm$	0.18}	&	Present	work		\\
152.92	&	{20}	&		C	&	FRAM-ORM	&	{16.86	$\pm$	0.10}	&	Present	work		\\
178.84	&	{20}	&		C	&	FRAM-ORM	&	{16.73	$\pm$	0.09}	&	Present	work		\\
204.76	&	{20}	&		C	&	FRAM-ORM	&	{16.67	$\pm$	0.09}	&	Present	work		\\
230.68	&	{20}	&		C	&	FRAM-ORM	&	{16.72	$\pm$	0.09}	&	Present	work		\\
256.60	&	{20}	&		C	&	FRAM-ORM	&	{16.73	$\pm$	0.10}	&	Present	work		\\
283.39	&	{20}	&		C	&	FRAM-ORM	&	{16.96	$\pm$	0.12}	&	Present	work		\\
309.31	&	{20}	&		C	&	FRAM-ORM	&	{16.74	$\pm$	0.10}	&	Present	work		\\
335.23	&	{20}	&		C	&	FRAM-ORM	&	{16.63	$\pm$	0.09}	&	Present	work		\\
361.15	&	{20}	&		C	&	FRAM-ORM	&	{16.90	$\pm$	0.11}	&	Present	work		\\
387.07	&	{20}	&		C	&	FRAM-ORM	&	{16.67	$\pm$	0.09}	&	Present	work		\\
412.99	&	{20}	&		C	&	FRAM-ORM	&	{16.79	$\pm$	0.10}	&	Present	work		\\
438.91	&	{20}	&		C	&	FRAM-ORM	&	{16.77	$\pm$	0.11}	&	Present	work		\\
464.83	&	{20}	&		C	&	FRAM-ORM	&	{16.75	$\pm$	0.10}	&	Present	work		\\
490.75	&	{20}	&		C	&	FRAM-ORM	&	{16.93	$\pm$	0.12}	&	Present	work		\\
516.67	&	{20}	&		C	&	FRAM-ORM	&	{17.01	$\pm$	0.13}	&	Present	work		\\
542.59	&	{20}	&		C	&	FRAM-ORM	&	{17.09	$\pm$	0.14}	&	Present	work		\\
605.66	&	{98}	&		C	&	FRAM-ORM	&	{17.40	$\pm$	0.09}	&	Present	work		\\
709.34	&	{98}	&		C	&	FRAM-ORM	&	{17.64	$\pm$	0.09}	&	Present	work		\\
813.88	&	{98}	&		C	&	FRAM-ORM	&	{17.65	$\pm$	0.09}	&	Present	work		\\
917.56	&	{98}	&		C	&	FRAM-ORM	&	{17.95	$\pm$	0.12}	&	Present	work		\\
1022.11	&	{98}	&		C	&	FRAM-ORM	&	{18.19	$\pm$	0.15	}	&	Present	work		\\
40.60	&	11	&		C	&	BOOTES-1B	&	17.49	$\pm$	0.47	&	Present	work	\\		
103.68	&	30	&		C	&	BOOTES-1B	&	17.68	$\pm$	0.52	&	Present	work	\\		
162.43	&	15	&		C	&	BOOTES-1B	&	16.59	$\pm$	0.29	&	Present	work	\\		
194.40	&	10	&		C	&	BOOTES-1B	&	16.41	$\pm$	0.35	&	Present	work	\\		
221.18	&	10	&		C	&	BOOTES-1B	&	16.58	$\pm$	0.38	&	Present	work	\\		
247.10	&	10	&		C	&	BOOTES-1B	&	16.54	$\pm$	0.37	&	Present	work	\\		
273.88	&	10	&		C	&	BOOTES-1B	&	16.20	$\pm$	0.30	&	Present	work	\\		
304.12	&	14	&		C	&	BOOTES-1B	&	17.04	$\pm$	0.49	&	Present	work	\\		
329.18	&	10	&		C	&	BOOTES-1B	&	16.96	$\pm$	0.31	&	Present	work	\\		
340.41	&	10	&		C	&	BOOTES-1B	&	16.69	$\pm$	0.25	&	Present	work	\\		
351.64	&	10	&		C	&	BOOTES-1B	&	16.71	$\pm$	0.25	&	Present	work	\\		
362.88	&	10	&		C	&	BOOTES-1B	&	17.12	$\pm$	0.34	&	Present	work	\\		
374.97	&	10	&		C	&	BOOTES-1B	&	17.09	$\pm$	0.32	&	Present	work	\\		
386.20	&	10	&		C	&	BOOTES-1B	&	16.55	$\pm$	0.23	&	Present	work	\\		
398.30	&	10	&		C	&	BOOTES-1B	&	16.57	$\pm$	0.23	&	Present	work	\\		
408.67	&	10	&		C	&	BOOTES-1B	&	16.97	$\pm$	0.29	&	Present	work	\\		
420.76	&	10	&		C	&	BOOTES-1B	&	16.89	$\pm$	0.28	&	Present	work	\\		
432.00	&	10	&		C	&	BOOTES-1B	&	17.07	$\pm$	0.34	&	Present	work	\\		
444.09	&	10	&		C	&	BOOTES-1B	&	16.78	$\pm$	0.26	&	Present	work	\\		
455.32	&	10	&		C	&	BOOTES-1B	&	17.09	$\pm$	0.33	&	Present	work	\\		
467.42	&	10	&		C	&	BOOTES-1B	&	17.19	$\pm$	0.36	&	Present	work	\\	\hline	
477.79	&	10	&		C	&	BOOTES-1B	&	16.90	$\pm$	0.28	&	Present	work	\\		
500.25	&	30	&		C	&	BOOTES-1B	&	18.20	$\pm$	0.64	&	Present	work	\\		
535.68	&	30	&		C	&	BOOTES-1B	&	17.39	$\pm$	0.30	&	Present	work	\\		
569.37	&	30	&		C	&	BOOTES-1B	&	16.98	$\pm$	0.23	&	Present	work	\\		
604.80	&	30	&		C	&	BOOTES-1B	&	16.99	$\pm$	0.24	&	Present	work	\\		
650.59	&	50	&		C	&	BOOTES-1B	&	17.51	$\pm$	0.29	&	Present	work	\\		
707.61	&	50	&		C	&	BOOTES-1B	&	17.07	$\pm$	0.22	&	Present	work	\\		
764.64	&	50	&		C	&	BOOTES-1B	&	17.50	$\pm$	0.29	&	Present	work	\\		
851.90	&	100	&		C	&	BOOTES-1B	&	17.75	$\pm$	0.27	&	Present	work	\\		
966.81	&	100	&		C	&	BOOTES-1B	&	17.70	$\pm$	0.28	&	Present	work	\\		
1080.86	&	100	&		C	&	BOOTES-1B	&	17.87	$\pm$	0.30	&	Present	work	\\		
1195.77	&	100	&		C	&	BOOTES-1B	&	17.89	$\pm$	0.31	&	Present	work	\\		
1368.57	&	200	&		C	&	BOOTES-1B	&	18.22	$\pm$	0.50	&	Present	work	\\		
1656.28	&	300	&		C	&	BOOTES-1B	&	18.08	$\pm$	0.21	&	Present	work	\\		
2016.57	&	300	&		C	&	BOOTES-1B	&	18.60	$\pm$	0.26	&	Present	work	\\		
2399.32	&	300	&		C	&	BOOTES-1B	&	18.72	$\pm$	0.16	&	Present	work	\\		
2928.09	&	600	&		C	&	BOOTES-1B	&	19.24	$\pm$	0.21	&	Present	work	\\		
3889.72	&	900	&		C	&	BOOTES-1B	&	19.01	$\pm$	0.24	&	Present	work	\\		
5011.20	&	900	&		C	&	BOOTES-1B	&	19.92	$\pm$	0.32	&	Present	work	\\		
6096.38	&	840	&		C	&	BOOTES-1B	&	19.63	$\pm$	0.27	&	Present	work	\\
72316.80	&	{300}	&		I	&	{DOT}	&	{21.88	$\pm$	0.09}	&	Present	work	\\
52099.20	&	{300}	&		R	&	{DOT}	&	{22.14	$\pm$	0.08}	&	Present	work	\\
73008.00	&	{300}	&		R	&	{DOT}	&	{21.94  $\pm$	0.12}	&	Present	work	\\
52790.40	&	{300}	&		V	&	{DOT}	&	{22.70	$\pm$	0.08}	&	Present	work	\\
73267.20	&	{300}	&		V	&	{DOT}	&	{22.63	$\pm$	0.09}	&	Present	work	\\
74217.60	&	{300}	&		B	&	{DOT}	&	{23.31	$\pm$	0.10}	&	Present	work	\\
\hline    
\end{longtable}
\FloatBarrier

\begin{table*}
\begin{center}
\caption{Our optical afterglow observations of \thisgrbb with FRAM-ORM. The quoted magnitude values are in the AB system and are not corrected for foreground extinction.}
\label{table:16C_optical}
\begin{tabular}{|c|c|c|c|c|c|c|c|c|c|c|c|c|}
\hline
\multicolumn{2}{|c|}{\bf time (s)}&\multicolumn{2}{|c|}{\bf Exposure (s)}&\multicolumn{2}{|c|}{\bf Filter}& \multicolumn{2}{|c|}{\bf Telescope}&\multicolumn{2}{|c|}{\bf Magnitude}& \bf Reference\\ [1ex]\hline
\multicolumn{2}{|c|}{47.410} & \multicolumn{2}{|c|}{20}  &  \multicolumn{2}{|c|}{C}  &  \multicolumn{2}{|c|}{FRAM-ORM } &  \multicolumn{2}{|c|}{18.339 $\pm$ 0.295  }& Present work\\
\multicolumn{2}{|c|}{73.770} & \multicolumn{2}{|c|}{20}  &  \multicolumn{2}{|c|}{C}  &  \multicolumn{2}{|c|}{FRAM-ORM } &  \multicolumn{2}{|c|}{$>$ 18.38           }& Present work\\
\multicolumn{2}{|c|}{112.40} & \multicolumn{2}{|c|}{45}  &  \multicolumn{2}{|c|}{C}  &  \multicolumn{2}{|c|}{FRAM-ORM } &  \multicolumn{2}{|c|}{19.101 $\pm$ 0.421  }& Present work\\
\multicolumn{2}{|c|}{154.26} & \multicolumn{2}{|c|}{20}  &  \multicolumn{2}{|c|}{C}  &  \multicolumn{2}{|c|}{FRAM-ORM } &  \multicolumn{2}{|c|}{18.036 $\pm$ 0.251  }& Present work\\
\multicolumn{2}{|c|}{180.23} & \multicolumn{2}{|c|}{20}  &  \multicolumn{2}{|c|}{C}  &  \multicolumn{2}{|c|}{FRAM-ORM } &  \multicolumn{2}{|c|}{18.034 $\pm$ 0.253  }& Present work\\
\multicolumn{2}{|c|}{218.19} & \multicolumn{2}{|c|}{46}  &  \multicolumn{2}{|c|}{C}  &  \multicolumn{2}{|c|}{FRAM-ORM } &  \multicolumn{2}{|c|}{18.269 $\pm$ 0.252  }& Present work\\
\multicolumn{2}{|c|}{293.39} & \multicolumn{2}{|c|}{98}  &  \multicolumn{2}{|c|}{C}  &  \multicolumn{2}{|c|}{FRAM-ORM } &  \multicolumn{2}{|c|}{18.563 $\pm$ 0.247  }& Present work\\
\multicolumn{2}{|c|}{398.52} & \multicolumn{2}{|c|}{98}  &  \multicolumn{2}{|c|}{C}  &  \multicolumn{2}{|c|}{FRAM-ORM } &  \multicolumn{2}{|c|}{18.775 $\pm$ 0.299  }& Present work\\
\multicolumn{2}{|c|}{526.91} & \multicolumn{2}{|c|}{150} &  \multicolumn{2}{|c|}{C}  &  \multicolumn{2}{|c|}{FRAM-ORM } &  \multicolumn{2}{|c|}{18.665 $\pm$ 0.239  }& Present work\\
\multicolumn{2}{|c|}{813.95} & \multicolumn{2}{|c|}{467} &  \multicolumn{2}{|c|}{C}  &  \multicolumn{2}{|c|}{FRAM-ORM } &  \multicolumn{2}{|c|}{19.226 $\pm$ 0.227  }& Present work\\
\multicolumn{2}{|c|}{1643.47}& \multicolumn{2}{|c|}{ 21x60}  &  \multicolumn{2}{|c|}{R}  &  \multicolumn{2}{|c|}{FRAM-ORM } &  \multicolumn{2}{|c|}{19.927 $\pm$ 0.346  }& Present work\\ \hline
\multicolumn{2}{|c|}{7884.00}& \multicolumn{2}{|c|}{ 3x40}  &  \multicolumn{2}{|c|}{r}  &  \multicolumn{2}{|c|}{VLT} &  \multicolumn{2}{|c|}{21.81 $\pm$ 0.05  }& \cite{2020GCN.29066....1I}\\
\hline
\end{tabular}
\end{center}
\end{table*}

\begin{table}
\addtolength{\tabcolsep}{0.001pt}
\tiny
\begin{center}
\caption{Closure relations obtained from the best fit values of spectral and temporal indices for \thisgrba and \thisgrbb using \swift-XRT and our optical observations from FRAM-ORM, BOOTES, and 3.6m DOT.}
\label{tab:closure_relations}
\begin{tabular}{|c c c c c c c c|}\hline
\multicolumn{8}{|c|}{\textbf{\thisgrba}} \\[1ex]
\bf Time interval (s)&\boldmath $\alpha_{o}$&\boldmath $\alpha_x$&\boldmath $\beta_o$&\boldmath $\beta_x$&\boldmath$p$ & \bf Spectral regime& \bf Medium\\[1ex]\hline
3300 - 4800 &$- 0.92_{-0.09}^{+0.08}$&$-2.36_{-0.26}^{+0.17}$&-&-1.07$^{+0.31}_{-0.30}$&2.72$\pm$0.44 & $\nu_m$ $<$ $\nu_o$ $<$ $\nu_c$ $<$ $\nu_x$ &ISM\\[1ex]
(1.0 - 180) $\times 10^{4}$ &$- 0.92_{-0.09}^{+0.08}$&$-2.36_{-0.26}^{+0.17}$&$-1.13_{-0.30}^{+0.30}$&-1.38$^{+0.49}_{-0.47}$&2.93$\pm$0.38& $\nu_m$ $<$ $\nu_o$ $<$ $\nu_c$ $<$ $\nu_x$&ISM\\[1ex]\hline
\multicolumn{8}{|c|}{\textbf{\thisgrbb}} \\[1ex]
\bf Time interval (s)&\boldmath $\alpha_{o}$&\boldmath $\alpha_x$&\boldmath $\beta_o$&\boldmath $\beta_x$&\boldmath$p$ & \bf Spectral regime& \bf Medium\\[1ex]\hline
2900 - 17000&$-1.05^{+0.11}_{-0.10}$&$-2.21_{-0.11}^{+0.10}$&-&-0.97$^{+0.05}_{-0.05}$&2.42$\pm$0.48 &$\nu_m$ $<$  $\nu_o$ $<$ $\nu_c$ $<$  $\nu_x$ &wind\\ [1ex]\hline
\end{tabular}
\end{center}
\end{table}

\FloatBarrier

\end{document}